\documentclass[11pt]{article}
\usepackage{jheppub}
\usepackage{slashed,dsfont,bm}
\usepackage{mathtools}

\edef\restoreparindent{\parindent=\the\parindent\relax}
\usepackage{parskip}
\restoreparindent
\DeclarePairedDelimiter{\ceil}{\lceil}{\rceil}
\DeclarePairedDelimiter{\floor}{\lfloor}{\rfloor}

\usepackage{tabularx}
\newcolumntype{Y}{>{\centering\arraybackslash}X}

\usepackage{booktabs}
\usepackage{multirow}

\usepackage[whole]{bxcjkjatype} 

\usepackage[usenames,dvipsnames,svgnames,table]{xcolor}

\usepackage{tikz}
\usetikzlibrary{decorations.pathmorphing,decorations.markings,positioning}
\tikzset{snake it/.style={decorate, decoration=snake}}
\usetikzlibrary{decorations,decorations.pathmorphing}
\usetikzlibrary{patterns, shapes, arrows.meta}

\usepackage{colortbl}

\def\tr{\mathrm{tr}}

\def\d{\mathrm{d}}
\def\CA{\mathcal{A}}

\def\CM{\mathcal{M}}
\def\CO{\mathcal{O}}
\def\i{\textrm{i}}
\newcommand{\dd}{\mathrm{d}}
\def\SO{\mathrm{SO}}
\def\BHS{\mathbb{HS}}
\def\BH{\mathbb{H}}
\def\BR{\mathbb{R}}
\def\BS{\mathbb{S}}
\def\BZ{\mathbb{Z}}
\def\CD{\mathcal{D}}

\title{
Free energy and defect $C$-theorem in free scalar theory
}

\author[a]{Tatsuma Nishioka}
\author[b]{and Yoshiki Sato}

\affiliation[a]{
Yukawa Institute for Theoretical Physics, Kyoto University,\\
Kitashirakawa Oiwakecho, Sakyo-ku, Kyoto 606-8502, Japan
}
\affiliation[b]{Physics Division, National Center for Theoretical Sciences, National Tsing-Hua University, \\
Hsinchu 30013, Taiwan}

\abstract{
We describe conformal defects of $p$ dimensions in a free scalar theory on a $d$-dimensional flat space as boundary conditions on the conformally flat space $\BH^{p+1}\times \BS^{d-p-1}$.
We classify two types of boundary conditions, Dirichlet type and Neumann type, on the boundary of the subspace $\BH^{p+1}$ which correspond to the types of conformal defects in the free scalar theory.
We find Dirichlet boundary conditions always exist while Neumann boundary conditions are allowed only for defects of lower codimensions.
Our results match with a recent classification of the non-monodromy defects, showing Neumann boundary conditions are associated with non-trivial defects.
We check this observation by calculating the difference of the free energies on $\BH^{p+1}\times \BS^{d-p-1}$ between Dirichlet and Neumann boundary conditions.
We also examine the defect RG flows from Neumann to Dirichlet boundary conditions and provide more support for a conjectured $C$-theorem in defect CFTs.
}

\preprint{YITP-21-04}

\begin{document}

\maketitle
\pagenumbering{roman}

\pagenumbering{arabic}
\section{Introduction}
A conventional view of quantum field theories (QFTs) relies on particle picture of quantum fields as a fundamental description of the theories, but the importance of non-local objects has been increasingly recognized in recent studies to discriminate theories from those having the same local descriptions but  different global structures \cite{Kapustin:2014gua,Gaiotto:2014kfa}.
Extended observables such as Wilson-'t Hooft loops rarely have concrete realizations in terms of fundamental fields in Lagrangian and are typically defined as boundary conditions, hence called defects in general.
Various types of defects are pervasive in physics: 
loop and surface operators in condensed matter and high energy physics, cosmic string and domain wall in cosmology, D-branes in string theory, to name just a few. (Refer to \cite{Andrei:2018die} for recent progress in diverse fields.)

Theoretical aspects of defects are less scrutinized as opposed to local operators due to the lack of their fundamental descriptions in QFT as well as their intricate dependence on the shapes.
On the other hand, focusing on a class of defects with a large amount of symmetry we have a better understanding of their universal characters.
In particular the kinematics of planar and spherical defects (conformal defects) in conformal field theories (CFTs) are highly constrained by a large subgroup of conformal group \cite{McAvity:1995zd,Liendo:2012hy,Billo:2016cpy,Gadde:2016fbj,Fukuda:2017cup,Kobayashi:2018lil,Guha:2018snh,Isachenkov:2018pef,Lauria:2018klo,Herzog:2020bqw}.
To be concrete let $\CD^{(p)}$ be a $p$-dimensional conformal defect in Euclidean CFT on $\BR^d$.
The conformal group $\SO (1, d+1)$ is broken by the presence of the defect to the subgroup $\SO (1, p+1) \times \SO(q)$ where $\SO (1, p+1)$ is the conformal group on the $p$-dimensional worldvolume of $\CD^{(p)}$ while $\SO(q)$ is the rotation group around $\CD^{(p)}$ with $q = d - p$.
As a special case CFTs on a manifold with boundary (BCFTs) are also regarded as defect CFTs (DCFTs) with $q=1$.

It should be noted that there are two kinds of local operators in DCFT: local operators $\CO$ in the bulk CFT and defect local operators $\hat\CO$ with support only on the worldvolume of $\CD^{(p)}$.
This is easily seen in a simple example of DCFT consisting of a bulk CFT$_d$ and a lower-dimensional CFT$_p$ without interaction in between.
The relation between $\CO$ and $\hat \CO$ is determined by the bulk-to-defect operator product expansion, which thereby defines possible types of defects in a given bulk CFT.

CFTs occupy distinguished positions as fixed points of renormalization group (RG) flows where scale invariance is believed to enhance to conformal invariance \cite{Nakayama:2013is}.
One can perturb a CFT by a relevant operator $\CO$ and let it flow to another CFT at the IR fixed point.
RG flows can be geometrized by adding a ``height" function which measures the degrees of freedom of the theories on a space of QFTs.
Then theories are expected to flow only from the UV to the IR.
A $C$-theorem elevates this belief to the statement for the existence of such a monotonic function known as a $C$-function.
Zamolodchikov proved the $c$-theorem for the first time in two dimensions \cite{Zamolodchikov:1986gt}, which was generalized to the
$a$-theorem as a conjecture in four dimensions \cite{Cardy:1988cwa} and proved more recently \cite{Komargodski:2011vj}.
These theorems state the type $A$ central charges for the conformal anomalies play the role of a $C$-function in even dimensions.
On the other hand, the $F$-theorem asserts that the sphere free energy be a $C$-function in odd dimensions \cite{Jafferis:2011zi,Klebanov:2011gs,Myers:2010tj,Myers:2010xs}.
Despite the difference of the structures in even and odd dimensions, 
the dimensional dependence of the $C$-functions is beautifully unified by the generalized $F$-theorem \cite{Giombi:2014xxa} that proposes an interpolating function between the type $A$ anomaly and sphere free energy
\begin{align}\label{interpolating_F}
    \tilde F \equiv \sin \left( \frac{\pi\,d}{2}\right)\, \log\,Z[\BS^d] \ ,
\end{align}
decreases along any RG flow:
\begin{align}\label{generalized_F}
    \tilde F_\text{UV} \ge \tilde F_\text{IR} \ .
\end{align}
An information theoretic proof of the theorem was given by \cite{Casini:2004bw,Casini:2012ei,Casini:2017vbe} for $d\le 4$, but at the moment of writing it remains open whether the generalized $F$-theorem holds in higher dimensions.

Now for DCFTs one can trigger an RG flow by perturbing the theory using a defect local operator $\hat \CO$ in addition to $\CO$, thus DCFTs allow for a wider class of deformation than CFTs without defect.
If one is concerned with the dynamics of defect operators it will be convenient to focus on defect RG flows triggered by defect localized operators while keeping a bulk CFT fixed.
Given the success of the $C$-theorems in CFTs it is tempting to ask if there exists a monotonic function which decreases along any defect RG flow in DCFTs.
A few concrete proposals were put forwarded in the case of BCFTs and named the $g$-theorems which employ either the boundary entropy \cite{Affleck:1991tk} or the hemisphere free energy \cite{Nozaki:2012qd,Gaiotto:2014gha} as a $C$-function.
In BCFT$_2$, the $g$-theorem was given two proofs: one by a field theoretic method \cite{Friedan:2003yc} and the other by an information-theoretic method \cite{Casini:2016fgb}.
In BCFT$_3$ two proofs are given by \cite{Jensen:2015swa} and \cite{Casini:2018nym} with different means, showing the boundary central charge $b$ for conformal anomaly becomes a $C$-function.
In higher-dimensional BCFTs, there are no general proofs of the $g$-theorem, but some holographic calculations based on a probe brane model \cite{Yamaguchi:2002pa}, the AdS/BCFT construction \cite{Takayanagi:2011zk,Fujita:2011fp,Miao:2017gyt}
and supergravity solutions \cite{Estes:2014hka}, support the validity of the proposal.\footnote{In contrast, the boundary free energy does not necessarily decrease under a bulk RG flow \cite{Green:2007wr,Sato:2020upl}.
}
The situation is less clear for general DCFTs, but when $p=2$ (and for any $d$) the $b$-theorem \cite{Jensen:2015swa} states that the central charge of surface operators is shown to be a $C$-function. (See also \cite{Jensen:2018rxu,Estes:2018tnu,Rodgers:2018mvq,Chalabi:2020iie,Wang:2020xkc} for further investigations.)

In the previous work \cite{Kobayashi:2018lil}, we proposed a $C$-theorem in DCFT stating that the defect free energy on the sphere $\BS^d$ defined by an increment of the sphere free energy due to the defect
\begin{align}
    \log\, \langle \mathcal{D}^{(p)} \rangle  \equiv \log Z^{\text{DCFT}}[\BS^d] - \log Z^{\text{CFT}}[\BS^d] \ ,
\end{align}
is a $C$-function.
More precisely, we introduced an interpolating function of the defect free energy in an analogous way to \eqref{interpolating_F} by
\begin{align}
    \tilde{D} = \sin \left( \frac{\pi\, p}{2}  \right) \log |\langle \mathcal{D}^{(p)} \rangle| \ ,
\end{align}
and conjectured $\tilde D$ decreases under any defect RG flow:
\begin{align}\label{D_theorem}
    \tilde{D}_{\text{UV}} \geq \tilde{D}_{\text{IR}} \ .
\end{align}
When defects do not interact with the bulk theory our proposal simply reduces to the generalized $F$-theorem \eqref{generalized_F} for the $p$-dimensional defect theories while it incorporates the $g$-theorem for BCFTs when $p=d-1$.
The relation \eqref{D_theorem} passes several checks for defects in field theories \cite{Kobayashi:2018lil,Beccaria:2017rbe,Estes:2018tnu} and holographic models \cite{Kumar:2016jxy,Kumar:2017vjv,Rodgers:2018mvq}.

The main purpose of this paper is to examine the conjectured relation \eqref{D_theorem} in more detail in the simplest theory: a free conformally coupled scalar field.
To this end we will describe conformal defects in the theory on flat space as boundary conditions on a conformally equivalent space, where a defect RG flow is caused by changing the boundary condition by the double trace deformation as is familiar in the AdS/CFT setup \cite{Witten:2001ua,Berkooz:2002ug,Gubser:2002zh,Gubser:2002vv,Hartman:2006dy,Diaz:2007an,Giombi:2013yva}.
The idea of mapping BCFTs on flat space to the hyperbolic space $\BH^d$ and studying the boundary RG flow has appeared in the recent works \cite{Herzog:2019bom,Giombi:2020rmc}.
Similarly, line operators in four dimensions (i.e., $p=1$ and $q=3$) can also be characterized as boundary conditions on $\BH^2\times \BS^2$ \cite{Kapustin:2005py}, and monopole operators which are codimension three defects can be characterized as boundary conditions on $\BH^{d-2}\times \BS^2$ \cite{Chester:2015wao}.
We will extend these ideas to more general DCFTs by employing a conformal map from flat space to $\BH^{p+1}\times \BS^{q-1}$ where defects are located at the boundary of $\BH^{p+1}$ (see figure \ref{fig:conf_map}).

We will introduce Neumann type boundary conditions on $\BH^{p+1}\times \BS^{q-1}$ and 
consider the defect RG flow from Neumann to Dirichlet type.
For BCFT on the hemisphere $\BHS^d$, Neumann and  Dirichlet boundary conditions can be realized by imposing parity conditions on the eigenmodes.
Then, the free energies for each boundary condition can be obtained from those on the sphere by truncation (see e.g.\,\cite{Gaiotto:2014gha,Jensen:2015swa} for the detail).
On the other hand, the Neumann/Dirichlet boundary condition on $\BH^d$, which is conformally equivalent to $\BHS^d$ though, cannot be described by a parity condition as the spectrum of the eigenfunctions is continuous.
The boundary conditions on $\BH^d$ are rather dictated by the asymptotic behavior of the field near the boundary as in the AdS/CFT.
This approach has an advantage 
that we can view the defect theory as a ``holographic" dual of the bulk field on the Euclidean AdS space $\BH^d$, which allows us to classify types of conformal defects through the boundary conditions on $\BH^{p+1}\times \BS^{q-1}$ in the conformally coupled free scalar theory.
We will show it is always possible to impose Dirichlet boundary conditions for any $p$ and $q$ while Neumann boundary conditions are allowed only for special cases if we require the defect theory to be unitary.
Reassuringly our results conform with the classification of the non-monodromy defects for a free massless scalar theory carried out in \cite{Lauria:2020emq} by other means. 
It leads us to speculate that Dirichlet boundary condition corresponds to trivial (or no) defects while Neumann boundary condition to non-trivial defects.

The free energy of a conformally coupled scalar field on $\BH^{p+1} \times \BS^{q-1}$ has been calculated in literature \cite{Mann:1997hm,Solodukhin:2010pk,Klebanov:2011gs,Rodriguez-Gomez:2017kxf,Rodriguez-Gomez:2017aca,Belin:2013uta} (see also \cite{Lewkowycz:2012qr,Gustavsson:2019zwm} for a related work) 
and shown to have a logarithmic divergence:
\begin{align}
    F[\BH^{p+1} \times \BS^{q-1}] = \cdots - A[\BH^{p+1} \times \BS^{q-1}]\, \log \left( \frac{R}{\epsilon}\right) + \cdots\ ,
\end{align}
where $R$ is the radius of the hyperbolic space and the sphere. 
The small parameter $\epsilon$ serves as a UV cutoff for the sphere as well as an IR cutoff for the hyperbolic space.
The explicit values of the coefficients $A[\BH^{p+1} \times \BS^{q-1}]$ are obtained for some $p$ and $q$ either using the heat kernel method or by summing over eigenvalues.
It is however not straightforward to apply one of these methods to the cases when both $p$ and $q$ are odd \cite{Rodriguez-Gomez:2017kxf}.\footnote{When both $p$ and $q$ are odd, it is expected that only bulk anomaly exists and the bulk anomaly is the same as that of $\BS^{p+q}$.
However, it is technically difficult to confirm this expectation.}
To overcome this difficulty, we will use the zeta function regularization throughout this paper and complete the calculation of the free energy on $\BH^{p+1} \times \BS^{q-1}$.
As we will see in the main text, our approach is not only applicable to any $p$ and $q$, but also makes it easy to compare the free energies on conformally equivalent spaces.
For instance, we will check the equality between the universal parts of the free energies on $\BH^d$ and $\BHS^d$ for Dirichlet boundary condition conjectured by \cite{Rodriguez-Gomez:2017aca} by combining numerical and analytic ways.\footnote{It is not clearly specified what type of boundary conditions is imposed on the hyperbolic space in \cite{Rodriguez-Gomez:2017aca}, but their boundary condition is of Dirichlet type in our terminology.}
We will also verify a few other relations for the free energies and prove or conjecture new ones which will be summarized below.
Another advantage of our approach than the other methods is to make manifest the difference between the anomalies from the bulk theory and defect.
Actually, there are two sources of logarithmic divergences: one from the bulk anomaly when $d=p+q$ even and the other from the defect anomaly when $p$ even.
In our approach, the bulk anomaly depends on the cutoff introduced for the zeta regularization while the defect anomaly depends on another cutoff that arises from the renormalized volume of the hyperbolic space $\BH^{p+1}$ for even $p$.

We will leverage our results to test if the conjectured relation \eqref{D_theorem} holds for the defect RG flow from Neumann to Dirichlet when the former is allowed.
Strictly speaking, we will not directly check our proposal that employs the defect free energy on $\BS^d$ as a $C$-function.
Instead we assume that the difference of the free energy is invariant under the conformal map from $\BS^d$ to $\BH^{p+1}\times \BS^{q-1}$.\footnote{When defects have conformal anomaly the free energy may not be invariant, but when defects are spherical the anomaly is of type $A$ which depends only on the Euler characteristic of the worldvolumes.
In our setup, defects on $\BS^d$ and $\BH^d$ are always spherical, so should have the same anomaly.
When there are no defect anomalies, we need not be worried about this issue as the free energy is invariant by definition.}
We will calculate the difference of the free energies on $\BH^{p+1}\times \BS^{q-1}$ between Neumann and Dirichlet boundary conditions in two ways: the residue method \cite{Diaz:2007an,Giombi:2013yva} and analytic continuation method.
We find both methods give the same result consistent with the defect $C$-theorem \eqref{D_theorem}.

The organization of this paper is as follows.
In section \ref{sec2}, we review several useful coordinates for DCFT and conformal maps among them.
Furthermore, we discuss boundary conditions of Dirichlet type and Neumann type for a conformally coupled scalar field on  $\BH^{p+1} \times \BS^{q-1}$.
The Neumann type boundary conditions fall into two classes, free boundary condition and the other.
We show that the Dirichlet boundary condition always exists but the Neumann boundary condition exists only in $q=1,2,3,4$ while the free boundary condition appears when $q=p+2$ for $q\ge 3$.
Section \ref{sec3} begins as a warm-up with the calculation of the free energy on $\mathbb{S}^d$ and $\mathbb{HS}^d$.
The purpose of this section is twofold: to illustrate the zeta regularization method and to provide analytic results for the boundary free energy on $\BHS^d$ in arbitrary dimensions.
In section \ref{sec4}, we proceed to compute the free energies on $\BH^d$ and $\BH^{p+1}\times \BS^{q-1}$ with the Dirichlet boundary conditions.
Along the way we find various identities for the free energies on the conformally equivalent spaces.
In section \ref{sec5}, we calculate the difference of the free energies between the Neumann and Dirichlet boundary conditions on $\BH^{p+1} \times \BS^{q-1}$ in two ways and confirm \eqref{D_theorem} holds for all the cases.
Finally section \ref{ss:discussion} is devoted to discussion and future directions.
Appendices include the lists of the free energies on $\BS^d, \BHS^d, \BH^d$ and $\BH^{p+1}\times \BS^{q-1}$ obtained in the main text, various formulas and technical details of some calculations.

\subsection{Summary of the paper}

Since the body of the paper is rather lengthy and technical, in what follows we will summarize the main results.

In section \ref{sec2} we classify the boundary condition of a conformally coupled scalar field theory on $\BH^{p+1} \times \BS^{q-1}$, which preserves a defect conformal symmetry.
Neumann boundary condition is allowed only for $q=1,2,3,4$ while free boundary condition exists when $q=p+2$ for $q\ge 3$.
Our result matches the classification of non-monodromy defects in a free scalar theory given by \cite{Lauria:2020emq} by other means.

In section \ref{sec3}, using the zeta-function regularization, we compute the free energies on $\BS^d$ and $\BHS^d$. 
        \begin{itemize}
            \item For $\BS^d$, the renormalized free energy takes the following form:
            \begin{align}
            \begin{aligned}
                F_\text{ren} [\BS^{d}] = \begin{dcases}
                      F_{\text{fin}} [\BS^{d}] & \quad d: \text{odd}\ ,\\  
                      -A [\BS^{d}]\, \log (\Lambda R) + F_{\text{fin}} [\BS^{d}] & \quad d: \text{even}\ ,
                \end{dcases} 
            \end{aligned}
            \end{align}
            where $\Lambda$ is a UV cutoff introduced in the zeta  regularization.
            We reproduce known anomaly coefficients $A[\BS^{d}] $ for even $d$ and known universal finite terms $F_{\text{fin}} [\BS^{d}]$ for odd $d$ \cite{Quine1996,Kumagai1999,Cappelli:2000fe,Klebanov:2011gs,Giombi:2014xxa}.
            \item For $\BHS^d$, we find the renormalized free energy takes the form:
        \begin{align}
            \begin{aligned}
                F_\text{ren} [\BHS^{d}] = \begin{dcases}
                      -A [\BHS^{d}]\, \log (\Lambda R) + F_{\text{fin}} [\BHS^{d}] & \quad d: \text{odd} \ , \\  
                      -\frac{1}{2}\,A [\BS^{d}]\, \log (\Lambda R) + F_{\text{fin}} [\BHS^{d}] & \quad d: \text{even} \ .
                \end{dcases} 
            \end{aligned}
            \end{align}
            After subtracting half of the free energy on $\BS^d$, we obtain the boundary free energy with the Dirichlet or Neumann boundary condition. This reproduces known results in  \cite{Jensen:2015swa,Nozaki:2012qd,Gaiotto:2014gha,Dowker:2014rva,Rodriguez-Gomez:2017aca}.
        \end{itemize}

In section \ref{sec4} we examine the case for $\BH^d$ and $\BH^{p+1} \times \BS^{q-1}$ with the Dirichlet boundary condition for the hyperbolic space in the zeta regularization.
\begin{itemize}
    \item For $\BH^d$, we find the renormalized free energy takes the following form:
            \begin{align}
            \begin{aligned}
                F_\text{ren} [\BH^{d}] = \begin{dcases}
                     - \mathcal{A} [\BH^{d} ]\, \log \left( \frac{R}{\epsilon}\right) & \quad d: \text{odd} \ ,\\ 
                      -A [\BH^{d} ]\, \log (\Lambda R) + F_{\text{fin}} [\BH^{d}]  & \quad d: \text{even} \ ,
                \end{dcases} 
            \end{aligned}
            \end{align}
where $\log (R/\epsilon)$ arises from the regularized volume of $\BH^{d}$ and only appears for odd $d$.
We obtain the universal parts of the free energy and reproduce known results in  \cite{Giombi:2008vd,Rodriguez-Gomez:2017kxf,Rodriguez-Gomez:2017aca,Bytsenko:1995ak}.
We confirm the equivalence of the free energies between $\BH^d$ with the Dirichlet boundary condition and  $\BHS^d$ with the Dirichlet boundary condition:
\begin{align}
            \begin{aligned}
                 \mathcal{A}[\BH^{d} ] &= A [\BHS^{d} ]  &  & \quad d: \text{odd} \\ 
                F_\text{ren} [\BH^{d}] &= F_\text{ren} [\BHS^{d}]  & & \quad d: \text{even} 
            \end{aligned}
\end{align}
(We also confirm similar results hold for the Neumann boundary conditions.)
    \item For $\BH^{p+1} \times \BS^{q-1}$ we find the renormalized free energy takes the form:
            \begin{align}\label{fren_general_form}
            \begin{aligned}
                F_\text{ren} [\BH^{p+1} \times \BS^{q-1}] = \begin{dcases}
                      -\mathcal{A} [\BH^{p+1} \times \BS^{q-1}]\, \log \left( \frac{R}{\epsilon}\right) & \quad p: \text{even} \ ,\\  
                      -A [\BH^{p+1} \times \BS^{q-1}]\, \log (\Lambda R) + F_{\text{fin}} [\BH^{p+1} \times \BS^{q-1}] & \quad p: \text{odd} \ .
                \end{dcases} 
            \end{aligned}
            \end{align}
We obtain the following results:
\begin{enumerate}
    \item For even $p$ and even $q$, $\mathcal{A}[\BH^{p+1} \times \BS^{q-1}] \neq 0$, indicating the presence of defect anomaly.
    \item For even $p$ and $q=p+2$, we numerically check
    \begin{align}
        A[\BS^{2p+2}] = 2\, \mathcal{A}[\BH^{p+1} \times \BS^{p+1}] \ .
    \end{align}
    \item For even $p$ and odd $q$, $\mathcal{A}[\BH^{p+1} \times \BS^{q-1}] =0$ or equivalently $F_\text{ren}[\BH^{p+1} \times \BS^{q-1}]= 0$.
    \item For odd $p$, we numerically verify the relation:\footnote{For $\BS^1$, we can compute the both sides analytically but we can not prove the equality for arbitrary $d$.}
    \begin{align}
        F_\text{ren}[\BS^d] = F_\text{ren} [\BH^{2k} \times \BS^{d-2k}] \ .
    \end{align}

\end{enumerate}
These relations were conjectured in \cite{Rodriguez-Gomez:2017kxf} from the calculations in free scalar and holographic theories.
Our results provide more evidence for their conjectures at least in the conformally coupled free scalar in arbitrary dimensions.
In particular, we perform systematic computations in the zeta regularization including the cases with odd $p$ and odd $q$ which were missing in \cite{Rodriguez-Gomez:2017kxf} due to some technical difficulties.
   
These results lead us to speculate that $A[\BH^{p+1} \times \BS^{q-1}]$ is associated with the bulk anomaly while $\mathcal{A}[\BH^{p+1} \times \BS^{q-1}]$ is with the defect anomaly.
\end{itemize}

In section \ref{sec5} we obtain the free energies with the Neumann boundary condition for the hyperbolic space using two different methods: (1) an analytic continuation, and (2) the residue method which is conjectured in \cite{Giombi:2013yva}.
We give a proof of the conjecture and apply it to $\BH^{p+1} \times \BS^{q-1}$.
We confirm that the interpolated defect free energy $\tilde D$ with the Dirichlet boundary condition is always smaller than that with Neumann boundary condition, and our results are consistent with the conjectured $C$-theorem \eqref{D_theorem} in DCFT.
Specifically, the difference of the free energies on $ \BH^{p+1} \times \BS^{q-1}$ between the two boundary conditions equals the free energy on $\BS^p$ for $q=2,4$ or the difference of the free energies on $ \BH^{p+1}$ between the two boundary conditions for $q=3$, and the monotonicity follows from the positivity of the interpolated sphere free energy $\tilde F$ or the monotonicity of the free energy on $\BH^{p+1}$.

\section{Classification of boundary conditions}
\label{sec2}

We first review coordinate systems and conformal maps between them which are suitable for describing conformal defects in DCFTs.
We then proceed to classify conformal boundary conditions for a conformally coupled free scalar field, and show that they correspond to a classification of conformal (non-monodromy) defects in the same theory. 

\subsection{Coordinate systems for DCFT}
\begin{figure}[!h]
\centering
\begin{tikzpicture}

\draw[fill=gray!20] (-9.5, -2) --  (-9.5, 1) -- (-8.5, 2) -- (-8.5, -1) --cycle;
\node at (-10, 0) {$\CD^{(p)}$};
\node at (-9, -2.5) {$\BR^d$};

\node at (-6.5, 0) {$\xrightarrow{ \text{\large conformal map}}$};

\node[draw, ellipse, fill=gray!20, minimum width=1cm, minimum height=3cm, outer sep=0, label=180:$\CD^{(p)}$] (ell) at (-3,0) {};

\draw (ell.85) to[out=-45,in=175] (0,0.2)
    to[out=-45, in=45] (0,-0.2)
    to[out=-175, in=45] (ell.-85);

\node at (0.6, 0) {\Large $\times$};
\draw (2.7,0) circle (1.5cm);
\draw[dashed] (2.7,0) ellipse (1.5cm and 0.5cm);

\node at (-1.5,-2.5) {$\BH^{p+1}$};
\node at (2.7,-2.5) {$\BS^{q-1}$};

\end{tikzpicture}
\caption{Conformal map from flat space $\BR^d$ with a $p$-dimensional planar defect to $\BH^{p+1}\times \BS^{q-1}$.}
\label{fig:conf_map}
\end{figure}
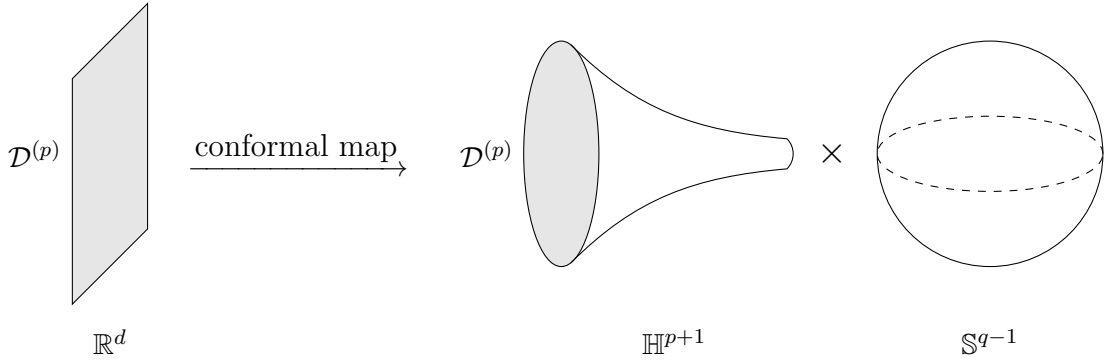

Let us consider DCFT$_d$ on flat space with the metric
\begin{align}
    \dd s^2 = \dd x_a^2 + \dd y_i^2 \,, \qquad (a=1,\cdots, p,\, i = p+1,\cdots,d) \ ,
\end{align}
where a $p$-dimensional defect sits at the origin $y_i = 0$ in the transverse directions.
For later convenience we introduce $q=d-p$, which represents a codimension of the defect.
By using the polar coordinate for the $y_i$-coordinates, 
\begin{align}
    \dd y_i^2 = \dd z^2 + z^2 \dd s_{\BS^{q-1}}^2 \ ,
\end{align}
with the metric $\dd s_{\BS^{q-1}}^2$ for a unit $(q-1)$-sphere, the flat space metric becomes
\begin{align}\label{Flat_defect_map}
    \dd s^2 
     = z^2 \left( \frac{\dd x_a^2 + \dd z^2}{z^2} + \dd s_{\BS^{q-1}}^2 \right) \ .
\end{align}
By a Weyl transformation, the above metric reduces to the geometry $\BH^{p+1} \times \BS^{q-1}$ with radius $R$,
\begin{align} \label{metric_poincare}
    \dd s^2 = R^2 \left( \frac{\dd x_a^2 + \dd z^2}{z^2} +  \dd s_{\BS^{q-1}}^2 \right) \ .
\end{align}
Now the defect is located at the boundary of the hyperbolic space.
We can also use the global coordinate for the hyperbolic space part:
\begin{align} \label{metric_hyperbolic}
   \dd s^2 = R^2 \left( \dd \rho^2 +\sinh^2 \rho \, \dd s_{\BS^p}^2 + \, \dd s_{\BS^{q-1}}^2 \right) \ ,
\end{align}
where the defect becomes a $p$-sphere at $\rho=\infty$.
Introducing a new variable $\varphi$ by $\tan \varphi = \sinh \rho$, the metric \eqref{metric_hyperbolic} becomes
\begin{align} 
    \dd s^2 = \frac{R^2}{\cos^2 \varphi} \left( \dd \varphi ^2 +\sin^2 \varphi \, \dd s_{\BS^p}^2+ \cos ^2 \varphi \, \dd s_{\BS^{q-1}}^2 \right) \ ,
\end{align}
which the metric can be mapped by a further Weyl transformation to the $d$-sphere metric :
\begin{align} \label{metric_sphere}
    \dd s^2 = R^2 \left( \dd \varphi ^2 +\sin^2 \varphi \, \dd s_{\BS^p}^2+ \cos ^2 \varphi \, \dd s_{\BS^{q-1}}^2 \right) \ ,
\end{align}
where the defect is mapped to a $p$-sphere at $\varphi = \pi/2$.
See figure \ref{fig:conf_map} for the illustration of the resulting conformal map.

It will also be convenient to introduce the standard representation of the sphere metric,
\begin{align} \label{metric_sphere_standard}
    \dd s^2 = R^2 \left( \dd \varphi ^2 +\sin^2 \varphi \, \dd s_{\BS^{d-1}}^2 \right) \ ,
\end{align}
where $0\leq \varphi <\pi$ for the sphere and  $0\leq \varphi \leq \pi/2$ for the hemisphere.

\subsection{Conformally coupled scalar field}
Next, let us consider a conformally coupled real scalar field on \eqref{metric_poincare} (or \eqref{metric_hyperbolic}).
The action is given by 
\begin{align}
    I = - \frac{1}{2} \int \! \dd^d x \, \sqrt{g}\left[ (\partial_\mu \phi)^2 + \xi\, \mathcal{R}\, \phi^2 \right]  \ ,
\end{align}
with the parameter $\xi$ and the Ricci scalar $\mathcal{R}$:
\begin{align}
    \xi = \frac{d-2}{4(d-1)} \ , \qquad \mathcal{R}= \frac{(q-1)(q-2)- p(p+1)}{R^2} \ .
\end{align}
Now we would like to investigate the boundary condition for the scalar field near the boundary, $z=0$ (or $\rho=\infty$), of the hyperbolic space.
For this purpose, we decompose the scalar field into  eigenfunctions by the spherical harmonics on $\BS^{q-1}$:
\begin{align}
    \phi (z,x,\theta) = \sum_{\ell} \phi_{\BH^{p+1}} (z,x)\, Y_{\ell,\BS^{q-1}} (\theta) \ ,
\end{align}
where $(z,x)$ are the coordinates of the hyperbolic space in Poincar\'{e} coordinate and $\theta$ stands for those of the sphere $\BS^{q-1}$.
The spherical harmonics $Y_{\ell,\BS^{q-1}} (\theta)$ satisfies the equation:
\begin{align} \label{laplace_eq_sphere}
    -\nabla_{\BS^{q-1}}^2 Y_{\ell,\BS^{q-1}} (\theta) = \frac{\ell (\ell +q-2)}{R^2}\, Y_{\ell,\BS^{q-1}} (\theta) \ .
\end{align}
Here $\ell$ is an integer whose range is from $-\infty$ to $\infty$ for $q=2$ and from $0$ to $\infty$ for $q\geq 3$.

With this decomposition, the equation of motion of the scalar field on $\BH^{p+1}\times \BS^{q-1}$
\begin{align}
    \left( -\nabla_{\BH^{p+1}}^2 - \nabla_{\BS^{q-1}}^2 + \xi\, \mathcal{R}\right) \phi (z,x,\theta) = 0 \ ,
\end{align}
reduces to the equation of motion of a massive scalar field on $\BH^{p+1}$:
\begin{align}
    \left( -\nabla_{\BH^{p+1}}^2 + M^2 \right) \phi_{\BH^{p+1}} (z,x) = 0\ ,
\end{align}
with the mass given by
\begin{align}
    M^2 R^2= \ell (\ell +q-2) + \frac{(q-2)^2 -p^2}{4} \ .
\end{align}
Then the solution to the equation of motion behaves as 
\begin{align}
    \phi_{\BH^{p+1}} \sim z^{\Delta_\pm^\ell} \ ,
\end{align}
near the boundary, $z=0$, as is well known in the AdS/CFT correspondence.
Here $\Delta_\pm^\ell$ are the roots of the equation:
\begin{align}\label{Delta_equation}
    \Delta (\Delta - p ) = M^2 R^2 \ ,
\end{align}
and are explicitly given by 
\begin{align}\label{defect_op_dimension}
    \Delta_\pm^\ell = 
    \begin{dcases}
            \frac{p}{2} \pm |\ell|  &\qquad (q= 2) \ , \\
            \frac{p}{2} \pm \left( \ell + \frac{q-2}{2}\right) &\qquad (q>2) \ .\\
        \end{dcases} 
\end{align}
For $q=1$, the spherical part does not exist, so there is no $\ell$-dependence in $\Delta_\pm$:
\begin{align}\label{Delta_mass}
    \Delta_\pm = 
    \frac{p}{2} \pm \frac{1}{2} \ , \qquad (q=1) \ .
\end{align}

While we are only concerned with a QFT of a scalar field on $\BH^{p+1}\times \BS^{q-1}$, it may be viewed as a bulk system in a holographic setup as shown by the above consideration.
The parameters $\Delta_\pm^\ell$ can be understood as the conformal dimensions of operators localizing on a $p$-dimensional conformal defect at the boundary of $\BH^{p+1}$.
(See also section 2.3 in \cite{Dorn:2003au} for a related discussion.)
Then not all the operators with dimensions \eqref{defect_op_dimension} (or \eqref{Delta_mass} for $q=1$) are allowed to exist due to the unitarity bound in $p$ dimensions:\footnote{Restricting to normalizable boundary conditions on the hyperbolic space $\BH^{p+1}$ the mass of the scalar field is subject to the so-called Breitenlohner-Freedman (BF) bound:
\begin{align}
    M^2 R^2 \ge - \frac{p^2}{4} \ .
\end{align}
When this condition is met there are two real solutions to \eqref{Delta_equation}, $\Delta_\pm$.
While the solution with the larger root $\Delta_+$ is always square-integrable, the solution with the smaller root $\Delta_-$ is not necessarily so with respect to the Klein-Gordon inner product.
Thus, requiring the square integrability leads to the bound for $\Delta_-$:
\begin{align}
    \Delta_- \ge \frac{p}{2} - 1 \ .
\end{align}
From the viewpoint of the AdS/CFT correspondence, this matches with the unitarity bound for scalar primary operators in a $p$-dimensional CFT.}
\begin{align}\label{unitarity_bound}
\begin{aligned}
            &\Delta_\pm^\ell\geq \frac{p}{2} - 1\ ,  &&\qquad (p> 2)\ , \\
            &\Delta_\pm^\ell > 0 \ , &&\qquad (p\le 2)\ .     
\end{aligned}
\end{align}
It follows that $\Delta_+^\ell$ is always above the bound, while $\Delta_-^\ell$ is not necessarily so unless
\begin{align}\label{bound_ell}
    \begin{aligned}
    &|\ell| \le 1\ , &\quad &(q=2) \ , \\
    &\ell \leq 2- \frac{q}{2} \ , &\qquad &(q>2) \ .
    \end{aligned}
\end{align}
Hence the modes with small $\ell$ are allowed to have sensible boundary conditions corresponding to $\Delta_-^\ell$.
For clarity we define the Dirichlet and Neumann boundary conditions for $q\ge 2$ as follows:
\begin{align}
    \begin{aligned}
        \text{Dirichlet b.\,c.}:& \quad \Delta_{\text{D}} = \Delta_+^\ell \quad \text{for all }\ell\ , \\
        \text{Neumann b.\,c.}:& \quad \Delta_\text{N} = 
        \begin{dcases}
        \Delta_-^\ell > 0 \quad &\text{for some } \ell \ , \\
        \Delta_+^{\ell} \quad &\text{otherwise} \ ,
        \end{dcases} 
    \end{aligned}
\end{align}
in accordance with the case for $q=1$ where $\Delta_\text{D} = \Delta_+$ and $\Delta_\text{N} = \Delta_-$.

In addition to them there are boundary conditions with a constant solution (zero mode) on $\BH^{p+1}$:\footnote{These modes should be treated with case as they are the source of the IR divergence in the free energy.}
\begin{align}
    \phi_{\BH^{p+1}} \sim \text{const} \ ,
\end{align}
which corresponds to defect operators of dimension $\Delta_-^\ell = 0$.
Among them is the special boundary condition $\Delta^{\ell = 0}_- = 0$ associated with the excitation of the identity operator on the defect.\footnote{The defect identity operators are taken into account in the v2 of \cite{Lauria:2020emq}.}
We call them ``free" boundary conditions following \cite{Kapustin:2005py}:
\begin{align}
        \begin{aligned}
        \text{Free b.\,c.}:& \quad \Delta_\text{F} = 
        \begin{dcases}
        \Delta_-^\ell = 0 \quad & \ell = 0 \ ,\\
        \Delta_+^{\ell} \quad &\ell \neq 0 \ ,
        \end{dcases} 
    \end{aligned}
\end{align}
for $q\ge 2$, and $\Delta_\text{F} = \Delta_- = 0$ for $p=q=1$.
On the other hand, the zero modes with $\ell\neq 0$ are termed charged dimension zero operators and excluded in \cite{Lauria:2020emq} on the basis of the cluster decomposition which assures the dimension zero mode must be the defect identity operator.
Thus, we will also take into account the free boundary conditions while excluding the charged zero modes $\Delta_-^{\ell\neq 0}=0$ from the classification.

It follows from \eqref{defect_op_dimension} and \eqref{Delta_mass} that the free boundary conditions are allowed only when $p=q=1$ and $q=p+2~(p\ge 1)$.
From the viewpoint of conformal defects, the free boundary condition on $\BH^{p+1}\times \BS^{q-1}$ is associated with a $p$-dimensional scalar Wilson surface in $d=2p+2$ dimension:
\begin{align}
    W_{\Sigma_p} = \mathrm{e}^{g\,\int_{\Sigma_p}\,\phi} \ ,
\end{align}
where $g$ is a dimensionless coupling, $\Sigma_p$ the worldvolume of a $p$-dimensional surface and $\phi$ a bulk scalar field of dimension $p$.

Having this caveat in mind, we obtain the classification of the boundary conditions for $\Delta_-^\ell$:
\paragraph{$q=1$ case:}
   It follows from \eqref{Delta_mass} that there exists the Neumann boundary condition for $p\ge 2$ and the free boundary condition for $p=1$.
\paragraph{$q=2$ case:}
    The bound \eqref{bound_ell} becomes $\ell \le 1$, so the Neumann boundary conditions with $\ell= 0,\pm 1$ are allowed to exist.
    The $\ell =0$ mode, however, does not give a new boundary condition as $\Delta_+^{\ell=0} = \Delta_-^{\ell=0} = \frac{p}{2}$.
    Note that $\Delta_-^{\ell=\pm 1}= \frac{p}{2} - 1$ saturate the unitarity bound for $p>2$.
    In \cite{Lauria:2020emq} it is argued that the $\ell=\pm 1$ modes do not give rise to nontrivial boundary conditions as the defect operators saturating the unitarity bound become free, and are excluded from the spectrum.
    On the other hand, we will keep them as nontrivial Neumann boundary conditions for completeness in the latter sections.

    \noindent
    Thus, there are two types of Neumann boundary conditions for $p>2$:
\begin{align} \label{bc_q=2}
\begin{aligned}
    \Delta_{\text{N}1} & = 
    \begin{dcases}
    \Delta_+^\ell & \text{for }\ell \neq 1 \ ,\\
    \Delta_-^\ell & \text{for }\ell = 1 \ ,
    \end{dcases} \\
    \Delta_{\text{N}2} & = 
    \begin{dcases}
    \Delta_+^\ell & \text{for }\ell \neq \pm 1 \ , \\
    \Delta_-^\ell & \text{for }\ell = \pm 1 \ .
    \end{dcases}
\end{aligned}
\end{align}
The mode with $(\Delta_+^{\ell\neq -1},\Delta_-^{\ell=-1})$ is essentially the same as the $\Delta_{\text{N}1}$ boundary condition because we can change the label of $\ell$ without changing physics.
\paragraph{$q=3$ case:}
    Only the $\ell=0$ mode satisfies the unitarity bound and gives us a nontrivial Neumann boundary condition with $\Delta_-^{\ell=0}= \frac{p-1}{2}$ for $p\ge 2$.
    Hence there is only one type of the Neumann boundary condition:
    \begin{align} \label{bc_q=3,4}
        \Delta_\text{N} = \begin{dcases}
                \Delta_+^\ell &\text{for } \ell \geq 1 \ , \\
                \Delta_-^\ell &\text{for } \ell = 0 \ .
            \end{dcases}
    \end{align}
    The free boundary condition can be imposed only when $p=1$, which describes a scalar Wilson loop in the four-dimensional free scalar field theory \cite{Kapustin:2005py}.

\paragraph{$q=4$ case:}
    Only the $\ell=0$ mode is allowed, resulting in the Neumann boundary condition with $\Delta^{\ell=0}_-= \frac{p}{2} - 1$ for $p\ge 3$ saturating the unitarity bound \eqref{unitarity_bound}.
    Thus, the boundary condition \eqref{bc_q=3,4} can be imposed.
    In \cite{Lauria:2020emq} these boundary conditions are attributed to ``trivial" ones as they saturate the unitarity bound and do not have interesting dynamics and excluded from the classification, but
    here we include them for completeness.
    The free boundary condition is allowed when $p=2$, which corresponds to a scalar Wilson surface in the six-dimensional free scalar field theory (see also \cite{Gustavsson:2004gj}).

\paragraph{$q\ge 5$ case:}
In this case, there are no Neumann boundary conditions satisfying the unitarity bound \eqref{unitarity_bound}, but
there still exists the free boundary condition when $q=p+2$ associated with a $p$-dimensional scalar Wilson surface in $d=2p+2$ dimensions.

\bigskip

Our results are consistent with the classification of the non-monodromy defects in a free scalar theory in \cite{Lauria:2020emq}, which are summarized in table \ref{tab:classification}.

\begin{table}[ht]
    \centering
    \begin{tabularx}{\linewidth}{cYYYYYYY}
        \toprule
          & $q=1$ & $q=2$ & $ q=3$ & $q=4$ & $q=5$ & $q=6$ & $\cdots$ \\ \hline
         $p=1$ &  $\Delta_\text{D}/\Delta_\text{F}$ & $\Delta_\text{D}$ & $\Delta_\text{D}/\Delta_\text{F}$ & $\Delta_\text{D}$ & $\Delta_\text{D}$ & $\Delta_\text{D}$ & $\cdots$ \\
         $p=2$ & \cellcolor{gray!10} $\Delta_\text{D}/\Delta_\text{N}$ & $\Delta_\text{D}$ & \cellcolor{gray!10} $\Delta_\text{D}/\Delta_-^{\ell=0}$ & $\Delta_\text{D}/\Delta_\text{F}$ & $\Delta_\text{D}$ & $\Delta_\text{D}$ & \\
         $p=3$ & \cellcolor{gray!10} $\Delta_\text{D}/\Delta_\text{N}$ & \cellcolor{gray!10} $\Delta_\text{D}/\Delta_-^{\ell=\pm 1}$ & \cellcolor{gray!10} $\Delta_\text{D}/\Delta_-^{\ell=0}$ & \cellcolor{gray!10} $\Delta_\text{D}/\Delta_-^{\ell=0}$ & $\Delta_\text{D}/\Delta_\text{F}$ & $\Delta_\text{D}$ & $\cdots$ \\
         $p=4$ & \cellcolor{gray!10} $\Delta_\text{D}/\Delta_\text{N}$ & \cellcolor{gray!10} $\Delta_\text{D}/\Delta_-^{\ell=\pm 1}$ & \cellcolor{gray!10} $\Delta_\text{D}/\Delta_-^{\ell=0}$ & \cellcolor{gray!10} $\Delta_\text{D}/\Delta_-^{\ell=0}$ & $\Delta_\text{D}$ & $\Delta_\text{D}/\Delta_\text{F}$ & \\
         $p=5$ & \cellcolor{gray!10} $\Delta_\text{D}/\Delta_\text{N}$ & \cellcolor{gray!10} $\Delta_\text{D}/\Delta_-^{\ell=\pm 1}$ & \cellcolor{gray!10} $\Delta_\text{D}/\Delta_-^{\ell=0}$ & \cellcolor{gray!10} $\Delta_\text{D}/\Delta_-^{\ell=0}$ & $\Delta_\text{D}$ & $\Delta_\text{D}$ & \\
         $\vdots$ & \cellcolor{gray!10} $\vdots$ & \cellcolor{gray!10}  &\cellcolor{gray!10}  & \cellcolor{gray!10}  $\vdots$ & & & $\ddots$ \\
         \bottomrule
    \end{tabularx}
    \caption{Classification of the allowed boundary conditions in the free scalar theory.
    The Neumann boundary conditions 
    exist in the shaded cells and the allowed modes differ from the Dirichlet ones are shown in the right side.
    Our table is the same as the classification of the non-monodromy defects in \cite{Lauria:2020emq} except that ours has additional column for $q=1$ and boundary conditions $\Delta_-^{\ell =\pm 1}$ for $q=2$ and $p\ge 3$, and $\Delta_-^{\ell =0}$ for $q=4$ and $p\ge 3$ saturating the unitarity bound.
    The free boundary condition appears when $p=q=1$ and $q=p+2$.
    }
    \label{tab:classification}
\end{table}

\section{Free energy on hemisphere}
\label{sec3}

The aim of this section is to demonstrate the zeta function regularization through the calculation of free energies on the sphere $\BS^d$ and the hemisphere $\BHS^d$.
They have been extensively studied in literature in various methods (see e.g., \cite{Camporesi:1990wm,Bytsenko:1994bc} for reviews), and we do not claim any originality of our results except for giving their explicit expressions.
The main results are \eqref{free_energy_finite_sphere_odd}, \eqref{sphere_anom_coeff} and \eqref{sp_even_finite} for $\BS^d$, and \eqref{anomaly_hemi_odd}, \eqref{finite_hemi_odd}, \eqref{anomaly_hemi_even}, \eqref{fin_hemi_even} for $\BHS^d$.

\subsection{Free energy on \texorpdfstring{$\BS^d$}{Sd}}

Let us first consider the free energy on $\BS^d$ as a warm-up.
For a conformally coupled scalar on $\BS^d$, 
the free energy is given by\footnote{\label{footnote_ambiguity}When we decompose $\log \left[( (\nu_\ell^{(d)})^2 -1/4)/(\tilde \Lambda R)^2\right]$ into the two logarithmic functions in the third line, there is an ambiguity,
\begin{align}
    \log \left( \frac{\left(\nu_\ell^{(d)}\right)^2 -1/4}{(\tilde \Lambda R)^2} \right) = \log \left( \mathrm{e}^{\i \, \theta} \frac{\nu_\ell^{(d)} +1/2}{ (\tilde \Lambda R)^{1-\rho}} \right) + \log \left( \mathrm{e}^{-\i \, \theta} \frac{\nu_\ell^{(d)} -1/2}{(\tilde \Lambda R)^{1+\rho}} \right) \ ,
\end{align}
where $\rho$ is a real number and $0 \leq \theta < 2\pi$ (see also \cite{Monin:2016bwf}).
It leads to the ambiguities of the anomaly term and the finite term of the free energy.
We fix the ambiguity of the phase by demanding a good convergence in the Schwinger representation of the free energy \eqref{schwinger_rep} at large $\ell$. 
The ambiguity of the scale can also be fixed by requiring that the zeta function be independent of the parameter $\tilde \Lambda R$.}
\begin{align} \label{free_energy_sphere_def}
    \begin{aligned}
        F[\BS^d] 
            &= \frac{1}{2}\, \tr \log\left[\tilde\Lambda^{-2}\, \left( -\nabla_{\BS^{d}}^2 
            +  \frac{d(d-2)}{4R^2}  \right)\right] \\
            &= \frac{1}{2}\, \tr \log\left[\tilde\Lambda^{-2}\, \left( \frac{\ell (\ell + d-1)}{R^2}
            + \frac{d(d-2)}{4R^2}
            \right) \right] \\
            &=  \frac{1}{2} \sum_{\ell =0}^\infty g^{(d)}(\ell) \, \left[ \log \left( \frac{\nu_\ell^{(d)} + \frac{1}{2}}{\tilde\Lambda R}\right) + \log \left( \frac{\nu_\ell^{(d)} - \frac{1}{2}}{\tilde\Lambda R}\right) \right]\ ,
    \end{aligned}
\end{align}
where $R$ is the radius of $\BS^d$ and $\tilde\Lambda$ is the UV cutoff scale introduced to make the integral dimensionless.
The degeneracy $g^{(d)}(\ell)$ and the parameter $\nu_\ell^{(d)}$ are defined by
\begin{align} \label{degeneracy}
        g^{(d)}(\ell)  = (2\ell + d-1)\,\frac{\Gamma(\ell+ d-1)}{\Gamma(d)\,\Gamma(\ell+1)} \ , \qquad
        \nu_\ell^{(d)} = \ell + \frac{d-1}{2} \ .
\end{align}
One can rewrite the free energy \eqref{free_energy_sphere_def} in the Schwinger representation:
\begin{align} \label{schwinger_rep}
    F[\BS^d] = - \frac{1}{2}\,\int_0^\infty\,\frac{\dd t}{t}\,\sum_{\ell =0}^\infty\, g^{(d)}(\ell)\, \left[ \mathrm{e}^{-t \left(\nu_\ell^{(d)} + \frac{1}{2} \right)/(\tilde\Lambda R)} + \mathrm{e}^{-t \left(\nu_\ell^{(d)} - \frac{1}{2} \right)/(\tilde\Lambda R)} \right]\ .
\end{align}
This is divergent, implying the UV divergence of the free energy.
To make the integral finite we introduce the regularized free energy \cite{Vassilevich:2003xt}:
\begin{align}
    \begin{aligned}
        F_{s}[\BS^d] 
            &=
            - \frac{1}{2}\,\int_0^\infty\,\frac{\dd t}{t^{1-s}}\,\sum_{\ell =0}^\infty \,g^{(d)}(\ell) \, \left[ \mathrm{e}^{-t \left(\nu_\ell^{(d)} + \frac{1}{2} \right)/(\tilde\Lambda R)} + \mathrm{e}^{-t \left(\nu_\ell^{(d)} - \frac{1}{2} \right)/(\tilde\Lambda R)} \right]\\
            &=
            - \frac{1}{2}\, (\tilde\Lambda R)^{s}\,\Gamma(s)\, \zeta_{\BS^d} (s) \ ,
    \end{aligned}
\end{align}
where the zeta function $\zeta_{\BS^d} (s)$ is defined by
\begin{align}\label{zeta_sphere}
    \zeta_{\BS^d}(s) \equiv \sum_{\ell =0}^\infty \,g^{(d)}(\ell)\,\left[ \left( \nu_\ell^{(d)} + \frac{1}{2}\right)^{-s} + \left( \nu_\ell^{(d)} - \frac{1}{2}\right)^{-s} \right] \ .
\end{align}
Then the (unrenormarized) free energy is obtained in the $s\to 0$ limit:
\begin{align}
    F_s[\BS^d] = - \frac{1}{2}\,\left( \frac{1}{s} - \gamma_\text{E} + \log(\tilde\Lambda R)\right)\, \zeta_{\BS^d}(0) - \frac{1}{2}\,\partial_s \zeta_{\BS^d}(0) + O(s) \ ,
\end{align}
which is divergent due to the pole at $s=0$.
After removing the pole term, the remaining part becomes the renormalized free energy
\begin{align}\label{Fren_sphere}
    F_\text{ren}[\BS^d] \equiv - \frac{1}{2}\,\zeta_{\BS^d}(0)\,\log(\Lambda R) - \frac{1}{2}\,\partial_s \zeta_{\BS^d}(0)   \ ,
\end{align}
where $\Lambda = \mathrm{e}^{-\gamma_\text{E}}\,\tilde\Lambda$.

In calculating the zeta function \eqref{zeta_sphere}, we find it convenient to expand the degeneracy as:
\begin{align}\label{degen_expand}
    g^{(d)}(\ell) = \frac{2}{\Gamma(d)}\,\sum_{k=0}^{d-1}\,\gamma_{k,d}(c)\,\left(\nu_\ell^{(d)} + c\right)^{k} \ .
\end{align}
To fix $\gamma_{k,d}(c)$ we introduce coefficients $\alpha_{n,d}$ and $\beta_{n,d}$ as follows \cite{Camporesi:1990wm}:
\begin{align}\label{degen_expand_alphabeta}
    g^{(d)}(\ell)
    =
    \begin{dcases}
        \frac{2}{\Gamma(d)}\,\prod_{j=0}^{\frac{d-3}{2}}\left[\left(\nu_\ell^{(d)}\right)^2 - j^2\right]
        = \frac{2}{\Gamma(d)}\,\sum_{n=0}^{\frac{d-1}{2}}(-1)^{\frac{d-1}{2}+n}\,\alpha_{n, d}\,\left(\nu_\ell^{(d)}\right)^{2n} & d: \text{odd} \ ,  \\
        \frac{2\nu_\ell^{(d)}}{\Gamma(d)}\,\prod_{j=\frac{1}{2}}^{\frac{d-3}{2}}\left[\left(\nu_\ell^{(d)}\right)^2 - j^2\right]
        = \frac{2}{\Gamma(d)}\,\sum_{n=0}^{\frac{d}{2}-1}(-1)^{\frac{d}{2}-1+n}\,\beta_{n, d}\,\left(\nu_\ell^{(d)}\right)^{2n+1} & d: \text{even} \ .
    \end{dcases}
\end{align}
Note that we use a slightly different notation from \cite{Camporesi:1990wm} and include the $n=0$ contribution in odd $d$ although $\alpha_{0,d}=0$ for some convenience.
Comparing \eqref{degen_expand} with \eqref{degen_expand_alphabeta} we find
\begin{align}\label{gamma_exp}
    \gamma_{k,d}(c)
        &=
        \begin{dcases}
            \sum_{n = \ceil{\frac{k}{2}}}^{\frac{d-1}{2}}
            (-1)^{\frac{d-1}{2}+n+k}\,\binom{2n}{k}
            \,\alpha_{n,d}\,c^{2n-k} & d:\text{odd} \ , \\
            \sum_{n = \floor{\frac{k}{2}}}^{\frac{d}{2}-1}
            (-1)^{\frac{d}{2}+n+k}\,\binom{2n+1}{k}
            \,\beta_{n,d}\,c^{2n+1-k} & d:\text{even} \ .
        \end{dcases}
\end{align}
With this expansion we can perform the summation over $\ell$ in \eqref{zeta_sphere} and obtain a summation of the Hurwitz zeta functions:
\begin{align}\label{zeta_sphere_rewrite}
    \zeta_{\BS^d}(s) =    
        \frac{2}{\Gamma(d)}\,\sum_{k=0}^{d-1}\,
        \left[ 
            \gamma_{k,d}\left(\frac{1}{2}\right)\,\zeta_\text{H}\left( s - k, \frac{d}{2}\right) 
            + 
            \gamma_{k,d}\left(-\frac{1}{2}\right)\,\zeta_\text{H}\left( s - k, \frac{d}{2}-1\right)
        \right] \ .
\end{align}
It remains to determine the coefficients $\alpha_{n,d}$ and $\beta_{n,d}$ to calculate the renormalized free energy.
We fix them by comparing the two representations, \eqref{degeneracy} and \eqref{degen_expand_alphabeta}, of $g^{(d)}(\ell)$.
Using the asymptotic expansion in \cite[5.11.14]{NIST:DLMF},
\begin{align}
    \frac{\Gamma(x + a)}{\Gamma(x + b)} = \sum_{k=0}^\infty\,\left(x + \frac{a + b - 1}{2}\right)^{a-b-2k}\,\binom{a-b}{2k}\,B_{2k}^{(a-b+1)}\left(\frac{a-b+1}{2}\right) \ ,
\end{align}
where $B_k^{(m)}(x)$ is the generalized Bernoulli polynomial which reduces to the Bernoulli polynomial $B_k(x) = B^{(1)}_k(x)$ when $m=1$, and comparing both sides, we find
\begin{align}\label{coeff_alpha}
    \alpha_{n,d} = (-1)^{\frac{d-1}{2}+n}\, \binom{d-2}{d-1-2n}\, B^{(d-1)}_{d-1-2n} \left( \frac{d-1}{2}\right) \ ,
\end{align}
for odd $d$ and
\begin{align}\label{coeff_beta}
    \beta_{n, d} = (-1)^{\frac{d-2}{2} + n}\, \binom{d-2}{d-2 - 2n}\,B^{(d-1)}_{d-2-2n} \left( \frac{d-1}{2}\right) \ ,
\end{align}
for even $d$.

\subsubsection{Odd \texorpdfstring{$d$}{d}}
\label{sec3.1.1}

When $d$ is odd the zeta function \eqref{zeta_sphere_rewrite} reduces to
\begin{align}
\begin{aligned}
    \zeta_{\BS^d} (s) &=  \frac{2}{\Gamma (d)} \sum_{k=0}^{d-1} \sum_{n=\ceil{\frac{k}{2}}}^{\frac{d-1}{2}} (-1)^{\frac{d-1}{2}+n+k} \alpha_{n,d}\, \binom{2n}{k}\, 2^{k-2n} \, 
    \left[ \zeta_\text{H} \left(s-k,\frac{d}{2} \right) + (-1)^{k}\,\zeta_\text{H} \left(s-k,\frac{d}{2}-1 \right) \right]\ .
    \end{aligned}
\end{align}
Using the identity \eqref{hurwitz_identity_1} for the Hurwitz zeta functions the terms in the bracket become
\begin{align}
    \begin{aligned}
    & \zeta_\text{H} \left(s-k,\frac{d}{2} \right) + (-1)^{k}\,\zeta_\text{H} \left(s-k,\frac{d}{2}-1 \right) \\
    & \quad = \zeta_\text{H} \left(s-k,\frac{1}{2} \right) + (-1)^{k}\,\zeta_\text{H} \left(s-k,\frac{1}{2} \right)
    -
    \sum_{m=0}^{\floor{\frac{d}{2}} -1}
    \left( m+\frac{1}{2}\right)^{k-s} 
    - 
    \sum_{m=0}^{\floor{\frac{d}{2}}-2}\,(-1)^k\,\left( m+\frac{1}{2}\right)^{k-s} \ .
    \end{aligned}
\end{align}
Rearranging the summations $\sum_{k=0}^{d-1} \sum_{n=\ceil{\frac{k}{2}}}^{\frac{d-1}{2}} =  \sum_{n=0}^{\frac{d-1}{2}}\sum_{k=0}^{2n}$ 
the zeta function becomes
\begin{align}
\begin{aligned}
    \zeta_{\BS^d} (s) 
        &= 
        \frac{2}{\Gamma (d)} \sum_{n=0}^{\frac{d-1}{2}}\sum_{k=0}^{2n} (-1)^{\frac{d-1}{2} + n+k} \, \alpha_{n,d}\,\binom{2n}{k}\, 2^{k-2n}  (1 + (-1)^{k})\,\zeta_\text{H} \left(s-k,\frac{1}{2} \right)  \\
        & \qquad - \frac{2}{\Gamma (d)} \sum_{m=1}^{\floor{\frac{d}{2}}-1} \left[\left( m+\frac{1}{2}\right)^{-s}+\left( m-\frac{1}{2}\right)^{-s}\right]\, \sum_{n=0}^{\frac{d-1}{2}}  (-1)^{\frac{d-1}{2} + n} \, \alpha_{n,d} \, m^{2n} \\
        & = 
        \frac{4}{\Gamma (d)} \sum_{n=0}^{\frac{d-1}{2}}\sum_{l=0}^{n} (-1)^{\frac{d-1}{2} + n} \, \alpha_{n,d}\,\binom{2n}{2l}\,   2^{-2n} (2^{s}-2^{2l}) \, \zeta (s-2l)
    \end{aligned}
\end{align}
where we used the identity \eqref{hurwitz_to_zeta_half} 
and removed the summation over $m$ by resorting to the definition \eqref{degen_expand_alphabeta} of $\alpha_{n,d}$ in the third equality.

Taking the $s\to 0$ limit we find $\zeta_{\BS^d}(0) = 0$ as  $(1-2^{2k}) \, \zeta (-2k)=0$ holds for a non-negative integer $k$.
Thus, there is no conformal anomaly in the free energy and only the universal finite part remains in the free energy \eqref{Fren_sphere}:
\begin{align} \label{free_energy_finite_sphere_odd}
    \begin{aligned}
    F_\text{ren} [\BS^d]
        &= F_\text{fin} [\BS^d] \\
        & = \frac{\Gamma (\frac{d}{2})}{2\,\Gamma (d)\, \Gamma (2-\frac{d}{2})}  \log 2  \\
        & \qquad + 
        \sum_{k=1}^{\frac{d-1}{2}} \left(\sum_{n=k}^{\frac{d-1}{2}}
        (-1)^{\frac{d+1}{2} + n + k} \, \alpha_{n,d}\,  (1 - 2^{2k}) \frac{(2n-2k+1)_{2k}}{2^{2k+2n}\,\pi^{2k}\,\Gamma(d)}\right) \zeta (2k+1) \ ,
    \end{aligned}
\end{align}
where we used \eqref{zeta_values} and \eqref{hurwitz_der_to_zeta}.
We also used the Pochhammer symbol $(n)_k \equiv \Gamma(n+k)/\Gamma(n)$ to simplify the expression.
The explicit values of $F_\text{fin}[\BS^d]$ up to $d=9$ are shown in table \ref{tab:Fin_p_q} in appendix \ref{app_table}.

\subsubsection{Even \texorpdfstring{$d$}{d}}
\label{sec3.1.2}

Performing a similar reduction for odd $d$ using the identity \eqref{hurwitz_identity_1} with $a=1$, the zeta function \eqref{zeta_sphere_rewrite} for even $d$ can be written as
\begin{align}
    \begin{aligned}
        \zeta_{\BS^d}(s)
            &=
            \frac{2}{\Gamma(d)}\,\sum_{k=0}^{d-1}\,\sum_{n = \floor{\frac{k}{2}}}^{\frac{d}{2}-1} \,(-1)^{\frac{d}{2} +n+k }\,2^{k-2n-1}\,\binom{2n+1}{k}\,
            \beta_{n,d} \\
            &\qquad \cdot
            \left[ (1+ (-1)^{k-1})\, \zeta \left(s - k \right) 
            + (-1)^{k-1}\, \delta_{d,2}\left(\zeta_\text{H} (s-k,0) - \zeta(s-k)\right)
            \right] \ .
    \end{aligned}
\end{align}
In contrary to the odd-dimensional case, 
there is a logarithmic divergent term associated with the conformal anomaly in the free energy \eqref{Fren_sphere}:
\begin{align}
    F_\text{ren}[\BS^d] 
        = 
         - A[\BS^d]\,\log(\Lambda R) + F_\text{fin}[\BS^d] \ ,
\end{align}
where the anomaly coefficient $A[\BS^d]$ can be read off from \eqref{zeta_sphere_rewrite} as\footnote{We used the Taylor expansion of the Bernoulli polynomials $B_n (x+y) = \sum_{k=0}^n \binom{n}{k} B_k (x)\, y^{n-k}$.}
\begin{align}\label{sphere_anom_coeff}
    A[\BS^d] 
        = \frac{1}{\Gamma(d)}\,\sum_{n = 0}^{\frac{d}{2}-1}\, \beta_{n,d}\, \frac{(-1)^{\frac{d}{2} +n}}{n+1} \left( B_{2n+2} \left( \frac{1}{2}\right) + \frac{2n+1}{2^{2n+2}}
        \right) + \frac{1}{2}\,\delta_{d,2}
\end{align}
while the universal term becomes
\begin{align}\label{sp_even_finite}
    \begin{aligned}
    F_\text{fin}[\BS^d] 
        &=
        \frac{1}{\Gamma(d)}\,\sum_{n = 0}^{\frac{d}{2}-1} \,\sum_{m=0}^{n}\,(-1)^{\frac{d}{2} +n }\,2^{2m-2n+1}\,\binom{2n+1}{2m+1}\,
            \beta_{n,d}\,\zeta' \left( - 2m-1 \right)\\
        &\qquad
        -
        \delta_{d,2}\,\sum_{k=0}^1\,2^{k-1}\,\left(\partial_s\zeta_\text{H}(-k,0) - \zeta'(-k)\right)
        \ .
    \end{aligned}
\end{align}
For $d=2$, $\partial_s\zeta_\text{H}(0,0)$ is ill-defined, which
reflects the IR divergence due to the zero mode.

Tables \ref{tab:F_p_q} and \ref{tab:Fin_p_q} in appendix \ref{app_table} show the anomaly coefficients and the finite parts of the free energies on $\BS^d$ for even $d$ up to $d=10$.\footnote{When we compare our results with \eqref{free_energy_sphere}, the divergent factor, $\left(\sin \left( \frac{\pi d}{2}\right)\right)^{-1}$, should be replaced with the logarithmic term according to \eqref{replacement}.
}
Our result \eqref{sphere_anom_coeff} correctly reproduces the conformal anomaly of the free scalar theory obtained in literature (e.g.\,\cite{Quine1996,Kumagai1999,Cappelli:2000fe,Klebanov:2011gs,Giombi:2014xxa,Giombi:2014iua}).
The finite part is less known as it depends on the regularization scheme (i.e., the choice of the UV cutoff $\Lambda$) when there exists a conformal anomaly. 
When $d=4$ \eqref{sp_even_finite} agrees with the result in \cite{Gaiotto:2014gha} which uses the same zeta regularization as ours.

\subsubsection{Interpolating \texorpdfstring{$a$}{a} and \texorpdfstring{$F$}{F}}

The finite parts of the free energy \eqref{free_energy_finite_sphere_odd} for odd $d$ and the anomaly parts of the free energy \eqref{sphere_anom_coeff} for even $d$ are universal in the sense that they are independent of the cutoff choice.
Thus it will be convenient to introduce the ``universal" free energy:
\begin{align} \label{free_energy_univ}
    F_{\text{univ}}[\BS^{d}]  =
    \begin{dcases}
            F_\text{fin}[\BS^{d}] &\qquad d: \text{odd} \ ,\\
            -A [\BS^{d}]\, \log \left( \frac{R}{\epsilon}\right)  &\qquad d: \text{even} \ ,
    \end{dcases}
\end{align}
where we use $\epsilon$ for the cutoff instead of $\Lambda$.
While the structure of the universal free energy appears to depend on the dimensionality it is shown in \cite{Giombi:2014xxa} to have an integral representation which smoothly interpolates between even and odd $d$:
\begin{align} \label{free_energy_sphere}
    F_{\text{univ}}[\BS^{d}] = -\frac{1}{\sin \left(\frac{\pi d}{2}\right)\Gamma (d+1)} \int_{0}^{1} \! \dd \nu \, \nu\, \sin (\pi \nu)\, \Gamma \left( \frac{d}{2}\pm  \nu \right) \ . 
\end{align}
Here we used the shorthand notation $\Gamma(x \pm y) \equiv \Gamma(x + y)\,\Gamma(x-y)$.
The prefactor is finite for odd $d$, but divergent for even $d$ due to the pole from the zeros of the sine function.
This divergence may be replaced with the logarithmic divergence by introducing a small cutoff parameter $\epsilon$:
\begin{align} 
    -\frac{1}{\sin\left(\frac{\pi d}{2}\right)} =
    \begin{dcases}
            (-1)^\frac{d+1}{2}  &\qquad d: \text{odd} \ , \\
            (-1)^\frac{d}{2}\, \frac{2}{\pi} \,\log \left( \frac{R}{\epsilon}\right) &\qquad d: \text{even} \ .
    \end{dcases}
\end{align}
See \cite{Dowker:2012rp,Dowker:2017cqe} for a proof of the equivalence of the two expressions
\eqref{free_energy_univ} and \eqref{free_energy_sphere}.\footnote{We thank J.\,S.\,Dowker for drawing our attention to these works.}

\subsection{Free energy on \texorpdfstring{$\BHS^d$}{HSd}}

The free energies on hemisphere are obtained in \cite{Jensen:2015swa,Gaiotto:2014gha,Dowker:2014rva,Rodriguez-Gomez:2017aca,Dowker:2010qy}. 
Here we extend them to higher dimensions by using the zeta function regularization.
See also \cite{Herzog:2019bom} for a related work.

In the coordinate system \eqref{metric_sphere_standard}, the Dirichlet boundary condition is given by\footnote{To derive the boundary condition from the action, one needs to add a boundary term to the action. See Appendix C in \cite{Jensen:2015swa}. }
\begin{align}
    \phi \left( \frac{\pi}{2} \right) = 0 \ , 
\end{align}
and the Neumann boundary condition is given by 
\begin{align}
    \partial_\varphi \phi \left( \frac{\pi}{2} \right) = 0 \ .
\end{align}
If we put the theory on the sphere \eqref{metric_sphere_standard} with the defect at $\varphi = \pi/2$, the Dirichlet boundary condition is equivalent to imposing an anti-symmetric condition at $\varphi = \pi/2$:
\begin{align}
    \phi (\varphi) = - \phi(\pi- \varphi) \ ,
\end{align}
and the Neumann boundary condition is equivalent to imposing a symmetric condition at $\varphi = \pi/2$:
\begin{align}
    \phi (\varphi) =  \phi(\pi- \varphi) \ .
\end{align}
If the scalar field is expanded into the spherical harmonics on $\BS^{d-1}$,
\begin{align}
    \phi (\varphi, \theta) = \sum_{m} f_m(\varphi)\, Y_{m,\BS^{d-1}} (\theta) \ ,
\end{align}
$Y_{m,\BS^{d-1}} (\theta)$ with odd (even) $\ell-m$ are odd (even) functions about $\varphi =\pi/2$ and satisfy the Dirichlet (Neumann) boundary condition. 
Then, the degeneracies for the Dirichlet and Neumann boundary condition are given by
\begin{align}
    \begin{aligned}
    \text{Dirichlet:} \qquad  g_+^{(d)}(\ell) &= \binom{\ell+d-2}{d-1} = \ell\, \frac{\Gamma (\ell +d-1) }{\Gamma (d) \Gamma (\ell+1)} \ , \\
    \text{Neumann:} \qquad g_-^{(d)}(\ell) &= \binom{\ell+d-1}{d-1} =(\ell+d-1)\, \frac{ \Gamma (\ell +d-1) }{\Gamma (d) \Gamma (\ell+1)} \ .
    \end{aligned}
\end{align}
We use a subscript $+$ ($-$) for the Dirichlet (Neumann) boundary condition.\footnote{We adopt this unusual convention because we use $+$ for the Dirichlet boundary condition on $\BH^d$.
}
The degeneracy can also be written as
\begin{align}
    g_\pm^{(d)}(\ell) 
    = \frac{1}{2}\, g^{(d)}(\ell) \mp \frac{1}{2\,\Gamma (d-1)}  \frac{\Gamma \left(\nu_\ell^{(d)} +\frac{d-1}{2} \right) }{ \Gamma \left(\nu_\ell^{(d)} - \frac{d-1}{2}+1 \right)} \ ,
\end{align}
by using $\nu_\ell^{(d)}$ defined by \eqref{degeneracy}.

For a conformally coupled scalar on $\BHS^d$, 
the free energy is given by \eqref{free_energy_sphere} with the degeneracy replaced by those for the Dirichlet/Neumann boundary conditions:\footnote{We choose this normalization for the free energy on $\BHS^d$. Hence our free energy satisfies \eqref{relation_of_free energy} contrary to Appendix C in \cite{Jensen:2015swa}.}
\begin{align} \label{free_energy_hemisphere_def}
        F[\BHS_\pm^d] 
            =  \frac{1}{2} \sum_{\ell =0}^\infty g_\pm^{(d)}(\ell)\, \left[ \log \left( \frac{\nu_\ell^{(d)} + \frac{1}{2}}{\tilde\Lambda R}\right) + \log \left( \frac{\nu_\ell^{(d)} - \frac{1}{2}}{\tilde\Lambda R}\right) \right]\ .
\end{align}
Here we added the suffix to manifest boundary conditions explicit.
Since the sum of the degeneracies satisfies the relation,
\begin{align}
    g_+^{(d)}(\ell)+ g_-^{(d)}(\ell) = g^{(d)}(\ell) \ ,
\end{align}
the sum of the free energies on a hemisphere equals
the free energy on a sphere, 
\begin{align} \label{relation_of_free energy}
    F[\BHS_+^d] + F[\BHS_-^d] =F[\BS^d] \ .
\end{align}
The zeta functions of each boundary condition are given by
\begin{align}
    \zeta_{\BHS_\pm^d} (s)
     & = \frac{1}{2}\, \zeta_{\BS^d} (s) \mp 
    \frac{1}{2\,\Gamma (d-1)} \sum_{\ell =0}^\infty \, \frac{\Gamma \left(\nu_\ell^{(d)} +\frac{d-1}{2}\right) }{ \Gamma \left(\nu_\ell^{(d)} - \frac{d-1}{2}+1\right)} \,\left[ \left( \nu_\ell^{(d)} + \frac{1}{2}\right)^{-s} + \left( \nu_\ell^{(d)} - \frac{1}{2}\right)^{-s} \right] \ ,
\end{align}
and the renormalized free energies become 
\begin{align} \label{Fren_hemisphere}
    F_\text{ren}[\BHS_\pm^d] \equiv - \frac{1}{2}\,\zeta_{\BHS_\pm^d}(0)\,\log(\Lambda R)  - \frac{1}{2}\,\partial_s \zeta_{\BHS_\pm^d}(0)  \ ,
\end{align}
as in the previous section.

In the following, we compute the zeta functions by using the relation, 
\begin{align}\label{hemi_degen_expand}
    \frac{\Gamma \left(\nu_\ell^{(d)} +\frac{d-1}{2} \right) }{ \Gamma \left(\nu_\ell^{(d)} - \frac{d-1}{2}+1 \right)} =
    \begin{dcases}
         \sum_{n=0}^{\frac{d-1}{2}}(-1)^{\frac{d-1}{2}+n}\,\alpha_{n, d}\, \sum_{k=0}^{2n-1} \binom{2n-1}{k}\, x^k\, c^{2n-1-k} & d: \text{odd} \ , \\
         \sum_{n=0}^{\frac{d}{2}-1}(-1)^{\frac{d}{2}-1+n}\,\beta_{n, d}\, \sum_{k=0}^{2n} \binom{2n}{k}\, x^k\, c^{2n-k}
         & d: \text{even} \ ,
    \end{dcases}
\end{align}
where $x=\nu_\ell^{(d)} - c$.\footnote{This follows from \eqref{degen_expand_alphabeta} and the binomial theorem for $\nu_\ell^{(d)}$.}

\subsubsection{Odd \texorpdfstring{$d$}{d}}

Using the expansion \eqref{hemi_degen_expand} and performing a similar computation in section \ref{sec3.1.1}, 
the zeta functions can be written as 
\begin{align}
\begin{aligned}
    &\zeta_{\BHS_\pm^d} (s) - \frac{1}{2}\, \zeta_{\BS^d} (s) \\
    & = \mp \frac{1}{\Gamma (d-1)} \sum_{n=1}^{\frac{d-1}{2}} \sum_{l=0}^{n-1} (-1)^{\frac{d-1}{2}+n}\,\alpha_{n, d}\,  \binom{2n-1}{2l+1}\,  2^{1-2n}\,  (2^{s}-2^{2l+1} )\, \zeta \left(s-2l-1 \right) \ .
    \end{aligned} 
\end{align}
Since $\zeta_{\BS^d} (0)=0$ for odd dimension, we find
\begin{align} \label{zeta_hemisphere_odd}
    \zeta_{\BHS_\pm^d} (0) 
    & = \pm
    \frac{1}{\Gamma (d-1)} \sum_{n=1}^{\frac{d-1}{2}}  (-1)^{\frac{d-1}{2}+n}\,\alpha_{n, d}\, \left( \frac{B_{2n}}{2n} - \frac{2^{-2n-1}}{n} \right) \ ,
\end{align}
where we again used the Taylor expansion of the Bernoulli polynomials.
The derivative of the zeta functions reduces to\footnote{The term including $\log 2$ can be simplified further to
\begin{align}
    \mp
    \frac{1}{\Gamma (d-1)} \left( \sum_{n=1}^{\frac{d-1}{2}}
    (-1)^{\frac{d-1}{2}+n}\,2^{-2n} \,\alpha_{n, d}\,  \frac{1-B_{2n}}{n}\, -  \frac{\Gamma \left(\frac{d}{2} \right)}{2\, \Gamma \left( 2- \frac{d}{2} \right)} \right) \log 2 \ .
\end{align}}
\begin{align} \label{delta_zeta_hemisphere_odd}
\begin{aligned}
    \partial_s \zeta_{\BHS_\pm^d} (0) - \frac{1}{2}\, \partial_s \zeta_{\BS^d} (0) 
    & = \mp
    \frac{1}{\Gamma (d-1)} \sum_{n=1}^{\frac{d-1}{2}} \sum_{l=0}^{n-1} (-1)^{\frac{d-1}{2}+n}\,2^{1-2n} \,\alpha_{n, d}\,  \binom{2n-1}{2l+1}  \\
    & \qquad \cdot \left[ (1-2^{2l+1} )\, \zeta' \left(-2l-1 \right) +  \zeta \left(-2l-1 \right)\,\log 2  \right] \ .
    \end{aligned} 
\end{align}
It follows that the renormalized free energies on  $\BHS^d$ is 
\begin{align}
    F_{\text{ren}}[\BHS_\pm^d] =  - A[\BHS_\pm^d] \log (\Lambda R) + F_{\text{fin}} [\BHS_\pm^d] \ ,
\end{align}
where 
\begin{align}\label{anomaly_hemi_odd}
    A[\BHS_\pm^d] 
    & = \pm
    \frac{1}{2\, \Gamma (d-1)} \sum_{n=1}^{\frac{d-1}{2}}  (-1)^{\frac{d-1}{2}+n}\,\alpha_{n, d}\, \left( \frac{B_{2n}}{2n} - \frac{2^{-2n-1}}{n} \right) \ ,
\end{align}
and 
\begin{align}\label{finite_hemi_odd}
    \begin{aligned}
    F_{\text{fin}} [\BHS_\pm^d] 
        &= \frac{1}{2}\, F_{\text{fin}} [\BS^d] \pm
        \frac{1}{2\,\Gamma (d-1)} \sum_{n=1}^{\frac{d-1}{2}} \sum_{l=0}^{n-1} (-1)^{\frac{d-1}{2}+n}\,2^{1-2n}\,\alpha_{n, d}\,\binom{2n-1}{2l+1}   \\
        & \qquad \cdot \left[ (1-2^{2l+1} )\, \zeta' \left(-2l-1 \right) +  \zeta \left(-2l-1 \right)\,\log 2 \right] \ .
    \end{aligned}
\end{align}

The renormalized free energies on $\BHS^d$ for odd $d$ have logarithmic divergences due to the presence of the boundary.
The anomaly parts of the boundary free energies for the Neumann boundary condition are always greater than those of the Dirichlet boundary condition, and this is consistent with $C$-theorems in BCFT$_d$ as we will see in section \ref{sec5.1}.
The anomaly parts and the finite parts of the renormalized free energies are listed in table \ref{tab:hemi} in appendix \ref{app_table}.
In the presence of the boundary anomaly the finite terms depend on the regularization scheme and are not universal.\footnote{It is still meaningful to consider the difference of the finite parts in the same regularization as it no longer depends on the scheme.}

\subsubsection{Even \texorpdfstring{$d$}{d}}

Repeating a similar computation to section \ref{sec3.1.2},
the zeta functions reduce to
\begin{align}
\begin{aligned}
 \zeta_{\BHS_\pm^d} (s) - \frac{1}{2}\, \zeta_{\BS^d} (s)
     & = \mp
    \frac{1}{\Gamma (d-1)}
    \sum_{n=0}^{\frac{d}{2}-1} \sum_{l=0}^{n}  (-1)^{\frac{d}{2}-1+n}\,\beta_{n, d}\, \binom{2n}{2l}\, 2^{2l-2n}\, 
     \zeta \left(s-2l \right) \\
     & \qquad  \mp \frac{1}{2} \,\delta_{d,2}\, \left(\zeta_\text{H} (s,0) - \zeta (s) \right) \ .    
\end{aligned}
\end{align}
Then, at $s= 0$ we find
\begin{align} \label{hemi_zeta}
    \begin{aligned}
    \zeta_{\BHS_\pm^d} (0) - \frac{1}{2}\, \zeta_{\BS^d} (0) 
        &= 
        0 \ , \\
    \partial_s \zeta_{\BHS_\pm^d} (0) - \frac{1}{2}\, \partial_s \zeta_{\BS^d} (0) 
        &=
        \mp
        \frac{1}{\Gamma (d-1)}
        \sum_{n=1}^{\frac{d}{2}-1} \sum_{l=1}^{n}  (-1)^{\frac{d}{2}-1+n}\,\beta_{n, d}\, \binom{2n}{2l}\, 2^{2l-2n} 
     \zeta' \left(-2l \right) \\
     & \qquad \mp \frac{1}{2}\, \delta_{d,2} \left(\partial_s \zeta_\text{H} (0,0) + \zeta' (0) \right) \ .        
        \end{aligned}
\end{align}
It follows that the renormalized free energies on  $\BHS^d$ are
\begin{align}
    F_{\text{ren}}[\BHS_\pm^d] =  - A[\BHS_\pm^d] \log (\Lambda R) + F_{\text{fin}} [\BHS_\pm^d] \ ,
\end{align}
where 
\begin{align}\label{anomaly_hemi_even}
    A[\BHS_\pm^d] = \frac{1}{2}\, A[\BS^d] \ ,
\end{align}
and 
\begin{align}\label{fin_hemi_even}
    \begin{aligned}
    F_\text{fin} [\BHS_\pm^d] 
     & = \frac{1}{2}\, F_\text{fin}[\BS^d] \pm \frac{1}{2\,\Gamma (d-1)}
    \sum_{n=1}^{\frac{d}{2}-1} \sum_{l=1}^{n}  (-1)^{\frac{d}{2}-1+n}\,\beta_{n, d}\, 2^{2l-2n} \, \binom{2n}{2l}\,
     \zeta' \left(-2l \right) \\
     & \qquad  \pm \frac{1}{4}\, \delta_{d,2} \left(\partial_s \zeta_\text{H} (0,0) + \zeta' (0) \right)  \ .
    \end{aligned}
\end{align}

The bulk anomaly of $\BHS^d$ is just a half of the bulk anomaly on $\BS^d$ as is consistent with the fact that the type $A$ anomaly coefficient is fixed by the Euler characteristic of the manifold.
The boundary free energy with Neumann boundary condition is always greater than that with Dirichlet boundary condition. 
The anomaly parts and finite parts of the free energies are listed in table \ref{tab:hemi} in appendix \ref{app_table}.

\section{Free energy for Dirichlet boundary condition}
\label{sec4}

We move onto the calculation of the free energies on the hyperbolic space $\BH^d$ and a product space $\BH^{p+1}\times \BS^{q-1}$ in the zeta regularization.
Most parts of the calculations are the same as before, but the only difference from section \ref{sec3} is the continuous spectrum of the conformal laplacian on the hyperbolic space.
The main results of this section can be found in \eqref{F_zeta_odd_a}, \eqref{anomaly_h_odd}, \eqref{F_zeta_even_a}, \eqref{anomaly_h_even} and \eqref{finite_h_even} for $\BH^d$, \eqref{fren_h_s1_even}, \eqref{anomaly_h_s1_even} and \eqref{free_energy_finite_HS_odd} for $\BH^{p+1}\times \BS^1$, and 
\eqref{FE_even_p_even_q}, \eqref{FE_even_p_odd_q} and \eqref{FE_odd_p}
for $\BH^{p+1}\times \BS^{q-1}$.

\subsection{Free energy on \texorpdfstring{$\BH^d$}{Hd}}
\label{sec4.1}

Next we consider the case on the hyperbolic space $\BH^d$.
Since $\BH^d$ is non-compact the conformal laplacian of the free scalar has a continuous spectrum $\omega = [0, \infty)$:
\begin{align} \label{laplace_eq_hyperbolic}
    -\nabla^2_{\BH^d}\, \phi_\omega = \left( \frac{\omega^2}{R^2} + \left( \frac{d-1}{2R} \right)^2 \right)\, \phi_\omega \ .
\end{align}
Thus, the free energy on $\BH^d$ of radius $R$ is given by\footnote{As noted in the footnote \ref{footnote_ambiguity}, there are ambiguities to decompose $\log (\omega^2+\nu^2)$ into a sum of logarithmic functions.}
\begin{align}\label{unrenom_F_al}
        F[\BH^d]
            =  \frac{1}{2} \int_0^\infty \dd\omega\, \mu^{(d)} (\omega)\, \left[ \log \left( \frac{\omega + \i\,\nu }{\tilde\Lambda R}\right) + \log \left( \frac{\omega- \i\,\nu  }{\tilde\Lambda R}\right) \right]\ ,
\end{align}
where we introduced the parameter $\nu$ as
\begin{align}\label{nu_def}
    \nu = \Delta - \frac{d-1}{2} \ , \qquad \Delta (\Delta - d+1) = - \frac{d(d-2)}{4} \ .
\end{align}
There are two solutions to the above equation, which correspond to the Dirichlet and Neumann boundary conditions on $\BH^d$.
They are given by $\Delta_+ = \frac{d}{2}$ and $\Delta_- = \frac{d}{2}-1$ respectively, or equivalently $\nu=\frac{1}{2}$ and $\nu=-\frac{1}{2}$ in terms of $\nu$.\footnote{The free energy \eqref{unrenom_F_al} appears to be invariant under $\nu \to -\nu$, but it should be understood to be defined only for $\nu>0$ as the Plancherel measure is well defined only for normalizable modes.
Hence the Dirichlet and Neumann boundary conditions have different free energies as we will see in section \ref{sec5}.
}

The Plancherel measure $\mu^{(d)} (\omega)$ on $\BH^d$ of unit radius takes the form \cite{Camporesi:1990wm,Bytsenko:1994bc}:
\begin{align} \label{Plancherel}
    \begin{aligned}
        \mu^{(d)} (\omega) 
            &= 
            c_d\, \bigg|\,\frac{\Gamma\left(\frac{d-1}{2} + \i\,\omega\right)}{\Gamma(\i\,\omega)}\,\bigg|^2 \\
            &=
            c_d\,
            \begin{dcases}
            \prod_{j=0}^\frac{d-3}{2}(\omega^2 + j^2) & \quad d: \text{odd} \ , \\
            \omega\,\tanh (\pi\omega)\,\prod_{j=\frac{1}{2}}^\frac{d-3}{2}(\omega^2 + j^2) & \quad d: \text{even} \ ,
            \end{dcases}
    \end{aligned}
\end{align}
where the product should be omitted for $d=2$.
The coefficient
\begin{align} 
    c_d 
        \equiv 
        \frac{\text{Vol}(\BH^d)}{2^{d-1}\,\pi^\frac{d}{2}\,\Gamma\left(\frac{d}{2}\right)} \ ,
\end{align}
is proportional to the volume $\text{Vol}(\BH^{d})$ of the hyperbolic space of unit radius, which may be given in dimensional regularization by
\begin{align} \label{volume_hyperbolic}
    \text{Vol}(\BH^{d}) = \pi^{\frac{d-1}{2}}\,\Gamma\left(\frac{1-d}{2}\right) = - \frac{\pi^{\frac{d+1}{2}}}{\sin\left(\pi\,\frac{d-1}{2}\right)\,\Gamma\left(\frac{d+1}{2}\right)}\ .
\end{align}
The hyperbolic volume is finite for even $d$, but divergent  for odd $d$ due to the pole in the gamma function, which may be replaced by the logarithmic divergence by introducing a small cutoff parameter $\epsilon$\,:
\begin{align} \label{replacement}
    -\frac{1}{\sin\left(\pi\,\frac{d-1}{2}\right)} =
    \begin{dcases}
            (-1)^\frac{d-1}{2} \frac{2}{\pi} \,\log \left( \frac{R}{\epsilon}\right) &\qquad d: \text{odd} \ , \\
            (-1)^\frac{d}{2}  &\qquad d: \text{even} \ . 
    \end{dcases}
\end{align}
After the regularization, the coefficient takes the form:
\begin{align}\label{cd_explicit}
    c_d 
        =
        -\frac{1}{\sin\left(\pi\,\frac{d-1}{2}\right)\,\Gamma(d)}
        =
        \begin{dcases}
            \frac{(-1)^\frac{d-1}{2}}{\Gamma(d)}\, \frac{2}{\pi} \,\log \left( \frac{R}{\epsilon}\right) &\qquad d: \text{odd} \ , \\
            \frac{(-1)^\frac{d}{2}}{\Gamma(d)}  &\qquad d: \text{even} \ .
    \end{dcases}
\end{align}

The free energy \eqref{unrenom_F_al} is divergent and we regularize it by introducing the renormalized free energy with the zeta regularization as in section \ref{sec3}:
\begin{align}\label{Fren}
    F_\text{ren}[\BH^d] = - \frac{1}{2}\, \zeta_{\BH^d} (0,\nu)\,\log(\Lambda R) - \frac{1}{2}\,\partial_s \zeta_{\BH^d} (0, \nu)  \ ,
\end{align}
where the zeta function is defined by
\begin{align}\label{zeta_hyp}
    \zeta_{\BH^d} (s, \nu) \equiv \int_0^\infty \dd\omega\, \mu^{(d)} (\omega)\,\left[ \left(\omega + \i\, \nu \right)^{-s} + \left(\omega -\i\, \nu \right)^{-s} \right] \ .
\end{align}

In what follows, we will compute the renormalized free energy for the Dirichlet boundary condition with $\nu=\frac{1}{2}$ by evaluating the zeta function \eqref{zeta_hyp} following the method in \cite{Camporesi:1994ga,Bytsenko:1995ak}.

\subsubsection{Odd \texorpdfstring{$d$}{d}}
\label{sec4.1.1}

When $d$ is odd we expand the Plancherel measure $\mu^{(d)}(\omega)$ in an analogous way as in \eqref{degen_expand_alphabeta}:
\begin{align}
	\prod_{j=0}^\frac{d-3}{2}\,(\omega^2 + j^2) \equiv \sum_{k=1}^\frac{d-1}{2}\,\alpha_{k,d}\,\omega^{2k} \ .
\end{align}
Then we can perform the integration over $\omega$ in the zeta function \eqref{zeta_hyp} which is convergent for $\text{Re}\,s > d/2$:\footnote{The integral with respect to $\omega$ produces a factor $\Gamma(s-2k-1)/\Gamma(s)$, which equals the product $\prod_{i=1}^{2k+1}(s-i)^{-1}$ for integer $k$.
}
\begin{align} \label{zeta_function_odd_d}
	\begin{aligned}
		\zeta_{\BH^d} (s, \nu) 
			&=
			c_d\, \sum_{k=1}^\frac{d-1}{2}\,\alpha_{k,d}\, \int_0^\infty \dd\omega\, \left[ \frac{\omega^{2k}}{\left(\omega   + \i\, \nu \right)^s} + \frac{\omega^{2k}}{\left(\omega  - \i\,\nu \right)^s}\right] \\
			&=
			2c_d\,\sum_{k=1}^\frac{d-1}{2}\,(-1)^k\,\alpha_{k,d}\, \nu^{2k+1 - s}\, \sin \left( \frac{\pi s}{2} \right)\,\Gamma\left( 2k+ 1\right)\,\prod_{i=1}^{2k+1} \frac{1}{s-i} \ .
	\end{aligned}
\end{align}
It follows that $\zeta_{\BH^d} (0, \nu)$ and $\partial_s \zeta_{\BH^d} (0, \nu)$ become
\begin{align} \label{zeta_function_odd}
	\begin{aligned}
		\zeta_{\BH^d} (0, \nu) &= 0 \ ,\\
		\partial_s \zeta_{\BH^d} (0, \nu)
			&=
			c_d\, \sum_{k=1}^\frac{d-1}{2}\,(-1)^{k+1}\,\frac{\pi\,\alpha_{k,d}\, \nu^{2k+1}}{2k+1} \ .
	\end{aligned}
\end{align}
Hence the renormalized free energy \eqref{Fren} does not have a logarithmic divergence depending on the UV cutoff $\Lambda$, but has a logarithmic divergence that arises from the regularized volume of the hyperbolic space \eqref{volume_hyperbolic}:
\begin{align}\label{F_zeta_odd_a}
    F_\text{ren}[\BH^d] = -\mathcal{A} [\BH^d]\, \log \left( \frac{R}{\epsilon}\right)  \ ,
\end{align}
where the anomaly coefficient is given by
\begin{align} \label{anomaly_h_odd}
    \mathcal{A}[\BH^d] = 
        \frac{(-1)^\frac{d-1}{2}}{\Gamma(d)}\,\sum_{k=1}^\frac{d-1}{2}\,(-1)^{k+1}\,\frac{\alpha_{k,d}}{2^{2k+1}(2k+1)} 
        \ .
\end{align}
Since there are no bulk anomalies when $d$ is odd, we interpret $\mathcal{A}[\BH^d]$ as defect anomaly from the boundary theory in $(d-1)$ dimensions.

The anomaly coefficients for $d\leq 9$ are listed in table \ref{tab:F_p_q} in appendix \ref{app_table}.

\subsubsection{Even \texorpdfstring{$d$}{d}}
\label{sec4.1.2}

When $d$ is even, we expand the product in the Plancherel measure as
\begin{align}
	\prod_{j=\frac{1}{2}}^\frac{d-3}{2}\,(\omega^2 + j^2) \equiv \sum_{k=0}^{\frac{d}{2}-1}\,\beta_{k,d}\,\omega^{2k} \ .
\end{align}
We decompose the zeta function \eqref{zeta_hyp} into two parts using the identity
\begin{align} \label{tanh_decomposition}
	\tanh(\pi\,\omega) = 1 - \frac{2}{\mathrm{e}^{2\pi\omega} + 1} \ ,
\end{align}
and perform a similar integration to \eqref{zeta_function_odd_d} for the first part to obtain
\begin{align}
	\begin{aligned}
		\zeta_{\BH^d} (s, \nu) 
			&=
			2\,c_d\, \sum_{k=0}^{\frac{d}{2}-1}\,(-1)^{k+1}\,\beta_{k,d}\, 
				\left[ \nu^{2k+2-s}\,\cos \left(\frac{\pi s}{2}\right)\,\Gamma(2k+2)\,\prod_{i=1}^{2k+2}\frac{1}{s-i}\right.\\
			&\qquad\qquad\qquad \left.
			    - \int_0^\infty \dd\omega\, \frac{\omega^{2k+1}}{\mathrm{e}^{2\pi\omega} + 1}
			    \left( \left( \omega + \i\, \nu\right)^{-s} + \left( \omega - \i\, \nu\right)^{-s} \right)
			    \right] \ .
	\end{aligned}
\end{align}
While we do not know how to perform the remaining integral in the square bracket it is analytic in $s$ and convergent in the $s\to 0$ limit to give
\begin{align}
	\begin{aligned}
		\zeta_{\BH^d} (0, \nu) 
			&=  c_d\,\sum_{k=0}^{\frac{d}{2}-1}\,(-1)^{k+1}\,\frac{\beta_{k,d}}{k+1}\,\left[ (\nu^2)^{k+1} + (1-2^{-2k-1})\,B_{2k+2} \right]\ , \\
		\partial_s \zeta_{\BH^d} (0, \nu)
			&=
			c_d\, \sum_{k=0}^{\frac{d}{2}-1}\,\beta_{k,d}\left[\frac{(-\nu^2)^{k+1}\,(H_{2k+2} -\log \nu)}{k+1} + 2\, f_{k} (\nu) \right] \ ,
	\end{aligned} \label{421}
\end{align}
where $B_{2k+2}\equiv B_{2k+2}(0)$ and $H_{2k+2}$ are Bernoulli and Harmonic numbers respectively and we introduced $f_k(\nu)$ by
\begin{align}\label{fknu_def}
    f_{k} (\nu) \equiv \int_0^\infty \dd\omega\, \frac{\omega^{2k+1}}{ \mathrm{e}^{2\pi\omega} + 1}\,\log(\omega^2 + \nu^2) \ ,
\end{align}
whose details may be found in appendix \ref{app1}.

For $\nu =1/2$, the derivative of the zeta function can be simplified to
\begin{align}\label{hyp_even_der_zeta}
		\partial_s \zeta_{\BH^d} \left(0, \frac{1}{2} \right)
			&=
			c_d\, \sum_{k=0}^{\frac{d}{2}-1}\,\beta_{k,d}\,(-1)^k  \sum_{j=1}^{2k+1}\,\frac{(-1)^j}{2^{2k-j}}\,\binom{2k+1}{j}\,  \zeta'(-j) - \delta_{d,2}\, \zeta'(0) \ ,
\end{align}
where we use an identity \eqref{conjecture1} which we conjecture to hold in appendix \ref{app1}.\footnote{The odd $j$ terms are the same as the half of the derivative of the zeta function on $\BS^d$, \eqref{sp_even_finite}.}

The renormalized free energy \eqref{Fren} has the UV logarithmic divergence which reflects the bulk conformal anomaly:
\begin{align}\label{F_zeta_even_a}
    F_\text{ren}[\BH^d]  = - A[\BH^d]\,\log (\Lambda R) + F_\text{fin}[\BH^d] \ ,
\end{align}
where the anomaly coefficient is given  by
\begin{align}\label{anomaly_h_even}
    A[\BH^d] =
        \frac{(-1)^\frac{d}{2}}{2\Gamma(d)}\, \sum_{k=0}^{\frac{d}{2}-1}\,(-1)^{k+1}\,\frac{\beta_{k,d}}{k+1}\,\left[ 2^{-2k-2} + (1-2^{-2k-1})\,B_{2k+2} \right]
\end{align}
The finite term follows from \eqref{hyp_even_der_zeta}:
\begin{align}\label{finite_h_even}
    F_\text{fin}[\BH^d] =
        \frac{(-1)^\frac{d}{2}}{\Gamma(d)}\,\sum_{k=0}^{\frac{d}{2}-1}\,\sum_{j=1}^{2k+1}\,\frac{(-1)^{k+j+1}}{2^{2k+1-j}}\,\beta_{k,d}\,  \binom{2k+1}{j}\,  \zeta'(-j) + \frac{1}{2}\, \delta_{d,2}\, \zeta'(0) \ .
\end{align}
The anomaly coefficients and finite parts for $d\leq 10$ are summarized in tables \ref{tab:F_p_q} and \ref{tab:Fin_p_q} in appendix \ref{app_table}.
From the table we argue without proof the following identity holds for even $d$:
\begin{align}
    A[\BH^d] = \frac{1}{2}\,A[\BS^d] \ .
\end{align}

\subsubsection{Comparison to 
\texorpdfstring{$\BHS^d$}{HSd}}

Let us compare the anomaly coefficients and finite parts of the free energies on $\BH^d$ with those on $\BHS^d$ with Dirichlet boundary condition.

\paragraph{Odd $d$}
From \eqref{anomaly_hemi_odd} and \eqref{anomaly_h_odd}, the difference of the anomaly coefficients is given by 
\begin{align}
\begin{aligned}
    \mathcal{A} [\BH^d] - A [\BHS_+^d] &=  \frac{(-1)^\frac{d-1}{2}}{\Gamma(d)}\,\sum_{k=1}^\frac{d-1}{2}\,(-1)^{k+1}\, \alpha_{k,d} \left[ \frac{1}{2^{2k+1}(2k+1)} + \frac{d-1}{4k} \left( B_{2k} - \frac{1}{ 2^{2k}} \right) \right]  \ .
\end{aligned}
\end{align}
We argue the right hand side always vanishes, thus
\begin{align}\label{anom_hemi_hyp_equality}
    \mathcal{A} [\BH^d] = A [\BHS_+^d] 
\end{align}
holds for arbitrary $d$.
This relation is difficult for us to prove analytically  but we check it explicitly for a number of $d$.
However, by employing an alternative expression for the anomaly coefficients for $\BHS^d_+$ in odd $d$ given by \cite{Dowker:2010yj},
\begin{align}
    A [\BHS_+^d] = \frac{1}{2\,\Gamma (d+1)}\,B^{(d)}_d\left(\frac{d}{2}-1\right) \ ,
\end{align}
one can prove the identity \eqref{anom_hemi_hyp_equality} using the integral representation (see e.g.\,\cite{Dowker:2010qy})
\begin{align}
    B^{(d)}_d \left(\frac{d}{2}-1\right) = -2\,d\,\int_0^\frac{1}{2}\,\d \nu\, \prod_{j=0}^\frac{d-3}{2}\left(\nu^2 - j^2\right) \ ,
\end{align}
and the expansion \eqref{degen_expand_alphabeta}.\footnote{We are indebted to J.\,S.\,Dowker for providing us a proof of the identity \eqref{anom_hemi_hyp_equality} and valuable correspondences.}

For the finite parts of the free energies, $F_\text{fin}[\BH^d]$ is different from $F_\text{fin}[\BHS_+^d]$.
However, the finite parts are not universal in the sense that they depend on the choice of the cutoffs $\Lambda$ and $\epsilon$ in the presence of the boundary anomalies, so we can always make $F_\text{fin}[\BH^d]$ equal $F_\text{fin}[\BHS^d]$ by tuning $\epsilon$ appropriately in comparing the two.

\paragraph{Even $d$}
For the anomaly parts, both anomaly coefficients \eqref{anomaly_hemi_even} and \eqref{anomaly_h_even} on $\BHS^d$ and $\BH^d$ are given by a half of the bulk anomaly on $\BS^d$.
Thus, the difference of the free energies equals to that of the finite parts \eqref{fin_hemi_even} and \eqref{finite_h_even}:
\begin{align}
    F_\text{ren}[\BH^d] - F_\text{ren}[\BHS_+^d] =
        -\frac{1}{\Gamma(d)}\,\sum_{k=1}^{\frac{d}{2}-1}\,\epsilon_k\,\zeta'(-2k)  \ ,
\end{align}
where we rearrange the ranges of $n$ and $k$, $\sum_{n=1}^{\frac{d}{2}-1}\sum_{k=1}^n = \sum_{k=1}^{\frac{d}{2}-1} \sum_{n=k}^{\frac{d}{2}-1}$, to find
\begin{align}
    \begin{aligned}
    \epsilon_k &\equiv \sum_{n=k}^{\frac{d}{2} -1}\,\beta_{n,d}\,\frac{(-1)^{\frac{d}{2} + n}}{2^{2(n-k) +1}}\,\left[\binom{2n+1}{2k}- (d-1)\,\binom{2n}{2k}\right]\\
        &= \gamma_{2k,d}\left( \frac{1}{2}\right) - (d-1)\,\sum_{n=k}^{\frac{d}{2} -1}\,\beta_{n,d}\,\frac{(-1)^{\frac{d}{2} + n}}{2^{2(n-k) +1}}\,\binom{2n}{2k} \ .
    \end{aligned}
\end{align}
We check that $\epsilon_{k\ge 1} =0$ for $4\le d \le 20$, and further speculate it holds for any even $d$.
To sum up we observe that the free energy on $\BH^d$ with Dirichlet boundary condition coincides with that on $\BHS^d$ with Dirichlet boundary condition.
The values of the free energies for $d \leq 8$ are summarized in table \ref{tab:hemi} in appendix \ref{app_table}.

From the agreements of the free energies, we observe the Dirichlet boundary condition on $\BH^d$ can be identified with the Dirichlet boundary condition on $\BHS^d$ for any $d$ as mentioned in \cite{Rodriguez-Gomez:2017aca}.

\subsection{Free energy on \texorpdfstring{$\BH^{p+1}\times \BS^{1}$}{Hp-1,S1}}
\label{sec4.2}

We treat the free energy on $\BH^{p+1}\times \BS^{1}$ separately from the case on $\BH^{p+1}\times \BS^{q-1}$ as the degeneracy of $\BS^1$ is different from $\BS^{q-1}$ with $q\geq 3$.
This space is associated with a codimension $q=2$ defect and has been a focus of research due to the relation to entanglement entropy \cite{Casini:2011kv,Klebanov:2011gs,Belin:2013uta}.

Expanding the eigenfunctions with respect to the angular modes $\ell \in \BZ$, the free energy $\BH^{p+1} \times \BS^1$ is given by
\begin{align}
    \begin{aligned}
        F[\BH^{p+1} \times \BS^1] &= \frac{1}{2}\, \tr \log \left[ \tilde{\Lambda}^{-2} \left( -\nabla_{\BH^{p+1}}^2-\nabla_{\BS^1}^2 - \frac{p^2}{4R^2} \right)  \right]\\
        &=  \frac{1}{2} \sum_{\ell=-\infty}^\infty \int_0^\infty \dd \omega \, \mu^{(p+1)} (\omega)\,   \log \left( \frac{\omega^2 + \ell^2}{\tilde \Lambda^2 R^2}  \right) \ ,
    \end{aligned}
\end{align}
with the Plancherel measure $\mu^{(p+1)} (\omega)$ \eqref{Plancherel}.
It will be convenient to decompose the logarithmic function into two logarithmic functions.
Here we employ two different decompositions depending on the ordering of the integral over $\omega$  and the  summation over $\ell$.

\subsubsection{Even \texorpdfstring{$p$}{p}}
\label{sec4.2.1}

Since the Plancherel measure $\mu^{(p+1)} (\omega)$ \eqref{Plancherel} for even $p$ is a polynomial, the integral over $\omega$ can be performed before the summation over $\ell$.
Requiring the convergence in the $\omega \to \infty$ limit fixes the decomposition of the free energy:
\begin{align}
    \begin{aligned}
        F[\BH^{p+1} \times \BS^1] 
        &=  \frac{1}{2} \sum_{\ell=-\infty}^\infty \int_0^\infty \dd \omega \, \mu^{(p+1)} (\omega)\, \left[ \log \left( \frac{\omega + \i \, |\ell|}{\tilde \Lambda R}  \right) + \log \left( \frac{\omega - \i \, |\ell|}{\tilde \Lambda R}  \right) \right] \\
        & = - \frac{1}{2}\,\int_0^\infty\,\frac{\dd t}{t}\, \sum_{\ell=-\infty}^\infty \int_0^\infty \dd \omega \, \mu^{(p+1)} (\omega)\,  \left[ \mathrm{e}^{-t (\omega + \i \, |\ell|)/(\tilde\Lambda R)} + \mathrm{e}^{-t (\omega - \i \, |\ell|)/(\tilde\Lambda R)} \right]\ .
    \end{aligned}
\end{align}
Repeating the same regularization as in section \ref{sec4.1.1}, the renormalized free energy on $\BH^{p+1}\times \BS^1$ is given by
\begin{align}\label{Fren_h_s1}
    F_{\text{ren}} [\BH^{p+1} \times \BS^1] \equiv - \frac{1}{2}\,\zeta_{\BH^{p+1} \times \BS^1} (0) \,\log(\Lambda R) - \frac{1}{2}\,\partial_s \zeta_{\BH^{p+1} \times \BS^1} (0) \ ,
\end{align}
where
\begin{align}
    \begin{aligned}
    \zeta_{\BH^{p+1} \times \BS^1}(s) & \equiv \sum_{\ell=-\infty}^\infty \, \zeta_{\BH^{p+1}} (s, |\ell|) \\
    & = 4c_{p+1} \,\sum_{k=1}^\frac{p}{2}\,\alpha_{k,p+1}\, (-1)^k \sin \left(\frac{\pi s}{2} \right)\, \Gamma\left( 2k+ 1\right) \zeta (s-2k-1) \prod_{i=1}^{2k+1} \frac{1}{s-i}\ .
    \end{aligned}
\end{align}
To regularize the divergence from the zero mode from $\ell =0$ we introduce a mass $m$ for the scalar.
This IR regularization amounts to replacing $|\ell|$ with $\sqrt{\ell^2 + (mR)^2}$ above.
Then it can be shown that the resulting zeta function for $\ell = 0$ vanishes in the $m\to 0$.
From this expression, we immediately obtain
\begin{align}\label{zeta_h_s1_even}
    \begin{aligned}
    \zeta_{\BH^{p+1} \times \BS^1} (0) &= 0 \ , \\
    \partial_s \zeta_{\BH^{p+1} \times \BS^1} (0) & = 2c_{p+1} \,\sum_{k=1}^\frac{p}{2}\,\alpha_{k,p+1}\, \frac{(-1)^{k+1} \pi}{2k+1}\, \zeta (-2k-1) \ .
    \end{aligned}
\end{align}
Since both bulk and defect are even dimensional when $p$ is even the renormalized free energy \eqref{Fren_h_s1} may have both types of anomalies:
\begin{align}\label{fren_h_s1_even}
    F_\text{ren}[\BH^{p+1} \times \BS^1] 
        = 
         - A[\BH^{p+1} \times \BS^1]\,\log(\Lambda R) - \mathcal{A}[\BH^{p+1} \times \BS^1]\,\log \left( \frac{R}{\epsilon}\right) \ ,
\end{align}
but from \eqref{zeta_h_s1_even} we find
\begin{align}\label{anomaly_h_s1_even}
    \begin{aligned}
    A[\BH^{p+1} \times \BS^1] 
        &= 0 \ ,\\
     \mathcal{A}[\BH^{p+1} \times \BS^1] 
        &= \frac{(-1)^{\frac{p}{2}}}{\Gamma (p+1)}  \,\sum_{k=1}^\frac{p}{2}\,\alpha_{k,p+1}\, \frac{2(-1)^{k+1}}{2k+1}\, \zeta (-2k-1) \ .
    \end{aligned}
\end{align}
Thus defect anomalies are there while bulk anomalies vanish in this case.

The explicit values of the anomaly coefficients for $d \leq 10$ are listed in table \ref{tab:F_p_q} in appendix \ref{app_table}.

\subsubsection{Odd \texorpdfstring{$p$}{p}}
\label{sec4.2.2}

For odd $p$ we use the identity \eqref{tanh_decomposition} to the Plancherel measure $\mu^{(p+1)} (\omega)$ \eqref{Plancherel} and apply different decompositions to each term to derive
\begin{align} \label{def_free_energy_HS}
    \begin{aligned}
        F[\BH^{p+1} \times \BS^1] 
            &=  \frac{c_{p+1}}{2}  \sum_{k=0}^{\frac{p-1}{2}} \beta_{k,p+1}  \sum_{\ell=-\infty}^\infty \int_0^\infty \dd \omega \, \omega^{2k+1} \, \left[ \log \left( \frac{\omega+ \i \, |\ell|}{\tilde \Lambda R}  \right) + \log \left( \frac{\omega - \i \, |\ell|}{\tilde \Lambda R}  \right) \right] \\
            & \qquad - c_{p+1} \sum_{k=0}^{\frac{p-1}{2}} \beta_{k,p+1} \int_0^\infty \dd \omega \, \frac{\omega^{2k+1}}{\mathrm{e}^{2\pi \omega}+1} \, \sum_{\ell=-\infty}^\infty \left[ \log \left( \frac{|\ell|+ \i \, \omega}{\tilde \Lambda R}  \right) + \log \left( \frac{|\ell| - \i \, \omega}{\tilde \Lambda R}  \right) \right] \ .
    \end{aligned}
\end{align}
Here the ordering between the integral and the summation is important in this expression.
In the Schwinger representation of \eqref{def_free_energy_HS}, the first term is convergent in the $\omega \to \infty$ limit, while the second term is convergent in the $|\ell| \to \infty$ limit.\footnote{For odd $p$, it is possible to use a different representation of the zeta function
\begin{align}
    \begin{aligned}
    \zeta_{\BH^{p+1} \times \BS^1} (s) 
        & = \int_0^\infty \dd\omega\, \mu^{(p+1)} (\omega)\, \sum_{\ell=-\infty}^{\infty} \left[ \left(|\ell| + \i\, \omega \right)^{-s} + \left(|\ell| -\i\, \omega \right)^{-s} \right] \\
        & = \int_0^\infty \dd\omega\, \mu^{(p+1)} (\omega)\,  \left[ \zeta_\text{H} (s,-\i\,\omega)+\zeta_\text{H} (s,1-\i\,\omega)+\zeta_\text{H} (s,\i\,\omega)+\zeta_\text{H} (s,1+\i\,\omega) \right] \ .
    \end{aligned}
\end{align}
In the $s\to 0$ limit, the zeta function and its derivative are given by 
\begin{align}
    \begin{aligned}
    \zeta_{\BH^{p+1} \times \BS^1} (0) & = 0 \ , \\
    \partial_s \zeta_{\BH^{p+1} \times \BS^1} (0) & = -2 \int_0^\infty \dd\omega\, \mu^{(p+1)} (\omega)\,  \log (2\sinh (\pi \omega)) \ ,
    \end{aligned}
\end{align}
where we use \eqref{hurwitz_zeta_sum} and \eqref{hurwitz_zeta_sum_limit}.
Hence we obtained the same expression of the free energy in \cite{Rodriguez-Gomez:2017kxf}.
However, the derivative of the zeta function still diverges.
Hence we need to use the same regularization as in \cite{Klebanov:2011gs,Rodriguez-Gomez:2017kxf}.
}

Repeating similar computations as in 
section \ref{sec3} and  section \ref{sec4.1.2}, the renormalized free energy on $\BH^{p+1}\times \BS^1$ is given by the same form as \eqref{Fren_h_s1}
with the zeta function consisting of two parts:
\begin{align}
    \begin{aligned}
    \zeta_{\BH^{p+1} \times \BS^1}(s) &= \zeta_{\BH^{p+1} \times \BS^1}^{(1)}(s) + \zeta_{\BH^{p+1} \times \BS^1}^{(2)} (s) \ , \\
     \zeta_{\BH^{p+1} \times \BS^1}^{(1)}(s)
        &= 
        c_{p+1}\, \sum_{k=0}^{\frac{p-1}{2}}\,\beta_{k,p+1}\, \sum_{\ell=-\infty }^\infty 
        \, \int_0^\infty \dd\omega\, \omega^{2k+1}\,\left[\left( \omega + \i\, |\ell|\right)^{-s} + \left( \omega - \i\, |\ell| \right)^{-s}\right] \ , \\
       \zeta_{\BH^{p+1} \times \BS^1}^{(2)}(s) &=
			-2 c_{p+1}\, \sum_{k=0}^{\frac{p-1}{2}}\,\beta_{k,p+1}\, \int_0^\infty \dd\omega\, \frac{\omega^{2k+1}}{\mathrm{e}^{2\pi\omega} + 1}
			\sum_{\ell=-\infty}^\infty \,\left[
			\left( |\ell| + \i\, \omega \right)^{-s}
			+
			\left( |\ell| - \i\, \omega \right)^{-s} 
	       \right] \ .
    \end{aligned}
\end{align}
The first term can be computed as 
\begin{align}
    \begin{aligned}
    \zeta_{\BH^{p+1} \times \BS^1}^{(1)}(s)
        & = 4c_{p+1}\, \sum_{k=0}^{\frac{p-1}{2}}\,\beta_{k,p+1}\, (-1)^{k+1}\, \cos \left( \frac{\pi s}{2} \right)\, \Gamma (2k+2)\, \zeta (s-2k-2)\, \prod_{i=1}^{2k+2}\frac{1}{s-i} \ ,
    \end{aligned}
\end{align}
where we regulate the $\ell=0$ mode in the same way as for the even $p$ case.
It follows that $\zeta_{\BH^{p+1} \times \BS^1}^{(1)}(s)$ and its derivative at $s=0$ are given by 
\begin{align}
    \begin{aligned}
    \zeta_{\BH^{p+1} \times \BS^1}^{(1)}(0) &= 0 \,, \\
    \partial_s \zeta_{\BH^{p+1} \times \BS^1}^{(1)}(0) &= 2c_{p+1}\, \sum_{k=0}^{\frac{p-1}{2}}\, \frac{\beta_{k,p+1}\, (-1)^{k+1}}{k+1} \,
    \zeta' (-2k-2) \ .
    \end{aligned}
\end{align}
The second one can be written as 
\begin{align}
      \begin{aligned}
        \zeta_{\BH^{p+1} \times \BS^1}^{(2)}(s) 
		   & = -2 c_{p+1}\, \sum_{k=0}^{\frac{p-1}{2}}\,\beta_{k,p+1}\, \int_0^\infty \dd\omega\, \frac{\omega^{2k+1}}{\mathrm{e}^{2\pi\omega} + 1} \\
	       & \qquad \qquad \cdot 
			\left[
			\zeta_\text{H} (s,-\i\,\omega)+\zeta_\text{H} (s,1-\i\,\omega)+\zeta_\text{H} (s,\i\,\omega)+\zeta_\text{H} (s,1+\i\,\omega)
	       \right] \ .
	   \end{aligned}
\end{align}
Although it is difficult to perform the integration over $\omega$, it is possible to compute $\zeta_{\BH^{p+1} \times \BS^1}^{(2)}(0)$ and $\partial_s \zeta_{\BH^{p+1} \times \BS^1}^{(2)}(0)$.
Since the combination of the Hurwitz zeta function vanishes,
\begin{align} \label{hurwitz_zeta_sum}
    \zeta_\text{H} (0,-\i\,\omega)+\zeta_\text{H} (0,1-\i\,\omega)+\zeta_\text{H} (0,\i\,\omega)+\zeta_\text{H} (0,1+\i\,\omega) = 0 \ ,
\end{align}
in the $s\to 0$ limit, we obtain
\begin{align}
    \zeta_{\BH^{p+1} \times \BS^1}^{(2)} (0) = 0 \ .
\end{align}
Using the derivative of the combination of the Hurwitz functions, 
\begin{align} \label{hurwitz_zeta_sum_limit}
    \lim_{s\to 0} \partial_s \left[ \zeta_\text{H} (s,-\i\,\omega)+\zeta_\text{H} (s,1-\i\,\omega)+\zeta_\text{H} (s,\i\,\omega)+\zeta_\text{H} (s,1+\i\,\omega) \right] = -2\log (2\sinh (\pi \omega)) \ ,
\end{align}
which is the same as the regularization of $\sum_{\ell} \log (\omega^2 + |\ell|^2)$ \cite{Klebanov:2011gs,Rodriguez-Gomez:2017kxf}, the derivative of $\zeta_{\BH^{p+1} \times \BS^1}^{(2)}(s)$ reduces to
\begin{align}
    \begin{aligned}
         \partial_s \zeta_{\BH^{p+1} \times \BS^1}^{(2)}(0) 
          &= 4 c_{p+1}\, \sum_{k=0}^{\frac{p-1}{2}}\,\beta_{k,p+1}\, \frac{\Gamma (2k+2)}{(2\pi)^{2k+2}}\left[ \sum_{m=1}^k (1-2^{2m-2k-2})\, \zeta (2m)\, \zeta (-2m+2k+3) \right. \\
          & \qquad \qquad \qquad \left. - (1-2^{-2k-3})\, \zeta (2k+3) + 2 (1-2^{-2k-2})  \, \zeta (2k+2)\,\log 2 \right] \ ,
    \end{aligned}
\end{align}
where we use formulas eq.\,(4) and eq.\,(14) in \cite{zhao2020logarithmic} after changing  a variable $u= \mathrm{e}^{- 2\pi \omega}$. 

Since $\zeta_{\BH^{p+1} \times \BS^1}(0) = \zeta_{\BH^{p+1} \times \BS^1}^{(1)}(0) +\zeta_{\BH^{p+1} \times \BS^1}^{(2)}(0) =0$, neither bulk nor defect anomaly appear:
\begin{align}
    A[\BH^{p+1} \times \BS^1] = \CA[\BH^{p+1} \times \BS^1] = 0 \ .
\end{align}
This is consistent with the fact that both $p+q$ and $p$ are odd.
Hence the renormalized free energy has only a finite term: 
\begin{align}\label{free_energy_finite_HS_odd}
    \begin{aligned}
    F_\text{ren} [\BH^{p+1} \times \BS^1] 
        &= -2 c_{p+1}\, \sum_{k=0}^{\frac{p-1}{2}}\,\beta_{k,p+1}\, \frac{\Gamma (2k+2)}{(2\pi)^{2k+2}}\left[ \sum_{m=1}^k (1-2^{2m-2k-2})\, \zeta (2m)\, \zeta (-2m+2k+3) \right. \\
          & \qquad \qquad \qquad \left. - \frac{1-2^{-2k-2}}{2}\, \zeta (2k+3) + 2 (1-2^{-2k-2}) \, \zeta (2k+2)\,\log 2 \right] \ .
    \end{aligned}
\end{align}

Let us compare the free energy on $\BH^{p+1} \times \BS^1$
with the free energy on $\BS^{p+2}$ \eqref{free_energy_finite_sphere_odd}.
We argue the equivalence between \eqref{free_energy_finite_HS_odd} and \eqref{free_energy_finite_sphere_odd} for arbitrary $p$:
\begin{align}
    F_\text{ren} [\BH^{p+1} \times \BS^1] = F_\text{ren} [\BS^{p+2}] \ .
\end{align}
We are not aware of an analytic proof of this identity, but we check it holds up to $p$ of order $O(100)$.
Given the equivalence of the free energies one can derive a number of mathematical identities for the Riemann zeta functions which appear to be unknown in literature.

\subsection{Free energy on \texorpdfstring{$\BH^{p+1}\times \BS^{q-1}$}{Hp-1,Sq-1}}

Expanding the scalar field by the spherical harmonics labeled by $\ell$ on $\BS^{q-1}$ the free energy for Dirichlet boundary condition on $\BH^{p+1}\times \BS^{q-1}$ is given by
\begin{align}
    \begin{aligned}
        F[\BH^{p+1}\times \BS^{q-1}] &= \frac{1}{2}\, \tr \log \left[ \tilde{\Lambda}^{-2} \left( -\nabla_{\BH^{p+1}}^2-\nabla_{\BS^{q-1}}^2 + \frac{(q-p-2)(d-2)}{4R^2} \right)  \right]\\
        & =  \frac{1}{2}\, \sum_{\ell=0}^\infty g^{(q-1)}(\ell) \int_0^\infty \dd \omega \, \mu^{(p+1)} (\omega)\,  \log \left( \frac{\omega^2 + \left(\nu_\ell^{(q-1)} \right)^2}{\tilde \Lambda^2 R^2}  \right) \ ,
    \end{aligned}
\end{align}
with the Plancherel measure $\mu^{(p+1)} (\omega)$ given by \eqref{Plancherel}, the degeneracy $g^{(q-1)}(\ell) $ and $\nu_\ell^{(q-1)}$ for $q > 2$ by \eqref{degeneracy}. 
Repeating the same regularization as in section \ref{sec4.2.1}, the renormalized free energy takes the form:
\begin{align}\label{FreeEnergy_p_q}
    F_{\text{ren}}[\BH^{p+1}\times \BS^{q-1}] \equiv - \frac{1}{2}\, \zeta_{\BH^{p+1}\times \BS^{q-1}} (0)\,\log(\Lambda R) - \frac{1}{2}\,\partial_s \zeta_{\BH^{p+1}\times \BS^{q-1}}(0) \ ,
\end{align}
where $\zeta_{\BH^{p+1}\times \BS^{q-1}}  (s)$ is the summation of the zeta function on $\BH^{p+1}$ over the angular modes:
\begin{align}
    \zeta_{\BH^{p+1}\times \BS^{q-1}}  (s) \equiv \sum_{\ell=0}^\infty\, g^{(q-1)}(\ell)\, \zeta_{\BH^{p+1}}(s, \nu_\ell^{(q-1)}) \ .
\end{align}
As in section \ref{sec4.2}, we use different expressions for the decompositions of the logarithmic functions depending on the cases so that the resulting forms have good convergent behaviors in the Schwinger representation.

\subsubsection{Even \texorpdfstring{$p$}{p}}
The regularized volume of $\BH^{p+1}$ \eqref{volume_hyperbolic} has a logarithmic divergence after the regularization \eqref{replacement} and the renormalized free energy \eqref{FreeEnergy_p_q} has two types of logarithmic divergences:
\begin{align}
    F_{\text{ren}}[\BH^{p+1}\times \BS^{q-1}] = - A[\BH^{p+1}\times \BS^{q-1}] \log (\Lambda R) - \mathcal{A}[\BH^{p+1}\times \BS^{q-1}] \log \left( \frac{R}{\epsilon}\right) \ .
\end{align}

Using the degeneracy \eqref{degen_expand} and the expansion \eqref{gamma_exp}, we can perform the sum over $\ell$ to get
\begin{align}
    \begin{aligned}
    \zeta_{\BH^{p+1}\times \BS^{q-1}} (s) 
        &=
        \frac{4\, c_{p+1}}{\Gamma(q-1)} \sum_{k=1}^\frac{p}{2}\,(-1)^k\,\alpha_{k,p+1}\,\sin \left( \frac{\pi s}{2}\right)\,\Gamma\left( 2k+ 1\right)\, \prod_{i=1}^{2k+1} \frac{1}{s-i} \,\\
        &\qquad \cdot
        \begin{dcases}
        \sum_{n=0}^{\frac{q}{2}-1}\,(-1)^{\frac{q}{2}-1+n}\,\alpha_{n, q-1}\,
        \zeta_\text{H}\left( s-2k-2n-1, \frac{q-2}{2}\right) & q: \text{even} \ , \\
        \sum_{n=0}^{\frac{q-3}{2}}(-1)^{\frac{q-3}{2}+n}\,\beta_{n, q-1}\,
        \zeta_\text{H}\left( s - 2k-2n-2, \frac{q-2}{2}\right) & q: \text{odd} \ .
        \end{dcases}
    \end{aligned}
\end{align}
Using the identities \eqref{hurwitz_identity_1}, \eqref{hurwitz_to_zeta_half}, \eqref{zeta_values} and the relations \eqref{degen_expand_alphabeta} we obtain
\begin{align}\label{zetaEvenp}
    \begin{aligned}
        \zeta_{\BH^{p+1}\times \BS^{q-1}} (0) 
            &= 0 \ , \\
        \partial_s \zeta_{\BH^{p+1}\times \BS^{q-1}} (0)
            &= 
            \frac{2\pi\,c_{p+1}}{\Gamma(q-1)} \sum_{k=1}^\frac{p}{2}\,\alpha_{k,p+1}\,\frac{(-1)^{k+1}}{2k+1}\\
            &\qquad \cdot
                \begin{dcases}
                    \sum_{n=0}^{\frac{q}{2}-1}\,\frac{(-1)^{\frac{q}{2}+n}}{2k+2n+2}\,\alpha_{n, q-1}\,
        \,B_{2k+2n+2} &\quad q: \text{even} \ , \\
                    0 & \quad q: \text{odd}  \ .
                \end{dcases}
    \end{aligned}
\end{align}
Hence we find the following:
\begin{itemize}
\item  For even $q$, the renormalized free energy has the logarithmic divergence 
\begin{align}
    F_\text{ren}[\BH^{p+1}\times \BS^{q-1}] = - \mathcal{A}[\BH^{p+1}\times \BS^{q-1}]\,\log \left( \frac{R}{\epsilon}\right) \ ,\qquad (p: \text{even},~ q: \text{even}) \ ,
\end{align}
where the anomaly coefficient is given by
\begin{align} \label{FE_even_p_even_q}
     \mathcal{A}[\BH^{p+1}\times \BS^{q-1}]
        = 
        \frac{(-1)^\frac{p+q}{2}}{\Gamma(p+1)\,\Gamma(q-1)} \sum_{k=1}^\frac{p}{2}\,\sum_{n=1}^{\frac{q}{2}-1}\,\alpha_{k,p+1}\,\alpha_{n, q-1}\,\frac{(-1)^{k +n+1}}{(2k+1)(k+n+1)}\,
        B_{2k+2n+2}\ .
\end{align}

\item For odd $q$, we find  $A[\BH^{p+1}\times \BS^{q-1}] \propto  \zeta_{\BH^{p+1}\times \BS^{q-1}} (0) = 0$ and $\mathcal{A}[\BH^{p+1}\times \BS^{q-1}] \propto \partial_s \zeta_{\BH^{p+1}\times \BS^{q-1}} (0) = 0$, and there are no conformal anomalies.
From \eqref{zetaEvenp} the finite term of the renormalized free energy also vanishes, so
\begin{align} \label{FE_even_p_odd_q}
    F_\text{ren}[\BH^{p+1}\times \BS^{q-1}] = 0 \ , \qquad (p: \text{even},~ q: \text{odd}) \ .
\end{align}
This is consistent with the results obtained by \cite{Rodriguez-Gomez:2017kxf}.

\end{itemize}

\subsubsection{Odd \texorpdfstring{$p$}{p}}
As in section \ref{sec4.2.2}, we decompose the zeta function into two parts:\footnote{We use the regularization scheme for $\zeta_{\BH^{p+1}\times \BS^{q-1}}^{(2)} (s)$ which is different from that of $\zeta_{\BH^{p+1}\times \BS^{q-1}}^{(1)} (s)$.}
\begin{align}
    \zeta_{\BH^{p+1}\times \BS^{q-1}}(s)  = \zeta_{\BH^{p+1}\times \BS^{q-1}}^{(1)}(s) + \zeta_{\BH^{p+1}\times \BS^{q-1}}^{(2)}(s) \ ,
\end{align}
where $\zeta_{\BH^{p+1}\times \BS^{q-1}}^{(1)}(s)$ and $\zeta_{\BH^{p+1}\times \BS^{q-1}}^{(1)}(s)$ are defined by
\begin{align}
\begin{aligned}
     \zeta_{\BH^{p+1}\times  \BS^{q-1}}^{(1)}(s)& = 
        c_{p+1}\, \sum_{k=0}^{\frac{p-1}{2}}\,\beta_{k,p+1}\, \sum_{\ell=0}^\infty g^{(q-1)}(\ell)   \\
    & \qquad \cdot \int_0^\infty \dd\omega\, \omega^{2k+1}\,\left[\left( \omega + \i\, \nu_\ell^{(q-1)}\right)^{-s} + \left( \omega - \i\, \nu_\ell^{(q-1)}\right)^{-s}\right] \ , \\
        \zeta_{\BH^{p+1}\times \BS^{q-1}}^{(2)} (s) & = -2\, c_{p+1}\, \sum_{k=0}^{\frac{p-1}{2}}\,\beta_{k,p+1}\, \int_0^\infty \dd\omega\, \frac{\omega^{2k+1}}{\mathrm{e}^{2\pi\omega} + 1} \\
			& \qquad \cdot \sum_{\ell=0}^\infty g^{(q-1)}(\ell) \,\left[
			\left( \nu_\ell^{(q-1)}+ \i\, \omega \right)^{-s}
			+
			\left( \nu_\ell^{(q-1)} - \i\, \omega \right)^{-s}
	       \right] \ .
\end{aligned}
\end{align}
By performing the integration first for $ \zeta_{\BH^{p+1}\times \BS^{q-1}}^{(1)}(s)$, we obtain 
\begin{align}\label{zeta1}
    \begin{aligned}
    \zeta_{\BH^{p+1}\times \BS^{q-1}}^{(1)}(s)
        &=
        \frac{4\,c_{p+1}}{\Gamma(q-1)}\,\sum_{k=0}^{\frac{p-1}{2}}\,(-1)^{k+1}\,\beta_{k,p+1}\,\cos \left(\frac{\pi s}{2}\right)\,\Gamma(2k+2)\,\prod_{i=1}^{2k+2}\frac{1}{s-i}\\
        &\qquad \cdot
                \begin{dcases}
                    \sum_{n=1}^{\frac{q}{2}-1}\,(-1)^{\frac{q}{2}-1+n}\,\alpha_{n, q-1}\,
        \zeta\left(s -2k-2n-2\right) &\quad q: \text{even} \ , \\
                    \sum_{n=0}^{\frac{q-3}{2}}(-1)^{\frac{q-3}{2}+n}\,\beta_{n, q-1}\,
        (2^{s-2k-2n-3}-1)\,\zeta\left(s  -2k - 2n -3\right) & \quad q: \text{odd} \ ,
                \end{dcases}
    \end{aligned}
\end{align}
where we used \eqref{degen_expand_alphabeta} with $d\to q-1$.
For $\zeta_{\BH^{p+1}\times \BS^{q-1}}^{(2)} (s)$ we sum over $\ell$ first and use the expansion \eqref{degen_expand} with $d\to q-1$ and $c\to \i\,\omega$ and the identity \eqref{hurwitz_identity_1} as in section \ref{sec3} to write
\begin{align} \label{zeta2_new}
    \begin{aligned}
        &\zeta_{\BH^{p+1}\times \BS^{q-1}}^{(2)} (s)  =
			-\frac{4\,  c_{p+1}}{\Gamma(q-1)}\, \sum_{k=0}^{\frac{p-1}{2}}\,\beta_{k,p+1}\, \int_0^\infty \dd\omega\, \frac{\omega^{2k+1}}{ \mathrm{e}^{2\pi\omega} + 1} \\
			& \qquad\qquad  \cdot \begin{dcases}
			\sum_{n=0}^{q-2} \,
			\left[
			\gamma_{n,q-1}(\i\,\omega)\,\zeta_\text{H} \left(s-n, \i\,\omega\right)
			+\gamma_{n,q-1}(-\i\,\omega)\,\zeta_\text{H} \left(s-n, 1-\i\,\omega\right)
	       \right]  & \quad q: \text{even} \ ,  \\
			\sum_{n=0}^{q-2} \,
			\left[
			\gamma_{n,q-1}(\i\,\omega)\,\zeta_\text{H} \left(s-n, \frac{1}{2}+\i\,\omega\right)
			+
			\gamma_{n,q-1}(-\i\,\omega)\,\zeta_\text{H} \left(s-n, \frac{1}{2}-\i\,\omega\right)
	       \right]  & \quad q: \text{odd} \ .
			\end{dcases}
	   \end{aligned}
\end{align}

For odd $p$ the regularized volume of $\BH^{p+1}$ is finite and the coefficient $c_{p+1}$ given by \eqref{cd_explicit} does not give rise to a logarithmic divergence in the zeta function.
It follows from \eqref{FreeEnergy_p_q} that the logarithmic divergence of the free energy is determined by
\begin{align}
    F_{\text{ren}}[\BH^{p+1}\times \BS^{q-1}] = - A[\BH^{p+1}\times \BS^{q-1}]\,\log(\Lambda R) + F_{\text{fin}}[\BH^{p+1}\times \BS^{q-1}] \ , 
\end{align}
where the anomaly part and the finite part are
\begin{align}
    \begin{aligned}
    A[\BH^{p+1}\times \BS^{q-1}] &= \frac{1}{2}\,\left( \zeta_{\BH^{p+1}\times \BS^{q-1}}^{(1)}(0) + \zeta_{\BH^{p+1}\times \BS^{q-1}}^{(2)}(0)\right) \ , \\
    F_{\text{fin}}[\BH^{p+1}\times \BS^{q-1}] &= -\frac{1}{2}\,\left( \partial_s\zeta_{\BH^{p+1}\times \BS^{q-1}}^{(1)} (0) + \partial_s\zeta_{\BH^{p+1}\times \BS^{q-1}}^{(2)}(0)\right) \ .
    \end{aligned}
\end{align}

From \eqref{zeta1} we read
\begin{align}
    \begin{aligned}
    \zeta_{\BH^{p+1}\times \BS^{q-1}}^{(1)}(0)
        &=        \frac{2\,c_{p+1}}{\Gamma(q-1)}\,\sum_{k=0}^{\frac{p-1}{2}}\,\frac{(-1)^{k+1}\beta_{k,p+1}}{k+1}\\
        &\qquad \cdot
                \begin{dcases}                 
                    0 &\quad q: \text{even} \ , \\
                    \sum_{n=0}^{\frac{q-3}{2}}\,(-1)^{\frac{q-3}{2}+n}\,\beta_{n, q-1}\,(2^{-2k-2n-3}-1)\,
        \zeta\left(-2k - 2n -3\right) & \quad q: \text{odd} \ ,
                \end{dcases}
    \end{aligned}
\end{align}
while from \eqref{zeta2_new} and after a bit of calculation, we obtain
\begin{align}
    \begin{aligned}
    &\zeta_{\BH^{p+1}\times \BS^{q-1}}^{(2)}(0) \\
        &\quad =
        \frac{8\,c_{p+1}}{\Gamma(q-1)}\,\sum_{k=0}^{\frac{p-1}{2}}\,\beta_{k,p+1}\\
        &\qquad \cdot
                \begin{dcases}                 
                    0 &\quad q: \text{even} \ , \\
                   \sum_{m=0}^{\frac{q-3}{2}}\,\sum_{n=0}^{2m+1}\,\frac{(-1)^{\frac{q-1}{2}+n}}{n+1}\,\binom{2m+1}{n}\,\beta_{m, q-1} & \\
                   \qquad \cdot \sum_{r=0}^{\floor{\frac{n+1}{2}}}\,(-1)^r\,(1-2^{1-2r})\,\binom{n+1}{2r}\,B_{2r} \\
                   \qquad \qquad   \cdot \frac{1-2^{-2m-2k-3+2r}}{(2\pi)^{2m+2k+4-2r}}\,\Gamma(2m+2k+4-2r)\,\zeta(2m+2k+4-2r)
                   & \quad q: \text{odd} \ .
                \end{dcases}
    \end{aligned}
\end{align}

From \eqref{zeta1} and \eqref{zeta2_new}, we read 
\begin{align}
    \begin{aligned}
        &\partial_s\zeta_{\BH^{p+1}\times \BS^{q-1}}^{(1)} (0) \\
            &=
            \frac{2\,c_{p+1}}{\Gamma(q-1)}\,\sum_{k=0}^{\frac{p-1}{2}}\,\frac{(-1)^{k+1}}{k+1}\,\beta_{k,p+1}\,\\
            &\qquad  \cdot
                \begin{dcases}
                  \sum_{n=1}^{\frac{q}{2}-1}\,(-1)^{\frac{q}{2}-1+n}\,\alpha_{n, q-1}\,
      \zeta' \left( -2(k+n+1)\right) &\quad q: \text{even} \ , \\
                    \sum_{n=0}^{\frac{q-3}{2}}(-1)^{\frac{q-3}{2}+n}\,\beta_{n, q-1} \bigg[ (2^{-2n-2k-3}-1)\,\zeta' \left(-2k - 2n -3\right) &\\
         \qquad +\left( (2^{-2n-2k-3}-1)\,H_{2k+2} + 2^{-2n-2k-3}\,\log 2\right)\,\zeta \left(-2k - 2n -3\right) \bigg] & \quad q: \text{odd} \ , 
                \end{dcases}
    \end{aligned}
\end{align}
and 
\begin{align}
    \begin{aligned}
        &\partial_s \zeta_{\BH^{p+1}\times \BS^{q-1}}^{(2)} (0) \\
			&\quad =
			-\frac{4\,  c_{p+1}}{\Gamma(q-1)}\, \sum_{k=0}^{\frac{p-1}{2}}\,\beta_{k,p+1}\, \int_0^\infty \dd\omega\, \frac{\omega^{2k+1}}{ \mathrm{e}^{2\pi\omega} + 1} \, 	\sum_{n=0}^{q-2} \\
			& \qquad \cdot  
	       \begin{dcases}
	       \sum_{m = \ceil{\frac{n}{2}}}^{\frac{q}{2}-1}
            (-1)^{\frac{q}{2}-1+m+n}\,\binom{2m}{n}
            \,\alpha_{m,q-1}\, (\i \, \omega)^{2m-n} \\
            \qquad \qquad \cdot 
			\left[  
			\,\partial_s  \zeta_\text{H} \left(-n, \i\,\omega\right)
			+ (-1)^n \partial_s  \zeta_\text{H} \left(-n, 1-\i\,\omega\right) 	       \right]  &\quad q: \text{even} \ , \\
	       \sum_{m = \floor{\frac{n}{2}}}^{\frac{q-3}{2}}
            (-1)^{\frac{q-1}{2}+m+n}\,\binom{2m+1}{n}
            \,\beta_{m,q-1}\, (\i \, \omega )^{2m+1-n} \\
            \qquad \qquad \cdot \left[ 
			\,\partial_s  \zeta_\text{H} \left(-n, \frac{1}{2}+\i\,\omega\right)
			+ (-1)^{n-1}\partial_s  \zeta_\text{H} \left(-n, \frac{1}{2}-\i\,\omega\right)
	       \right]  &\quad q: \text{odd} \ .
	       \end{dcases}
	   \end{aligned}
\end{align}
For even $q$, the bracket can be written by using the polylogarithm functions $\mathrm{Li}_{n} (x)$,
\begin{align}
    \begin{aligned}
        &\partial_s \zeta_{\BH^{p+1}\times \BS^{q-1}}^{(2)} (0) \\
			& \quad =
			-\frac{4\,  c_{p+1}}{\Gamma(q-1)}\, \sum_{k=0}^{\frac{p-1}{2}}\,\beta_{k,p+1}\, \int_0^\infty \dd\omega\, \frac{\omega^{2k+1}}{ \mathrm{e}^{2\pi\omega} + 1} \, 	\sum_{n=0}^{q-2}\, \sum_{m = \ceil{\frac{n}{2}}}^{\frac{q}{2}-1} (-1)^{\frac{q}{2}-1+n}\,\binom{2m}{n} \,\alpha_{m,q-1} \\
			& \qquad \cdot  \,  \omega^{2m-n}
			\left[  
			\frac{(-1)^{n}\,\Gamma (n+1) }{(2\pi)^n} \, \mathrm{Li}_{n+1} (\mathrm{e}^{-2\pi \omega}) + \sum_{l=0}^{\floor{\frac{n+1}{2}}}\, (-1)^{l-1}\, \frac{\pi}{n+1}\, \binom{n+1}{n+1-2l}\, B_{2l}\, \omega^{n+1-2l} \right] \ ,
	   \end{aligned}
\end{align}
where we use \eqref{hurwitz_to_polylog} and the facts $B_{n+1-r}$ vanishes for even $n-r>0$ and the sum over $m$ vanishes for $n=r$ in the third line.
Our method reproduces the known regularization of $\sum_\ell g^{(q-1)} (\ell) \log (\omega^2+(\nu_\ell^{(q-1)})^2)$ ((3.38) and (3.50) in \cite{Rodriguez-Gomez:2017kxf}) and is easily generalized to higher dimensions.

For odd $p$ and even $q$ the anomaly parts vanish
while for odd $p$ and odd $q$ they turn out to equal the bulk anomaly of $\BS^{p+q}$.
We thus find a set of identities relating the anomaly coefficients on the conformally equivalent spaces:
\begin{align}
    A[\BH^{2k} \times \BS^{d-2k}] = A[\BS^{d}] \ .
\end{align}
Combined with the result in section \ref{sec4.2.2}, the above relation holds for $k=1,\cdots ,\ceil{d/2}-1$.
A similar relation holds also for the finite parts of the free energies:
\begin{align}
    F_{\text{fin}}[\BH^{2k} \times \BS^{d-2k}] = F_{\text{fin}}[\BS^{d}] \ .
\end{align}
The equality between the finite parts is pointed out in \cite{Rodriguez-Gomez:2017kxf} for odd $d\leq 7$.

Furthermore, these relations can be extended to the equivalence of the renormalized free energies:
\begin{align} \label{FE_odd_p}
    F_{\text{ren}}[\BH^{2k} \times \BS^{d-2k}] = F_{\text{ren}}[\BS^{d}] \ ,\qquad (k=1,\cdots ,\ceil{d/2}-1) \ .
\end{align}
We checked them either analytically or numerically for $d$ of order $O(10)$.\footnote{While we do not write explicitly, this equality leads to the evaluated form of integrals including the polygamma function, which to our best knowledge has not appeared in literature.}
They should hold for any $d$ on physical ground as there are no defect anomalies when $p$ is odd while $\BH^{2k} \times \BS^{d-2k}$ has the same Euler characteristic as $\BS^d$, thus has the same bulk anomaly as $\BS^d$.

Substituting various values for $p$ and $q$ to the anomaly parts and the finite parts of the free energies, we obtain tables \ref{tab:F_p_q} and \ref{tab:Fin_p_q} in appendix \ref{app_table}.

\section{Free energy for Neumann boundary condition}
\label{sec5}

In section \ref{sec4}, we computed the free energies on $\BH^d$ and $\BH^{p+1}\times \BS^{q-1}$ for Dirichlet boundary condition.
In this section, we turn to the case for Neumann boundary condition.
We calculate the difference of the free energies between Neumann and Dirichlet boundary conditions in two ways.
First we calculate the free energy on $\BH^d$ for Neumann boundary condition from the result for Dirichlet boundary condition by analytically continuing the dimension $\Delta$ from the Dirichlet value $\Delta_+$ to the Neumann value $\Delta_-$.
Next we use the residue method for the same calculation.
These two methods turn out to give the same answer.
We then apply the residue method to the calculation of the free energy on $\BH^{p+1}\times \BS^{q-1}$.
Finally, we check the defect $C$-theorem \eqref{D_theorem} holds for all the cases.

\subsection{Analytic continuation}
\label{sec5.1}

The free energy for Neumann boundary condition can be derived from the Dirichlet value on $\BH^d$ obtained in section \ref{sec4} as the latter is analytic in $\nu$, so can be analytically continued from the positive $\nu$ to negative $\nu$ region.

\paragraph{Odd $d$}

From \eqref{zeta_function_odd}, we read off the free energy as a function of $\nu$,
\begin{align}
    F_\text{ren}[\BH^d] (\nu) 
    = \frac{ (-1)^\frac{d-1}{2} }{\Gamma (d)} \, 
    \sum_{k=1}^\frac{d-1}{2}\,(-1)^{k}\,\frac{\alpha_{k,d}\, \nu^{2k+1}}{2k+1} \log \left( \frac{R}{\epsilon} \right) \ .
\end{align}
The free energy with Neumann boundary condition,  $\nu=-1/2$, is given by
\begin{align}
    F_\text{ren}[\BH^d] (-1/2)
     = - F_\text{ren}[\BH^d] (1/2) \ .
\end{align}
The anomaly parts of the Neumann boundary condition are the minus of those for the Dirichlet boundary condition.
The difference of the free energies between the two boundary conditions is given by
\begin{align} \label{diff_odd_naive}
    F_{\Delta_+}[\BH^d] - F_{\Delta_-} [\BH^d]
    &= -2\, \mathcal{A} [\BH^d] \log \left(\frac{R}{\epsilon} \right) \ ,
\end{align}
where we introduced the new notations $F_{\Delta_+}[\BH^d] = F_\text{ren}[\BH^d](1/2)$ and $ F_{\Delta_-} [\BH^d]=F_\text{ren}[\BH^d](-1/2)$ to manifest the boundary conditions in the free energies.
It follows from the relations \eqref{anom_hemi_hyp_equality} and \eqref{anomaly_hemi_odd} that the anomaly part of the free energy on $\BH^d$ with the Neumann boundary condition coincides with that on $\BHS^d_-$ provided the two cutoffs $\epsilon$ and $\Lambda$ are appropriately identified.

\paragraph{Even $d$}

The free energy for even $d$ is 
\begin{align}
    F_\text{ren}[\BH^d] (\nu) = - \frac{1}{2}\, \zeta_{\BH^d} (0,\nu )\,\log(\Lambda R) - \frac{1}{2}\,\partial_s \zeta_{\BH^d}(0,\nu) \ .
\end{align}
The zeta function and its derivative defined by \eqref{421} are analytical functions of $\nu$, and they can be analytically continued to the $\nu <0$ region ($\log \nu$ should be understood as $(1/2) \log \nu^2$).
Then most of the terms are canceled out in the difference of the free energy except for $f_k(\nu)$, resulting in 
\begin{align}\label{diff_even_naive}
    \begin{aligned}
    F_{\Delta_+}[\BH^d] - F_{\Delta_-}[\BH^d] 
        & = -c_d \int_{0}^\frac{1}{2} \! \dd \mu \, \mu\, \sin (\pi \mu)\, \Gamma \left(\frac{ d-1}{2} \pm \mu \right)  \ ,
    \end{aligned}
\end{align}
where we use the identities $\psi (\mu +1/2) - \psi(-\mu +1/2) = \pi \tan (\pi \mu)$ and 
\begin{align}
    \sum_{k=0}^{\frac{d}{2}-1}(-1)^{k}\,\beta_{k, d}\, \mu^{2k+1} 
    = \frac{\mu}{\pi}\, \Gamma \left(\frac{ d-1}{2} \pm \mu \right)\, \cos (\pi \mu) \ ,
\end{align}
which follows from \eqref{degeneracy} and \eqref{degen_expand_alphabeta}.

In lower dimensions, the difference of the free energy becomes
\begin{align} 
    F_{\Delta_+}[\BH^d] - F_{\Delta_-} [\BH^d]
    =  \begin{cases}
            -\frac{1}{4}\,\log(2\pi) + \frac{1}{2}\,\partial_s \zeta_\text{H}(0,0) &\qquad d=2 \ , \\
            -\frac{1}{8\pi^2}\,\zeta (3)  &\qquad d=4 \ , \\
            \frac{1}{96\pi^2}\,\zeta (3) + \frac{1}{32\pi^4}\,\zeta (5) &\qquad d=6 \ ,  \\
            -\frac{1}{720\pi^2}\,\zeta (3) - \frac{1}{192\pi^4}\,\zeta (5) -\frac{1}{128\pi^6}\,\zeta (7) &\qquad d=8 \ , 
        \end{cases} 
\end{align}
where we use the identity that follows from \eqref{c9} for $d=2$.
We find perfect agreement with the difference of the free energies on $\BHS^d$ listed in table \ref{tab:hemi} while both of the differences on $\BH^2$ and $\BHS^2$ have the same IR divergences from $\partial_s\zeta_\text{H} (0,0)$.
More generally we argue the equality
\begin{align}
    F_{\Delta_+}[\BH^d] - F_{\Delta_-} [\BH^d] = F_\text{fin}[\BHS^d_+] - F_\text{fin} [\BHS^d_-] \ ,
\end{align}
should hold for even $d\ge 2$.\footnote{With \eqref{fin_hemi_even} and \eqref{diff_even_naive} this equality leads to the identity for even $d\ge 4$:
\begin{align}
    \int_{0}^\frac{1}{2} \! \dd \mu \, \mu\, \sin (\pi \mu)\, \Gamma \left(\frac{ d-1}{2} \pm \mu \right) 
        =  (d-1)\,
    \sum_{n=1}^{\frac{d}{2}-1} \sum_{l=1}^{n}  (-1)^{n}\,\beta_{n, d}\, 2^{2l-2n} \, \binom{2n}{2l}\,
     \zeta' \left(-2l \right)\ .
\end{align}
}

In total, we conclude from the agreement of the universal parts of the free energies that the Dirichlet/Neumann boundary conditions on $\BH^d$ are one-to-one correspondence with those on $\BHS^d$ for any $d$.
We will derive the same conclusion from a more indirect method in the following.

\subsection{Residue method}
\label{sec5.2}

In the previous section, we obtained the difference of the free energies via a naive analytical continuation in terms of the parameter $\nu$. 
The same result can be derived by using the residue method \cite{Giombi:2013yva,Giombi:2020rmc}, which argues that
the difference of the derivatives of the free energies is given by the residue of $\mu^{(d)} (\omega)/(2\omega (\omega -\i \, \nu))$ at $\omega = \i \, \nu$ with suitable normalization:
\begin{align}
    \partial_\nu F[\BH^d](\nu) -\partial_\nu F[\BH^d](-\nu) 
        &= 2\pi \i\, \nu \, \underset{\omega=\i \, \nu }{\text{Res}}\, \frac{\mu^{(d)}(\omega) }{2\omega (\omega -\i \, \nu)} \label{conjecture_residue} \\
        &= 
   - c_d \, 
    \nu\, \sin (\pi \nu)\, \Gamma \left( \frac{d-1}{2} \pm \nu \right) \ .
    \label{diff_free_energy}
\end{align}
By integrating the above expression \eqref{diff_free_energy} from $\nu =0$ to $\nu = \frac{1}{2}$, we obtain
\begin{align} \label{diff_FE_hyperbolic}
\begin{aligned}
     F_{\Delta_+}[\BH^d] - F_{\Delta_-}[\BH^d] 
    = \frac{1}{\sin (\pi\, \frac{d-1}{2})\,\Gamma (d)} \int_{0}^\frac{1}{2} \! \dd \nu \, \nu \,\sin (\pi \nu)\, \Gamma \left( \frac{d-1}{2} \pm  \nu \right)\ .
\end{aligned}
\end{align}
For even $d$, this expression is the same as \eqref{diff_even_naive} derived from the analytic continuation.
Also, by replacing the pole from the sine function for odd $d$ with the logarithmic divergence using \eqref{replacement} this also reproduces the boundary anomaly given in \eqref{diff_odd_naive}.

\paragraph{Derivation of the residue method}

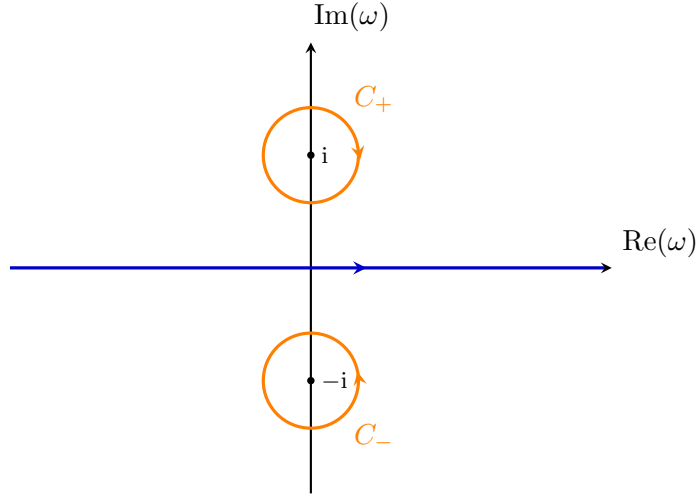
\begin{figure}
    \centering
    \begin{tikzpicture}[thick,>=stealth]
        \def\xr{3} \def\yr{2}
        \draw [->] (-\xr-1,0) -- (\xr+1,0) node [above right]  {$\text{Re}(\omega)$};
        \draw [->] (0,-\yr-1) -- (0,\yr+1) node [above right=0 and -0.1] {$\text{Im}(\omega)$};
        \draw[blue!80!black, very thick,decoration={markings,mark=at position 0.6 with {\arrow{>}}},postaction={decorate}] (-\xr-1,0) -- (\xr+0.9,0);
        \draw[fill] (0,1.5) circle (1pt) node [right] {\footnotesize $\i$};
        \draw[fill] (0,-1.5) circle (1pt) node [right] {\footnotesize $-\i$};
        \draw[orange, very thick, decoration={markings,mark=between positions 0.03 and 1.03 step 1 with \arrow{<}},postaction={decorate}] (0,1.5) circle (18pt) node [above right= 0.6 
        ] {$C_+$};
        \draw[orange, very thick, decoration={markings,mark=between positions 0.03 and 1.03 step 1 with \arrow{>}},postaction={decorate}] (0,-1.5) circle (18pt) node [below right = 0.6] {$C_-$};
\end{tikzpicture}
    \caption{The contours for the Green's functions for the Dirichlet (the blue real line) and Neumann boundary conditions (the blue real line and two orange circles $C_+$ and $C_-$).}
    \label{fig:contour}
\end{figure}

The Green's function of a massive scalar field with Dirichlet boundary condition has the integral representation \cite{Carmi:2018qzm}:
\begin{align}
    G_{\Delta_+} (x_1,x_2) = \frac{1}{R^{d-2}} \int_{-\infty}^\infty \! \dd \omega \,  \frac{1}{\omega^2+ \left(\Delta_+ - \frac{d-1}{2} \right)^2}\,\Omega_\omega^{(d)} (x_1,x_2) \ .
\end{align}
The Green's function with Neumann boundary condition can be obtained by changing the contour as in figure \ref{fig:contour}:
\begin{align}
    G_{\Delta_-} (x_1,x_2) = \frac{1}{R^{d-2}} \int_{\mathbb{R}+C_+ + C_-} \! \dd \omega \,  \frac{1}{\omega^2 + \left(\Delta_+ - \frac{d-1}{2} \right)^2}\,\Omega_\omega^{(d)} (x_1,x_2) \ ,
\end{align}
where $C_+$ is a clockwise circle around a pole at $\omega = \i \left(\Delta_+ - \frac{d-1}{2} \right)$ and $C_-$ is a counter-clockwise circle around a pole at $\omega =- \i \left(\Delta_+- \frac{d-1}{2} \right)$.

In the following, we need the expression of $\Omega_\omega^{(d)}(x_1,x_2)$ at the coincident point,
\begin{align}
    \Omega_\omega^{(d)} (0)= \Omega_\omega^{(d)} (x,x) = \frac{\Gamma \left(\frac{d-1}{2} \right)}{4 \pi^{\frac{d-1}{2} +1}\Gamma (d-1)} \frac{\Gamma \left(\frac{d-1}{2}  \pm \i \, \omega \right)}{\Gamma (\pm \, \i \, \omega )} \ .
\end{align}
The Plancherel measure \eqref{Plancherel} can be written by using $\Omega_\omega^{(d)}(0)$ as
\begin{align}
    \mu^{(d)} (\omega) = 2\, \text{Vol}(\BH^{d})\,    \Omega_\omega^{(d)}(0) \ .
\end{align}

Now the derivative of the free energy can be expressed as 
\begin{align}
\begin{aligned}
        \partial_\nu F[\BH^d](\nu)& = \text{Vol}(\BH^{d}) \, \nu \, G_{\Delta_+} (x,x) \ , \\
    \partial_\nu  F[\BH^d](-\nu) 
    & = \text{Vol}(\BH^{d}) \, \nu \,  G_{\Delta_-} (x,x) \ .
\end{aligned}
\end{align}
Then, the difference between them is given by
\begin{align}
\begin{aligned}
    \partial_\nu F[\BH^d] (\nu) -\partial_\nu F[\BH^d] (-\nu) &= -\frac{\nu}{2}   \int_{C_+ + C_-} \! \dd \omega \,  \frac{ \mu^{(d)} (\omega)}{\omega^2+\nu^2}  \\
    &=\frac{\pi \i \, \nu}{2}  \, \underset{\omega=\i \, \nu }{\text{Res}}\, \frac{ \mu^{(d)}(\omega)}{\omega(\omega - \i\,\nu )} - \frac{\pi \i \, \nu}{2}  \, \underset{\omega=-\i \, \nu  }{\text{Res}}\, \frac{\mu^{(d)}(\omega) }{\omega (\omega + \i\,\nu )} \\
    &=\pi \, \mu^{(d)}(\i \, \nu ) \ .
    \end{aligned}
\end{align}
In the last line, we used $\mu^{(d)}(\i \, \nu )= \mu^{(d)}(-\i \, \nu )$.
This completes the derivation of the residue method \eqref{conjecture_residue}.

\subsection{Application to \texorpdfstring{$\BH^{p+1} \times \BS^{q-1}$}{HS}}
\label{sec5.3}

Now let us apply the residue method \ref{sec5.2} to the free energy calculation on $\BH^{p+1} \times \BS^{q-1}$ with Neumann boundary condition.\footnote{It is also possible to apply the analytic continuation method in section \ref{sec5.1}, which gives the same result. Here we use the residue method due to its simplicity.
}
Since the Neumann boundary condition has negative $\nu_{\ell=\pm 1}^{(1)}$ for $q=2$ or negative $\nu_{\ell=0}^{(q-1)}$ for $q\geq 3$, it is convenient to express the free energy as a sum of each mode
\begin{align}
    & F[\BH^{p+1} \times \BS^{q-1}] = \sum_{\ell=0}^\infty\, g^{(q-1)}(\ell)\, F_\ell \left(\nu_\ell^{(q-1)}\right) \ , 
\end{align}
where $F_\ell \left(\nu_\ell^{(q-1)}\right)$ is the free energy for the $\ell$-th mode on $\BH^{p+1}$:
\begin{align}
    F_\ell \left(\nu_\ell^{(q-1)}\right) \equiv  \frac{1}{2}  \int_0^\infty \dd \omega \, \mu^{(p+1)} (\omega)\,  \log \left( \frac{\omega^2 + \left(\nu_\ell^{(q-1)}\right)^2}{\tilde \Lambda^2 R^2}  \right)  \ ,
\end{align}
and $\nu_\ell^{(q-1)} = \Delta^\ell - \frac{p}{2}$ as before.
For $q=2$, we have to take a summation from $\ell=-\infty$ to $\infty$.
By applying the residue method \eqref{diff_free_energy}, we obtain
\begin{align} \label{residue_HS}
    \partial_{\nu_\ell} F_\ell (\nu_\ell) -\partial_{\nu_\ell} F_\ell (-\nu_\ell) = 
   - c_{p+1}\, 
    \nu_\ell \, \sin (\pi \nu_\ell)\, \Gamma \left( \frac{d-1}{2} \pm \nu_\ell \right) \ .
\end{align}
Hereafter we omit the superscript $(q-1)$ in $\nu_\ell^{(q-1)}$ to simplify expressions. 
In the following, we will compute the difference of the free energies between $\Delta_\text{D}$ and $\Delta_\text{N}$.

\paragraph{$q=2$ case:}

The allowed boundary conditions are classified in \eqref{bc_q=2}.
The difference of the free energies between $\Delta_+$ and $\Delta_{\text{N}1}$ comes from the $\nu_1$ mode, where the former has $\nu_1 = 1$ ($\Delta_+^{\ell = 1} = \frac{p}{2} + 1$) while the latter has $\nu_1 = -1$ ($\Delta_-^{\ell=1} = \frac{p}{2} -1$).
By applying the residue method \eqref{residue_HS} to the $\nu_1$ mode and integrating from $\nu_1 = 0$ to $\nu_1 = 1$, we obtain
\begin{align}
\begin{aligned}
         F_1(\nu_1=1) - F_1(\nu_1=-1) & = \frac{1}{\sin \left( \frac{\pi\,p}{2}\right)\,\Gamma (p+1)} \int_{0}^{1} \! \dd \nu_1 \, \nu_1 \, \sin (\pi \nu_1)\, \Gamma \left( \frac{p}{2} \pm \nu_1 \right)\\
         & = - F_\text{univ}[\BS^p] \ .    
\end{aligned}
\end{align}
In the last line, we use the integral expression of the sphere free energy \eqref{free_energy_sphere}.
This implies 
\begin{align} \label{dif_FE_q=2a}
         F_{\Delta_\text{D}}[\BH^{p+1} \times \BS^1] -  F_{\Delta_{\text{N}1} }[\BH^{p+1} \times \BS^1] = - F_\text{univ}[\BS^p] \ .
\end{align}
In the same way, the difference of the free energies between the $\Delta_{\text{N}1}$ and $\Delta_{\text{N}2}$ boundary conditions reads
\begin{align} \label{dif_FE_q=2b}
         F_{\Delta_{\text{N}1}}[\BH^{p+1} \times \BS^1] -  F_{\Delta_{\text{N}2} }[\BH^{p+1} \times \BS^1] = - F_\text{univ}[\BS^p] \ .
\end{align}
Note that the difference of the free energies equals the sphere free energy of a $p$-dimensional free scalar field.
This result conforms with the fact that the Neumann boundary conditions for $q=2$ are trivial in the sense that the defect operator saturates the unitarity bound and becomes a free field.

\paragraph{$q=3$ case:}

In this case the difference of the free energies between the two boundary conditions \eqref{bc_q=3,4} comes from the  $\nu_{0}$ mode only, where the Dirichlet boundary condition has $\nu_0 = \frac{1}{2}$ $(\Delta_+^{\ell = 0} = \frac{p+1}{2})$ while the Neumann boundary condition has $\nu_0 = -\frac{1}{2}$ ($\Delta_-^{\ell = 0} = \frac{p-1}{2}$).
By applying the residue method \eqref{residue_HS} to the $\nu_0$ mode and integration from $\nu_0 = 0$ to $\frac{1}{2}$, the difference of the free energies is 
\begin{align}
\begin{aligned}
        F_0\left(\nu_0=\frac{1}{2}\right) - F_0\left(\nu_0= -\frac{1}{2}\right) 
        &= \frac{1}{\sin \left( \frac{\pi\,p}{2}\right)\,\Gamma (p+1)} \int_{0}^{\frac{1}{2}} \! \dd \nu_{0} \, \nu_{0}\, \sin (\pi \nu_{0})\, \Gamma \left( \frac{p}{2} \pm \nu_{0} \right) \\
        & = F_{\Delta_+}[\BH^{p+1}] - F_{\Delta_-}[\BH^{p+1}] \ .
\end{aligned}
\end{align}
In the final line, we use the integral expression of the difference of the free energy on $\BH^{p+1}$ \eqref{diff_FE_hyperbolic}.
Hence we conclude that 
\begin{align} \label{dif_FE_q=3}
    F_{\Delta_\text{D}}[\BH^{p+1}\times \BS^2] - F_{ \Delta_\text{N}}[\BH^{p+1}\times \BS^2] = F_{\Delta_+}[\BH^{p+1}] - F_{\Delta_-}[\BH^{p+1}] \ .
\end{align}

\paragraph{$q=4$ case:}

The difference of the free energies between the two boundary conditions \eqref{bc_q=3,4} comes from the $\nu_{0}$ mode for $p\geq 2$, where the Dirichlet condition has $\nu_0 = 1$ $(\Delta_+^{\ell = 0} = \frac{p}{2} +1)$ while the Neumann condition has $\nu_0 = -1$ ($\Delta_-^{\ell = 0} = \frac{p}{2}-1$).
By applying the residue method \eqref{residue_HS} to the $\nu_0$ mode and integrating from $\nu_0 = 0$ to $1$, we find
\begin{align}
\begin{aligned}
        F_0(\nu_0=1) - F_0(\nu_0= -1) 
            &= \frac{1}{\sin \left( \frac{\pi\,p}{2}\right)\,\Gamma (p+1)} \int_{0}^{1} \! \dd \nu_{0} \, \nu_{0}\, \sin (\pi \nu_{0})\, \Gamma \left( \frac{p}{2} \pm  \nu_{0} \right)  \\
            &  = - F_\text{univ}[\BS^p] \ .    
\end{aligned}
\end{align}
Hence we conclude that 
\begin{align} \label{dif_FE_q=4}
    F_{\Delta_\text{D}}[\BH^{p+1}\times \BS^3] - F_{ \Delta_\text{N}}[\BH^{p+1}\times \BS^3] = - F_\text{univ}[\BS^p] \ .
\end{align}
As in $q=2$ the difference if given by the sphere free energy of a $p$-dimensional scalar field.
This also conforms with the saturation of the unitarity bound for the Neumann condition when $q=4$.

\paragraph{Free boundary condition:}

In this case we see from \eqref{defect_op_dimension} that the free boundary condition associated with a $p$-dimensional scalar Wilson loop exists for $q=p+2$ and $q\ge 3$ with $\Delta_-^{\ell=0} = 0$ while the Dirichlet condition has $\Delta_+^{\ell=0} = p$.
Thus, the difference of the free energies between the two is given by
\begin{align} \label{dif_FE_q>5}
\begin{aligned}
    F_{\Delta_\text{D}} [\BH^{p+1}\times \BS^{p+1}]- F_{\Delta_\text{F}} [\BH^{p+1}\times \BS^{p+1}] 
        & =F_0\left(\nu_0=\frac{p}{2}\right) - F_0\left(\nu_0= -\frac{p}{2}\right) \\
        & = \frac{1}{\sin \left(\frac{\pi\,p}{2}\right)\,\Gamma (p+1)} \int_{0}^\frac{p}{2} \! \dd \nu_{0} \, \nu_{0}\, \sin (\pi \nu_{0})\, \Gamma \left( \frac{p}{2}\pm  \nu_{0} \right)  \ .    
\end{aligned}
\end{align}
The integral of the right hand side suffers from the IR divergences due to the zero mode for the free boundary condition with $\nu_0 = -\frac{p}{2}$.\footnote{The defect free energy of a scalar Wilson loop in four dimension has a similar IR divergence, but is shown to be zero after an IR regularization in \cite{Lewkowycz:2013laa}. We do not know if such a regularization can be applied to our case.}

\subsection{Evidence for defect \texorpdfstring{$C$}{C}-theorem}\label{ss:testing_conjecture}

Let us compare the results in sections \ref{sec5.1}, \ref{sec5.2} and \ref{sec5.3} with the proposed $C$-theorem \eqref{D_theorem} in DCFT.
We anticipate that the difference of the defect free energies is invariant under Weyl transformations.
Since one can trigger the defect RG flow from the Neumann to Dirichlet boundary condition by the double trace deformation \cite{Witten:2001ua,Berkooz:2002ug,Gubser:2002zh,Gubser:2002vv,Hartman:2006dy,Diaz:2007an,Giombi:2013yva}
the difference of the free energies between the UV and IR fixed points is given by
\begin{align}
    \tilde{D}_\text{UV} - \tilde{D}_\text{IR} = - \sin \left( \frac{\pi\, p}{2}  \right) \left( F_{\Delta_\text{N}}[\BH^{p+1}\times \BS^{q-1}] - F_{\Delta_\text{D}}[\BH^{p+1}\times \BS^{q-1}] \right) \ .
\end{align}

For the free boundary condition, however, the double trace deformation does not lead to a defect RG flow to the Dirichlet boundary condition as the double trace operator has dimension zero and is proportional to the defect identity operator.
Presumably there are defect RG flows between the free and Dirichlet boundary conditions which we are not aware of, but
we concentrate on our consideration to the flow driven by the double trace deformation of the Neumann boundary condition with positive dimension $\Delta_\text{N} > 0$.
This leaves us the cases with $q\le 4$ in the following.
We will discuss the implication of the defect $C$-theorem \eqref{D_theorem} for the case with the free boundary conditions in section \ref{ss:discussion}.

\paragraph{$q=1$ case:}
In this case we consider the defect RG flow on the boundary of $\BH^d$ ($p=d-1$).
From \eqref{diff_FE_hyperbolic} we find
\begin{align}
    \tilde{D}_{\text{UV}} - \tilde{D}_{\text{IR}} =  \frac{1}{\Gamma (d)} \int_{0}^\frac{1}{2} \! \dd \nu \, \nu\, \sin (\pi \nu)\, \Gamma \left( \frac{d-1}{2}\pm \nu \right) \ , 
\end{align}
which is always positive and the monotonicity of the defect free energy holds for $p\ge 2$ ($d\ge 3$).

\paragraph{$q=2$ case:}

From \eqref{dif_FE_q=2a} and \eqref{dif_FE_q=2b} we obtain
\begin{align}
    \tilde{D}_{\Delta_{\text{N}1}} - \tilde{D}_{\Delta_\text{D}} 
        =
        \tilde{D}_{\Delta_{\text{N}2}} - \tilde{D}_{\Delta_{\text{N}1}} 
        = \tilde F[\BS^p] \ ,
\end{align}
where we used $\Delta_{\text{N}1},\Delta_{\text{N}2},\Delta_{\text{D}}$ instead of UV and IR and 
\begin{align}
    \tilde F[\BS^p] \equiv  \frac{1}{\Gamma (p+1)} \int_{0}^{1} \! \dd \nu \, \nu \, \sin (\pi \nu)\, \Gamma \left( \frac{p}{2} \pm \nu \right)
\end{align}
is positive for any $p\ge 2$ \cite{Giombi:2014xxa}.
From these equations, the monotonicity of the free energies for $p\ge 3$ follows from the positivity of the right hand side. 
Hence our $C$-theorem is satisfied in this case.

\paragraph{$q=3$ case:}

The relation \eqref{dif_FE_q=3} can be translated to 
\begin{align}
    \tilde D_\text{UV} [\BH^{p+1}\times \BS^2] - \tilde D_\text{IR}[\BH^{p+1}\times \BS^2] = \tilde D_\text{UV}[\BH^{p+1}] - \tilde D_\text{IR}[\BH^{p+1}] \ ,
\end{align}
where we make explicit the dependence of $\tilde D$ on the space.
Hence the monotonicity of the defect free energy amounts to that of the $q=1$ case.

\paragraph{$q=4$ case:}
Since \eqref{dif_FE_q=4} can be translated to
\begin{align}
    \tilde{D}_{\text{UV}} - \tilde{D}_{\text{IR}} 
    = \tilde F[\BS^p]\ ,
\end{align}
the monotonicity of the defect free energy holds for the same reason as in the $q=2$ case for $p \ge 3$.

\section{Discussion}\label{ss:discussion}

In this paper we classified a certain class of conformal defects in the free scalar theory as boundary conditions on $\BH^{p+1} \times \BS^{q-1}$.
Our results are consistent with the classification of the non-monodromy defects in \cite{Lauria:2020emq} carried out by other means.
We believe our methods for characterizing conformal defects as boundary conditions on $\BH^{p+1}\times \BS^{q-1}$ can be applied to the monodromy defects classified in \cite{Lauria:2020emq} as well.
As a special case twist operators of codimension-two were studied as a boundary condition on $\BH^{p+1} \times \BS^1$ in \cite{Belin:2013uta}.
It is also worthwhile to revisit the $O(N)$ model discussed in a recent paper \cite{Giombi:2020rmc} that admits various non-trivial boundary conditions and supersymmetric theories with defects \cite{Gukov:2006jk,Gukov:2008sn,Gomis:2009ir,Gomis:2009xg,Gomis:2011pf,Kapustin:2012iw,Drukker:2012sr,Nishioka:2013haa,Gaiotto:2014ina,Nishioka:2016guu,Hosomichi:2017dbc,Drukker:2017dgn,Hosseini:2019and,Bianchi:2019sxz,Agmon:2020pde,Drukker:2020atp,Chalabi:2020iie,Goto:2020per,Wang:2020seq,Gupta:2020eev} from the viewpoint of this paper. (See also \cite{David:2016onq,Gava:2016oep,David:2018pex,David:2019ocd,Pittelli:2018rpl,Longhi:2019hdh} for related works.)

It should be possible to extend our analysis to fields with spin.
For fermion, a non-trivial boundary condition is allowed, and we can consider an RG flow from Neumann to Dirichlet boundary condition.
We will report this result in \cite{Sato:2021eqo}.
For a symmetric traceless tensor with spin $s$, $\Delta_+$ always satisfies the unitarity bound 
\begin{align}
    \Delta \geq d+s-2 \ ,
\end{align}
and $\Delta_+$ for $\BH^d$ also saturates the unitarity bound. 
However, $\Delta_-$ always violates the unitarity bound, which implies that a non-trivial Neumann boundary condition never exists for higher spin fields.

By comparing our results with the classification by \cite{Lauria:2020emq} we observe that Dirichlet boundary condition corresponds to trivial (or no) defects while Neumann boundary condition to non-trivial defects.
Indeed, we verified this observation through the free energy calculations in some cases, which leads us to speculate that defects with Dirichlet boundary condition have a least $\tilde D$ under any RG flow.

In section \ref{sec5.3} we examined the defect RG flow triggered by the double trace deformation of the Neumann boundary condition with $\Delta_\text{N}>0$.
This restriction excludes the flows from the free boundary condition with zero mode $(\Delta_\text{F}=0)$ from our consideration.
While we are not aware of any defect RG flow between the free and Dirichlet boundary conditions one can still calculate the difference of the defect free energy from \eqref{dif_FE_q>5}:
\begin{align}
        \tilde D_\text{F} - \tilde D_\text{D}
        =\frac{1}{\Gamma (p+1)} \int_{0}^\frac{p}{2} \! \dd \nu_{0} \, \nu_{0}\, \sin (\pi \nu_{0})\, \Gamma \left( \frac{p}{2}\pm \nu_{0} \right) \ .
\end{align}
The integral diverges for odd $p$ due to the IR divergence from the zero mode while it is positive and finite for $p=4m - 2~(m= 1,2,\cdots)$ and negative and finite for $p=4m$.
In this case the defect $C$-theorem implies that the free boundary condition may be a UV fixed point of some defect RG flow for $p=4m - 2$ while it may be an IR fixed point for $p = 4m$. (One cannot draw any implication for odd $p$ due to the IR divergence.) 
We leave further investigation on this direction as a future work. 

The entanglement entropy of a spherical region on flat space $\mathbb{R}^d$ can be mapped to the free energy on $\BH^{d-1} \times \BS^1$ by the Casini-Huerta-Myers map \cite{Casini:2011kv}.
In this context, the boundary condition on the entangling surface, or equivalently the boundary condition on $\BH^{d-1} \times \BS^1$, has not been clarified explicitly.
However, our results show that the boundary condition on $\BH^{d-1} \times \BS^1$ changes the universal parts of the free energy, and this implies that we should be careful in the boundary condition in the entanglement entropy.
In \cite{Kobayashi:2018lil} we derived a universal relation between the defect free energy and defect entropy.
They differ by a term proportional to the one-point function of the stress tensor in the presence of defects, so one can calculate the defect entropy from our results in this paper given the one-point function, without relying on conventional methods such as the replica trick.

The free energies for the Neumann boundary conditions were obtained somewhat indirectly as differences from the Dirichlet cases.
This was enough for us to check if the defect $C$-theorem \eqref{D_theorem} holds under the defect RG flow as we assumed the difference of the free energies is invariant under conformal maps from $\BS^d$ to $\BH^{p+1}\times \BS^{q-1}$.
Nonetheless it is desirable to have a precise relation between the defect free energy on $\BS^d$ and the free energy on $\BH^{p+1}\times \BS^{q-1}$.
A most naive guess would be
    \begin{align}\label{wrong_relation}
        D = \log\,|\langle \CD^{(p)}\rangle| \overset{?}{=} F[\BH^{p+1}\times \BS^{q-1}] - F[\BS^d] \ .
    \end{align}
However, this relation does not hold in general as the bulk anomalies are canceled out in the left hand side while there can remain a bulk anomaly term in the right hand side.
For instance, $F[\BS^d]$ should have bulk anomalies when $d$ is even for any $d$.
On the other hand, there are no bulk anomalies in $F[\BH^{p+1}\times \BS^{q-1}]$ when $p$ is even as seen from \eqref{fren_general_form}  (only defect anomalies are there).\footnote{The absence of the bulk anomaly on $\BH^{p+1}\times \BS^{q-1}$ follows from the fact that the anomaly coefficient is proportional to the Euler characteristic of the manifold and $\chi[\BH^{p+1}\times \BS^{q-1}] = 0$ as $\chi[\BH^{p+1}]=0$ for even $p$.}
Hence there remains the bulk anomaly in the right hand side of \eqref{wrong_relation}.
Finding a correct relation between $D$ and $F[\BH^{p+1}\times \BS^{q-1}]$ should be of interest.

In BCFT$_d$ with even $d$, the conformal anomalies in the bulk theory have boundary terms dictated by boundary central charges \cite{Fursaev:2015wpa,Solodukhin:2015eca,Herzog:2015ioa,Fursaev:2016inw,Herzog:2017xha,Herzog:2019bom}.
The free energy has a logarithmic divergence whose coefficient is completely fixed by the geometry of the boundary such as the extrinsic curvature at least in lower dimensions \cite{Fursaev:2015wpa,Solodukhin:2015eca}.
In DCFT, we regard conformal defects as boundary conditions on $\BH^{p+1}\times \BS^{q-1}$, so we may view $(\partial \BH^{p+1})\times \BS^{q-1}$ as the codimension-one boundary and are tempted to apply the boundary anomaly formula \cite{Fursaev:2015wpa,Solodukhin:2015eca} to the present case.
In a few cases, we computed the defect anomaly coefficients from the boundary anomaly formula, but we were not able to reproduce our results correctly.
We suspect the boundary anomaly formula may not be applicable to manifolds with boundary which is a product manifold.
We hope to address this issue in future.

\acknowledgments

We would like to thank A.\,O'Bannon with his research group, C.\,P.\,Herzog, D.\,Rodriguez-Gomez, J.\,G.\,Russo and M.\,Watanabe for valuable discussion and correspondences.
We also thank J.\,S.\,Dowker for providing us proofs of several identities conjectured in the earlier version of the paper.
The work of T.\,N. was supported in part by the JSPS Grant-in-Aid for Scientific Research (C) No.19K03863 and the JSPS Grant-in-Aid for Scientific Research (A) No.16H02182. 
The work of Y.\,S. was supported by the National Center of Theoretical Sciences (NCTS).

\newpage 

\appendix

\section{List of tables}
\label{app_table}

\begin{table}[ht]
    \centering
    \begin{tabularx}{\linewidth}{cYYYYY}
        \toprule
          & $\cdot$ & $\BS^1$  & $\BS^2$             & $\BS^3$       & $\BS^4$           \\ \hline
         $\cdot$ &  & $0$    & $\frac{1}{3}$  &   $0$  & $-\frac{1}{90}$   \\
         $\BH^2$ & $\frac{1}{6}$ & $0$  & $-\frac{1}{90}$   & $0$       & $\frac{1}{756}$    \\
         \rowcolor{gray!10} $\BH^3$ & $-\frac{1}{48}$ & $-\frac{1}{360}$  & $0$               & $\frac{1}{1512}$ &   $0$  \\
         $\BH^4$ & $-\frac{1}{180}$ & $0$  & $\frac{1}{756}$   & $0$       & $-\frac{23}{113400}$     \\
         \rowcolor{gray!10} $\BH^5$ & $\frac{17}{11520}$ & $\frac{1}{3360}$  & $0$               & $-\frac{163}{1814400}$  &   $0$   \\
         $\BH^6$ & $\frac{1}{1512}$ & $0$   & $-\frac{23}{113400}$   & $0$       & $\frac{263}{7484400}$  \\
         \rowcolor{gray!10} $\BH^7$ & $-\frac{367}{1935360}$ & $-\frac{79}{1814400}$   &   $0$   &   $\frac{1753}{119750400}$   &   $0$  \\
         $\BH^8$ & $-\frac{23}{226800}$ & $0$   & $\frac{263}{7484400}$   & $0$       & $-\frac{133787}{20432412000}$   \\
         \rowcolor{gray!10} $\BH^9$ & $\frac{27859}{928972800}$ & $\frac{1759}{239500800}$    &   $0$   &   $-\frac{3436133}{1307674368000}$   &   $0$     \\
         $\BH^{10}$ & $\frac{263}{14968800}$ & $0$   & $-\frac{133787}{20432412000}$   & $0$       & $\frac{157009}{122594472000}$ \\
         \bottomrule
    \end{tabularx}
    
    \vspace{1cm}
    
    \begin{tabularx}{\linewidth}{cYYYY}
        \toprule
          & $\BS^5$ & $\BS^6$   & $\BS^7$ & $\BS^8$  \\ \hline
         $\cdot$ & $0$  & $\frac{1}{756}$ & $0$ & $-\frac{23}{113400}$  \\
         $\BH^2$  &  $0$   & $-\frac{23}{113400}$   &   $0$ & $\frac{263}{7484400}$  \\
         \rowcolor{gray!10} $\BH^3$  &   $-\frac{41}{362880}$ &   $0$  &  $\frac{491}{23950080}$ & $0$ \\
         $\BH^4$ & $0$   & $\frac{263}{7484400}$ &   $0$ & $-\frac{133787}{20432412000}$\\
         \rowcolor{gray!10} $\BH^5$  &  $\frac{263}{14968800}$  &   $0$ & $-\frac{323117}{93405312000}$ & $0$ \\
         $\BH^6$   & $0$   & $-\frac{133787}{20432412000}$ & $0$ & $\frac{157009}{122594472000}$ \\
         \rowcolor{gray!10} $\BH^7$  &  $-\frac{403873}{130767436800}$   &   $0$ &   $\frac{157009}{245188944000}$ & $0$ \\
         $\BH^8$ & $0$   & $\frac{157009}{122594472000}$  &   $0$ & $ -\frac{16215071}{62523180720000}$ \\
         \rowcolor{gray!10} $\BH^9$ & $\frac{9134093}{15692092416000}$   &   $0$ &   $-\frac{286034933}{2286562037760000}$ & $0$ \\
         $\BH^{10}$ & $0$   & $ -\frac{16215071}{62523180720000}$  &   $0$ & $\frac{2689453969}{49893498214560000}$\\
         \bottomrule
    \end{tabularx}
    \caption{The bulk anomalies $A[\BH^{p+1}\times \BS^{q-1}]$ and the defect anomalies $\mathcal{A}[\BH^{p+1}\times \BS^{q-1}]$  (shaded) on $\BH^{p+1}\times \BS^{q-1}$ with Dirichlet boundary conditions.  }
     \label{tab:F_p_q}
\end{table}

\begin{table}[ht]
    \centering
        \begin{tabularx}{\linewidth}{cY}
        \toprule
        $\CM$ & $F_\text{fin}[\CM]$ \\
        \hline
        $\BS^2$  & $-\frac{1}{4}\,\log (2\pi) -2\, \zeta'(-1) - \frac{1}{2}\,\partial_s \zeta_\text{H}(0,0)$\qquad (\text{IR divergent}) \\ 
        $\BS^3$ & $\frac{1}{8}\,\log 2-\frac{3}{16 \pi^2}\, \zeta(3)$ \\ 
        $\BS^4$ & $-\frac{1}{6}\,\zeta'(-1) - \frac{1}{3}\,\zeta'(-3)$ \\
        $\BS^5$ & $-\frac{1}{128}\,\log 2 -\frac{1}{128\pi ^2}\,\zeta (3)+\frac{15}{256 \pi ^4}\,\zeta(5)$\\
        $\BS^6$ & $\frac{1}{60}\,\zeta'(-1) - \frac{1}{60}\,\zeta'(-5)$\\
        $\BS^7$ & $\frac{1}{1024}\,\log 2 + \frac{41}{30720 \pi ^2}\,\zeta(3)-\frac{5}{2048 \pi^4}\,\zeta (5)-\frac{63}{4096 \pi^6}\,\zeta(7)$ \\
        $\BS^8$ & $-\frac{1}{420}\,\zeta'(-1) + \frac{1}{720}\,\zeta'(-3) + \frac{1}{720}\,\zeta'(-5) - \frac{1}{2520}\,\zeta'(-7)$\\
        $\BS^9$ & $-\frac{5}{32768}\,\log 2 -\frac{397}{1720320 \pi^2}\, \zeta(3) +\frac{1}{32768\pi ^4}\,\zeta (5) +\frac{63}{32768 \pi^6}\, \zeta(7)+\frac{255}{65536 \pi ^8}\, \zeta(9)$ \\
        $\BS^{10}$ & $\frac{1}{2520}\, \zeta'(-1) - \frac{31}{90720}\, \zeta'(-3) - \frac{1}{8640}\,\zeta'(-5) + \frac{1}{15120}\,\zeta'(-7)- \frac{1}{181440}\,\zeta'(-9)$ \vspace*{0.2cm}
        \\
        \hline
        $\BH^2$ & $-\frac{1}{4}\,\log(2\pi) - \zeta'(-1)$ \\
        $\BH^3$ & $0$\\
        $\BH^4$ & $-\frac{1}{12}\,\zeta'(-1) + \frac{1}{4}\,\zeta'(-2) - \frac{1}{6}\,\zeta'(-3)$ \\
        $\BH^5$ & $0$\\
        $\BH^6$ & $\frac{1}{120}\,\zeta'(-1) - \frac{1}{48}\,\zeta'(-2) + \frac{1}{48}\,\zeta'(-4) - \frac{1}{120}\,\zeta'(-5)$ \\
        $\BH^7$ & $0$\\
        $\BH^8$ & 
        \begin{minipage}{15cm}
            \centering
            $-\frac{1}{840}\,\zeta'(-1) + \frac{1}{360}\,\zeta'(-2) +
              \frac{1}{1440}\,\zeta'(-3)
              - \frac{1}{288}\,\zeta'(-4)$\\
              \vspace*{0.2cm}
              $+ \frac{1}{1440}\,\zeta'(-5) + \frac{1}{1440}\,\zeta'(-6) - \frac{1}{5040}\,\zeta'(-7)$ 
        \end{minipage}
        \\
        $\BH^9$ & $0$ \\
        $\BH^{10}$ & 
            \begin{minipage}{15cm}
                \centering
            $\frac{1}{5040}\, \zeta'(-1) - \frac{1}{2240}\, \zeta'(-2) -  \frac{31}{181440}\, \zeta'(-3) + \frac{7}{11520}\, \zeta'(-4)  - \frac{1}{17280}\, \zeta'(-5)$ \\
            \vspace*{0.2cm}
            $- \frac{1}{5760}\, \zeta'(-6) + \frac{1}{30240}\, \zeta'(-7)  + \frac{1}{80640}\, \zeta' (-8) - \frac{1}{362860}\,\zeta'(-9)$
            \end{minipage}
            \vspace*{0.3cm}
            \\ \hline
        Even $p$ & $F_\text{fin}[\BH^{p+1}\times \BS^{q-1}]=0$ \\
        Odd $p$ & $F_{\text{fin}}[\BS^{d}] = F_{\text{fin}}[\BH^{2k} \times \BS^{d-2k}]$\quad 
for $k=1,\cdots ,\ceil{d/2}-1$\\
    \bottomrule
    \end{tabularx}
    \caption{Table of the finite parts of $F_\text{fin}[\BS^d]$, $F_\text{fin}[\BH^d]$, and $F_\text{fin}[\BH^{p+1}\times \BS^{q-1}]$ with Dirichlet boundary conditions.}
     \label{tab:Fin_p_q}
\end{table}

\clearpage

\begin{table}[ht]
    \centering
    \begin{tabularx}{\linewidth + 1cm}{ccY}
        \toprule
         $d$ & $A[\BHS_\pm^d]$ & $ F_\text{fin}[\BHS_\pm^d] - \frac{1}{2}\, F_\text{fin}[\BS^d]$ \\ \hline
         $2$ & $\frac{1}{6}$ & $\mp \left(\frac{1}{8}\,\log (2\pi) - \frac{1}{4}\,\partial_s \zeta_\text{H} (0,0)  \right)$\qquad (\text{IR divergent}) \\
         $3$ & $\mp \frac{1}{48}$ & $\mp \left( \frac{1}{48}\,\log 2 +\frac{1}{4}\,\zeta' (-1)   \right)$\\  
         $4$ & $-\frac{1}{180}$ & $\mp \frac{1}{16\pi^2}\, \zeta (3)$ \\
         $5$ &
         $\pm \frac{17}{11520}$ & $\pm \left( \frac{11}{11520}\,\log 2 +\frac{1}{96}\,\zeta' (-1)  -\frac{7}{96}\,\zeta' (-3)  \right)$ \\         
         $6$ &  $\frac{1}{1512}$ & $\pm \left( \frac{1}{192\pi^2}\, \zeta (3) + \frac{1}{64\pi^4}\, \zeta (5)\right)$\\
         $7$ &
         $\mp \frac{367}{1935360}$ & $\mp \left( \frac{211}{1935360}\,\log 2 +\frac{3}{2560}\,\zeta' (-1)  - \frac{7}{768}\,\zeta' (-3)  + \frac{31}{7680}\,\zeta' (-5) \right)$\\
         $8$ &
         $-\frac{23}{226800}$ & $\mp \left( \frac{1}{1440\pi^2}\, \zeta (3) + \frac{1}{384\pi^4}\, \zeta (5)+\frac{1}{256\pi^6}\, \zeta (7)\right)$ \\
         $9$ &
         $\pm \frac{27859}{928972800}$ & $\pm \left( \frac{15157}{928972800}\,\log 2 +\frac{5}{28672}\,\zeta' (-1)  -\frac{259}{184320}\,\zeta' (-3)  + \frac{31}{36864}\,\zeta' (-5) -\frac{127}{1290240}\,\zeta' (-7) \right)$\\
         \bottomrule
    \end{tabularx}
    \caption{The anomaly and the finite parts of the free energies on $\BHS^d$.}
    \label{tab:hemi}
\end{table}

\section{Useful formulas}\label{app:formula}

In this appendix, we summarize useful formulas of the zeta function and the Hurwitz zeta function.
Throughout this Appendix, we assume that $n$ be a non-negative integer $(n=0,1,2,3,\cdots)$ and $m$ be a positive integer $(m=1,2,3,4\cdots)$.

\paragraph{Zeta function}
At specific points the Riemann zeta function takes the values:
\begin{align}
    \zeta (0)& = - \frac{1}{2} \ , \\
    \zeta (-2m) &= 0 \ , \label{zeta_values} \\
    \zeta (2m) &= \frac{(-1)^{m-1}\,2^{2m-1}\,\pi^{2m}}{(2m)!}\, B_{2m} \ .\label{hurwitz_to_bernoulli}
\end{align}
More generally the zeta function satisfies the relation:
\begin{align}
    \zeta (s) = 2^s\, \pi^{s-1}\, \sin \left( \frac{\pi s}{2} \right)\, \Gamma (1-s)\, \zeta (1-s) \ .
\end{align}
The derivatives of the zeta function at non-positive integer points are given by
\begin{align}
    \zeta '(0) & = -\frac{1}{2}\,\log (2\pi) \ ,  \label{zeta_derivatice_zero} \\
    \zeta ' (-2m) &= \frac{(-1)^m}{2^{2m+1}\,\pi^{2m}}\, \Gamma (2m+1) \, \zeta (2m+1) \ , \label{hurwitz_der_to_zeta} \\
    \zeta' (1-2m) &= (-1)^{m+1}\, \frac{2\, \Gamma (2m)}{(2\pi)^{2m}} \left[ (\psi (2m)-\log (2\pi))\,\zeta (2m)+\zeta' (2m) \right] \ .
    \label{relation_derivative_hurwitz}
\end{align} 

\paragraph{Hurwitz zeta function}

The Hurwitz zeta function $\zeta_\text{H} (s,a)$ has two arguments, $s$ and $a$.
To distinguish derivatives of the Hurwitz zeta function, we explicitly write the differentiation variable such as $\partial_s \zeta_\text{H} (s,a)$. 

For specific values of $a$ the Hurwitz zeta function reduces to the Riemann zeta function:
\begin{align}\label{hurwitz_to_zeta}
    \begin{aligned}
    \zeta_\text{H} (s,1) &= \zeta (s) \ ,\\
    \zeta_\text{H} (s,0) &= 
    \begin{dcases}
             \zeta (s) &\qquad s<0 \ , \\
             \frac{1}{2} &\qquad s=0 \ .
    \end{dcases} 
    \end{aligned}
\end{align}
The argument $a$ of the Hurwitz zeta functions can be shifted by a positive integer $m$ by the relation:
\begin{align}\label{hurwitz_identity_1}
    \zeta_\text{H} (s, m+ a) = \zeta_\text{H} (s, a) - \sum_{k=0}^{m-1}(k+a)^{-s} \ .
\end{align}
At special values of $s$ it derives
\begin{align}\label{hurwitz_recursion1}
    \zeta_\text{H} (-n,m) & = \zeta (-n)-\sum_{k=1}^{m-1}  k^{n} \ , \\
    \zeta_\text{H} (-n,m-1) & = \zeta (-n)-\sum_{k=1}^{m-2} k^{n} + \delta_{m,1}\, \delta_{n,0} \ .
    \label{hurwitz_recursion2}
\end{align}
Other useful identities are
\begin{align}\label{hurwitz_to_zeta_half}
    \zeta_\text{H} \left(s, \frac{1}{2} \right) &= (2^s-1)\, \zeta (s) \ , \\
    \zeta_\text{H} (-n,a)& = -\frac{B_{n+1} (a)}{n+1} \ .
\end{align}
The derivative with respect to $s$ at special values are given by
\begin{align}
    \partial_s \zeta_\text{H} (0,a) & = \log\, \Gamma (a) -\frac{1}{2}\, \log (2\pi) \ , \\
      \partial_s \zeta_\text{H} \left( s, \frac{1}{2}\right) &= 2^s\, \log 2\, \zeta (s) +(2^s-1)\, \zeta' (s) \ ,\\
      \partial_s \zeta_\text{H} \left(1-2m,\tfrac{1}{2}\right)& =-\frac{B_{2m}}{m\cdot 4^{m}}\,\log 2-\frac{
2^{2m-1}-1}{2^{2m-1}}\,\zeta'\left(1-2m\right) \ .
\end{align}

\paragraph{Computation of the derivative of the Hurwitz zeta function}

From the formula \cite[25.12.13]{NIST:DLMF},
\begin{align}
    \mathrm{Li}_s (\mathrm{e}^{2\pi \i a}) + \mathrm{e}^{\pi \i s}\,\mathrm{Li}_s (\mathrm{e}^{-2\pi \i a}) = \frac{(2\pi)^s\, \mathrm{e}^{\frac{\pi \i s}{2}}}{\Gamma (s)}\, \zeta_\text{H} (1-s,a)\ ,
\end{align}
which holds for $\mathrm{Re}\, s>0$, $0<\mathrm{Re}\, a \leq 1$, $\mathrm{Im}\, a>0$, or $\mathrm{Re}\, s>1$, $0<\mathrm{Re}\, a \leq 1$, $\mathrm{Im}\, a=0$, the Hurwitz zeta function can be written as
\begin{align}
    \zeta_\text{H} (s,a) = \frac{\Gamma (1-s)}{(2\pi)^{1-s}} \, \mathrm{e}^{\frac{\pi \i (s-1)}{2}} \left( \mathrm{Li}_{1-s} (\mathrm{e}^{2\pi \i a}) + \mathrm{e}^{\pi \i (1-s)}\, \mathrm{Li}_{1-s} (\mathrm{e}^{-2\pi \i a})\right) \ ,
\end{align}
which holds for $\mathrm{Re}\, s<1$, $ \mathrm{Im}\, a>0$, or $\mathrm{Re}\, s<0$, $\mathrm{Im}\, a=0$.
By taking the derivative with respect to $s$ and replace $s$ with a negative integer $-n$ ($n\geq 0$), we obtain
\begin{align} \label{hurwitz_to_polylog}
    \partial_s \zeta_\text{H} (-n,a) + (-1)^n\, \partial_s \zeta_\text{H} (-n,1-a) 
    & = \frac{\Gamma (n+1) }{(2\pi \i)^n}\,  \mathrm{Li}_{n+1} (\mathrm{e}^{2\pi \i a}) + \pi \i\,  \frac{B_{n+1}(a)}{n+1} \ .
\end{align}

\section{Derivation of (\ref{hyp_even_der_zeta})}
\label{app1}

In this appendix, we give a detailed derivation of \eqref{hyp_even_der_zeta}.
Instead of performing the integral over $\omega$ directly, we take a derivative with respect to $\nu$ and integrate the obtained derivative in terms of $\nu$.
The same calculation can be found in Appendix A of \cite{Caldarelli:1998wk} (see also \cite{Camporesi:1991nw} and Appendix A of \cite{Giombi:2014iua}).
The derivative of $f_k (\nu)$ is given by
\begin{align} 
    \partial_\nu f_k (\nu) &= 2\, \nu g_{k} (\nu) \ , \\
    g_{k} (\nu) &\equiv \int_0^\infty \dd\omega\, \frac{\omega^{2k+1}}{ (\mathrm{e}^{2\pi\omega} + 1)(\omega^2 + \nu^2)} \ .
\end{align}
Since $g_{k} (\nu)$ satisfies the recursion relation,
\begin{align}
    g_{k} (\nu) &= -\nu^2\,  g_{k-1} (\nu) + \frac{2^{2k}-2}{(4\pi)^{2k}}\, \Gamma (2k)\, \zeta (2k) \ , \\
    g_{0} (\nu) & = \frac{1}{2} \psi \left( \nu +\frac{1}{2} \right) -\frac{1}{2}\, \log \nu \ , 
\end{align}
where $\psi (x)$ is a polygamma function, the general solution can be easily obtained:
\begin{align}
    g_{k} (\nu) = (-\nu^2)^{k} \left[ \frac{1}{2}\psi \left( \nu +\frac{1}{2} \right) -\frac{1}{2}\, \log \nu + \sum_{m=1}^{k} \frac{2^{2m}-2}{(4\pi)^{2m}}\,(-\nu^2)^{-m}\, \Gamma (2m)\, \zeta (2m) \right] \ .
\end{align}
By integrating $g_{k} (\nu)$ from $0$ to $\nu$, we obtain $f_k(\nu)$ as 
\begin{align} \label{fknu}
    \begin{aligned}
    f_k(\nu) 
    & = (-1)^k\,\int_0^\nu \! \dd \mu \,   \mu^{2k+1} \psi \left( \mu +\frac{1}{2} \right) + \frac{(-1)^{k+1}}{4(k+1)^2}\,\nu^{2k+2}\, \left(2(k+1)\log \nu-1\right)  \\
    & \qquad \qquad  + \sum_{m=1}^{k} \frac{(-1)^{k-m}}{k-m+1}  \, \frac{2^{2m}-2}{(4\pi)^{2m}}\, \Gamma (2m)\, \zeta (2m)\, \nu^{2k-2m+2}  + f_k (0) \ .
    \end{aligned}  
\end{align}
For $k=0$, the term involving the summation of $m$ should be omitted.
The remaining term $f_k(0)$ can be computed as 
\begin{align}
    \begin{aligned} \label{fk0}
    f_k (0) & = (-1)^{k}\,(1-2^{-2k-1})\, \zeta' (-2k-1) 
         + 2^{-2k-1}\, \log 2\,   \frac{|B_{2k+2}|}{2k+2} \ ,
    \end{aligned}
\end{align}
where we use \eqref{hurwitz_to_bernoulli}, \eqref{zeta_derivatice_zero},  \eqref{relation_derivative_hurwitz} and 
\begin{align}
   H_{2k+1} - \gamma -\psi (2k+2)=0 \ .
\end{align}
Using Theorem 4.3 in \cite{espinosa2001some}, the remaining integral in \eqref{fknu} can be performed 
\begin{align}
\label{c9}
\begin{aligned}
    \int_0^{\nu} \! &\dd \mu \,  \mu^{2k+1} \psi \left( \mu + \frac{1}{2} \right) \\
        & = \sum_{j=0}^{2k+1}\, (-1)^j \,\binom{2k+1}{j}\, \nu^{2k+1-j} \left( \zeta_\text{H}' \left(-j,\nu+\frac{1}{2} \right) - H_{j}\, \frac{B_{j+1} \left( \nu +\frac{1}{2} \right)}{j+1}\right) 
        \\
        & \qquad - 2^{-2k-1} \log 2 \, \frac{B_{2k+2}}{2k+2} -(1-2^{-2k-1}) \left( \zeta' (-2k-1) - H_{2k+1}\,  \frac{B_{2k+2}}{2k+2} \right) \ ,
\end{aligned}
\end{align}
where $\zeta_\text{H}' (-j,\nu+1/2) = \partial_s \zeta_\text{H}' (s,\nu+1/2)|_{s\to -j}$.
Specifically, the integral with $\nu =1/2$ becomes
\begin{align}
\begin{aligned}
    \int_0^\frac{1}{2} \! &\dd \mu \,  \mu^{2k+1} \psi \left( \mu + \frac{1}{2} \right) \\
        & = \sum_{j=0}^{2k+1}\,\frac{(-1)^j}{2^{2k+1-j}}\,\binom{2k+1}{j}\,\left( \zeta'(-j) + H_{j}\,\zeta(-j)\right) - 2^{-2k-1} \log 2 \, \frac{B_{2k+2}}{2k+2}
        \\
        & \qquad -(1-2^{-2k-1}) \left( \zeta' (-2k-1) - H_{2k+1} \, \frac{B_{2k+2}}{2k+2} \right) \ ,
\end{aligned}
\end{align}
and the derivative of the zeta function is given by
\begin{align} 
\begin{aligned}
		\partial_s \zeta_{\BH^d} \left(0, \frac{1}{2} \right)
			&=
			c_d\, \sum_{k=0}^{\frac{d}{2}-1}\,(-1)^k\beta_{k,d} \left[ -\frac{ 2^{-2k-2}}{k+1} H_{2k+1} - \sum_{m=1}^{k} \frac{2^{-2k-2}(2^{2m}-2)}{k-m+1}      \frac{B_{2m}}{2m} \right. \\
			&  \quad + \sum_{j=0}^{2k+1}\,\frac{(-1)^j}{2^{2k-j}}\,\binom{2k+1}{j}\,\left( \zeta'(-j) + H_{j}\,\zeta(-j)\right) 
        \left. + (1-2^{-2k-1})\,  H_{2k+1} \, \frac{B_{2k+2}}{k+1} 
    \right] \ .
    \end{aligned}
\end{align}
Now we would like to show a sum of the terms except $\zeta'(-j)$ in the bracket vanishes,
\begin{align} \label{conjecture1}
\begin{aligned}
    & -\frac{ 2^{-2k-2}}{k+1} H_{2k+1} - \sum_{m=1}^{k} \frac{2^{-2k-2}(2^{2m}-2)}{k-m+1}   \frac{B_{2m}}{2m}  \\
    & \qquad + \sum_{j=0}^{2k+1}\,\frac{(-1)^j}{2^{2k-j}}\,\binom{2k+1}{j}\, H_{j}\,\zeta(-j)
         + (1-2^{-2k-1})  H_{2k+1}  \frac{B_{2k+2}}{k+1} =0 \ .
    \end{aligned}
\end{align}
For $k=0$, the summation term $\sum_{m=1}^{k}$ should be omitted.
We confirmed \eqref{conjecture1} up to $k=100$ numerically. However, we do not know a proof of \eqref{conjecture1}.
The coefficient of $\zeta' (0)$ also vanishes for $d \geq 4$ because the coefficient is proportional to  $\Gamma (d/2)/\Gamma (2-d/2)$.

In total we obtain the derivative of the zeta function as 
\begin{align}
		\partial_s \zeta_{\BH^d} \left(0, \frac{1}{2} \right)
			&=
			c_d\, \sum_{k=0}^{\frac{d}{2}-1}\,(-1)^k\beta_{k,d}\, \sum_{j=1}^{2k+1}\,\frac{(-1)^j}{2^{2k-j}}\,\binom{2k+1}{j}\,  \zeta'(-j) - \delta_{d,2}\, \zeta'(0) \ . 
\end{align}

\bibliographystyle{JHEP}
\bibliography{Defect}

\end{document}